\documentclass[leqno,12pt,a4paper]{article}
\usepackage{setspace}
\usepackage{latexsym}
\usepackage{amsmath}
\usepackage{amssymb}
\usepackage{mathtools}
\usepackage{bbm}
\usepackage{amsthm}
\usepackage{siunitx} % \Si{}{} 
\usepackage[misc]{ifsym}

\usepackage[]{algpseudocode}
\usepackage[]{algorithm2e}
\SetKwComment{Comment}{/*}{*/}
\usepackage{comment}
\usepackage{graphicx}
\usepackage[colorlinks=true, allcolors=blue]{hyperref}
\usepackage{cancel}
\usepackage{mwe}
\usepackage{realhats}
\usepackage{amsbsy}
\usepackage{mathabx}
\usepackage{epsfig}
\usepackage{subcaption}
\usepackage{tabularx}
\usepackage{makecell}
\usepackage{array}
\usepackage{ulem}%Durchstreichen
\usepackage{float}
\usepackage{adjustbox}

\usepackage{import}
\usepackage{xcolor,colortbl}
\usepackage{caption}
\usepackage[utf8]{inputenc}
\usepackage{pgfgantt}
\usepackage{pdflscape}
\newcolumntype{L}[1]{>{\raggedright\arraybackslash}p{#1}} % MK: linksbündig mit Breitenangabe
\newcolumntype{C}[1]{>{\centering\arraybackslash}p{#1}} % MK: zentriert mit Breitenangabe
\newcolumntype{R}[1]{>{\raggedleft\arraybackslash}p{#1}} % MK: rechtsbündig mit Breitenangabe

\allowdisplaybreaks

\usepackage{tikz}
\usetikzlibrary{shapes,arrows,patterns,backgrounds,fit,arrows.meta}
\usetikzlibrary{spy,backgrounds}
\usetikzlibrary{decorations.markings}
\usetikzlibrary{arrows.meta, positioning, calc}
\usepackage{pgfplots}
\usepgfplotslibrary[groupplots]
\usepackage{pgfplotstable}
\usepackage{placeins}
\usepackage{todonotes}
\definecolor{LUHblue}{RGB}{0,80,155}
\definecolor{LUHblue40}{RGB}{153,185,216}
\definecolor{LUHblue20}{RGB}{204,220,235}
\definecolor{LUHblack40}{RGB}{153,153,153}
\definecolor{LUHblack20}{RGB}{204,204,204}
\definecolor{LUHgreen}{RGB}{200,211,23}
\definecolor{IKMgreen}{RGB}{175,218,0}

\definecolor{ANSYS_Red}{RGB}{255,0,0} 
\definecolor{ANSYS_Orange}{RGB}{255,189,0}
\definecolor{ANSYS_Yellow}{RGB}{233,255,0} 

\definecolor{ANSYS_Green}{RGB}{101,255,0} 
\definecolor{ANSYS_LightBlue}{RGB}{0,255,255} 
\definecolor{ANSYS_Blue}{RGB}{0,0,255}

\makeatletter
\renewenvironment{cases}[1][l]{\matrix@check\cases\env@cases{#1}}{\endarray\right.}
\def\env@cases#1{%
  \let\@ifnextchar\new@ifnextchar
  \left\lbrace\def\arraystretch{1.2}%
  \array{@{}#1@{\quad}l@{}}}
\makeatother

\usepackage{booktabs}

\makeatletter
\let\oldtheequation\theequation
\renewcommand\tagform@[1]{\maketag@@@{\ignorespaces#1\unskip\@@italiccorr}}
\renewcommand\theequation{(\oldtheequation)}
\makeatother

\usepackage{cleveref}

\RequirePackage{relsize}
\newfont{\Sf}{cmssbx10 scaled 2074}

\usepackage{multirow}
\usepackage{lineno} % Package for Linenumbers
\begin{document}
\newtheorem{thm}{Theorem}%[section]
\newtheorem{cor}[thm]{Corollary}
\newtheorem{lem}[thm]{Lemma}
\newtheorem{defi}[thm]{Definition}
\newtheorem{conj}[thm]{Conjecture}
\newtheorem{quest}[thm]{Question}

\newcommand{\argmax}{\operatornamewithlimits{argmax}}
\newcommand{\imax}{{i_\mathrm{m}}}

\newcommand{\boldface}[1]{\boldsymbol{#1}}
\newcommand{\bfa}{\boldface{a}}
\newcommand{\bfb}{\boldface{b}}
\newcommand{\bfc}{\boldface{c}}
\newcommand{\bfd}{\boldface{d}}
\newcommand{\bfe}{\boldface{e}}
\newcommand{\bff}{\boldface{f}}
\newcommand{\bfg}{\boldface{g}}
\newcommand{\bfh}{\boldface{h}}
\newcommand{\bfi}{\boldface{i}}
\newcommand{\bfj}{\boldface{j}}
\newcommand{\bfk}{\boldface{k}}
\newcommand{\bfl}{\boldface{l}}
\newcommand{\bfm}{\boldface{m}}
\newcommand{\bfn}{\boldface{n}}
\newcommand{\bfo}{\boldface{o}}
\newcommand{\bfp}{\boldface{p}}
\newcommand{\bfq}{\boldface{q}}
\newcommand{\bfr}{\boldface{r}}
\newcommand{\bfs}{\boldface{s}}
\newcommand{\bft}{\boldface{t}}
\newcommand{\bfu}{\boldface{u}}
\newcommand{\bfv}{\boldface{v}}
\newcommand{\bfw}{\boldface{w}}
\newcommand{\bfx}{\boldface{x}}
\newcommand{\bfy}{\boldface{y}}
\newcommand{\bfz}{\boldface{z}}
\newcommand{\bfA}{\boldface{A}}
\newcommand{\bfB}{\boldface{B}}
\newcommand{\bfC}{\boldface{C}}
\newcommand{\bfD}{\boldface{D}}
\newcommand{\bfE}{\boldface{E}}
\newcommand{\bfF}{\boldface{F}}
\newcommand{\bfG}{\boldface{G}}
\newcommand{\bfH}{\boldface{H}}
\newcommand{\bfI}{\boldface{I}}
\newcommand{\bfJ}{\boldface{J}}
\newcommand{\bfK}{\boldface{K}}
\newcommand{\bfL}{\boldface{L}}
\newcommand{\bfM}{\boldface{M}}
\newcommand{\bfN}{\boldface{N}}
\newcommand{\bfO}{\boldface{O}}
\newcommand{\bfP}{\boldface{P}}
\newcommand{\bfQ}{\boldface{Q}}
\newcommand{\bfR}{\boldface{R}}
\newcommand{\bfS}{\boldface{S}}
\newcommand{\bfT}{\boldface{T}}
\newcommand{\bfU}{\boldface{U}}
\newcommand{\bfV}{\boldface{V}}
\newcommand{\bfW}{\boldface{W}}
\newcommand{\bfX}{\boldface{X}}
\newcommand{\bfY}{\boldface{Y}}
\newcommand{\bfZ}{\boldface{Z}}
%
% boldface greek symbols
%
\newcommand{\bfalpha}{\boldsymbol{\alpha}}
\newcommand{\bfbeta}{\boldsymbol{\beta}}
\newcommand{\bfgamma}{\boldsymbol{\gamma}}
\newcommand{\bfdelta}{\boldsymbol{\delta}}
\newcommand{\bfvarepsilon}{\boldsymbol{\varepsilon}}
\newcommand{\bfzeta}{\boldsymbol{\zeta}}
\newcommand{\bfeta}{\boldsymbol{\eta}}
\newcommand{\bftheta}{\boldsymbol{\theta}}
\newcommand{\bfkappa}{\boldsymbol{\kappa}}
\newcommand{\bflambda}{\boldsymbol{\lambda}}
\newcommand{\bfmu}{\boldsymbol{\mu}}
\newcommand{\bfnu}{\boldsymbol{\nu}}
\newcommand{\bfpi}{\boldsymbol{\pi}}
\newcommand{\bfxi}{\boldsymbol{\xi}}
\newcommand{\bfrho}{\boldsymbol{\rho}}
\newcommand{\bfsigma}{\boldsymbol{\sigma}}
\newcommand{\bftau}{\boldsymbol{\tau}}
\newcommand{\bfphi}{\boldsymbol{\phi}}
\newcommand{\bfvarphi}{\boldsymbol{\varphi}}
\newcommand{\bfchi}{\boldsymbol{\chi}}
\newcommand{\bfomega}{\boldsymbol{\omega}}
\newcommand{\bfupsilon}{\boldsymbol{\upsilon}}
\newcommand{\bfeps}{\boldsymbol{\varepsilon}}
\newcommand{\bfGamma}{\boldsymbol{\Gamma}}
\newcommand{\bfDelta}{\boldsymbol{\Delta}}
\newcommand{\bfTheta}{\boldsymbol{\Theta}}
\newcommand{\bfLambda}{\boldsymbol{\Lambda}}
\newcommand{\bfPi}{\boldsymbol{\Pi}}
\newcommand{\bfXi}{\boldsymbol{\Xi}}
\newcommand{\bfSigma}{\boldsymbol{\Sigma}}
\newcommand{\bfPhi}{\boldsymbol{\Phi}}
\newcommand{\bfChi}{\boldsymbol{\Chi}}
\newcommand{\bfOmega}{\boldsymbol{\Omega}}
%
% caligraphic letters
%
\newcommand{\calA}{\mathcal{A}}
\newcommand{\calB}{\mathcal{B}}
\newcommand{\calC}{\mathcal{C}}
\newcommand{\calD}{\mathcal{D}}
\newcommand{\calE}{\mathcal{E}}
\newcommand{\calF}{\mathcal{F}}
\newcommand{\calG}{\mathcal{G}}
\newcommand{\calH}{\mathcal{H}}
\newcommand{\calI}{\mathcal{I}}
\newcommand{\calJ}{\mathcal{J}}
\newcommand{\calK}{\mathcal{K}}
\newcommand{\calL}{\mathcal{L}}
\newcommand{\calM}{\mathcal{M}}
\newcommand{\calN}{\mathcal{N}}
\newcommand{\calO}{\mathcal{O}}
\newcommand{\calP}{\mathcal{P}}
\newcommand{\calQ}{\mathcal{Q}}
\newcommand{\calR}{\mathcal{R}}
\newcommand{\calS}{\mathcal{S}}
\newcommand{\calT}{\mathcal{T}}
\newcommand{\calU}{\mathcal{U}}
\newcommand{\calV}{\mathcal{V}}
\newcommand{\calW}{\mathcal{W}}
\newcommand{\calX}{\mathcal{X}}
\newcommand{\calY}{\mathcal{Y}}
\newcommand{\calZ}{\mathcal{Z}}
% .. define more if needed
%
% double stroke
%
%\usepackage[bb=boondox]{mathalfa}

\newcommand{\dsA}{\mathbb{A}}
\newcommand{\dsB}{\mathbb{B}}
\newcommand{\dsC}{\mathbb{C}}
\newcommand{\dsD}{\mathbb{D}}
\newcommand{\dsE}{\mathbb{E}}
\newcommand{\dsF}{\mathbb{F}}
\newcommand{\dsG}{\mathbb{G}}
\newcommand{\dsH}{\mathbb{H}}
\newcommand{\dsI}{\mathbb{I}}
\newcommand{\dsJ}{\mathbb{J}}
\newcommand{\dsK}{\mathbb{K}}
\newcommand{\dsL}{\mathbb{L}}
\newcommand{\dsM}{\mathbb{M}}
\newcommand{\dsN}{\mathbb{N}}
\newcommand{\dsO}{\mathbb{O}}
\newcommand{\dsP}{\mathbb{P}}
\newcommand{\dsQ}{\mathbb{Q}}
\newcommand{\dsR}{\mathbb{R}}
\newcommand{\dsS}{\mathbb{S}}
\newcommand{\dsT}{\mathbb{T}}
\newcommand{\dsU}{\mathbb{U}}
\newcommand{\dsV}{\mathbb{V}}
\newcommand{\dsW}{\mathbb{W}}
\newcommand{\dsX}{\mathbb{X}}
\newcommand{\dsY}{\mathbb{Y}}
\newcommand{\dsZ}{\mathbb{Z}}
\newcommand{\be}{\begin{equation}}
\newcommand{\ee}{\end{equation}}
\newcommand{\bea}{\begin{eqnarray}}
\newcommand{\eea}{\end{eqnarray}}
\newcommand{\bes}{\begin{equation*}}
\newcommand{\ees}{\end{equation*}}
\newcommand{\beas}{\begin{eqnarray*}}
\newcommand{\eeas}{\end{eqnarray*}}
\newcommand{\D}{\displaystyle}
\newcommand{\tr}[1]{\mathrm{tr} \, #1}
\newcommand{\cof}[1]{\mathrm{cof}[#1]}
\newcommand{\inv}[1]{#1^{-1}}
\newcommand{\norm}[1]{\left\lVert#1\right\rVert}
\newcommand{\abs}[1]{\left\lvert#1\right\rvert}
\newcommand{\lap}{\mathop{}\!\mathbin\bigtriangleup}
\newcommand{\dd}{\ \mathrm{d}}

\newcommand{\re}[1]{{\color{blue}{#1}}}
\newcommand{\rev}[2]{\red{\textsuperscript{#1)}}\re{#2}}
\newcommand{\red}[1]{{\color{red}{#1}}}
\newcommand{\titem}{~~\llap{\textbullet}~~}

% Assembly operator with user-defined arguments
\newcommand{\Assemarg}[2]{{
  \underset{\mathsmaller{#1}}{\overset{\mathsmaller{#2}}%
  {\raisebox{-0.5ex}{\mbox{\Sf A}}}}\;%
  }}
\newcommand{\AssemargPJ}[2]{{
  \underset{#1}{\overset{\mathsmaller{#2}}%
  {\raisebox{-0.5ex}{\mbox{\Sf A}}}}\;%
  }}
  
\newcommand{\bb}[1]{\mathbb{#1}}
\newcommand{\vv}[1]{\pmb{#1}}
\newcommand{\bg}[1]{\pmb{#1}}
\newcommand{\nk}[1]{\mathrm{#1}}
\newcommand{\abl}[1]{\dot{#1}}
\newcommand{\pd}[2]{\frac{\partial{#1}}{\partial{#2}}}
\newcommand{\ablzeit}{\frac{\nk{d}}{\nk{d}t} }
\newcommand{\intok}[1]{\int\limits_{\Omega} #1 \, \nk{d}v }
\newcommand{\intdV}[1]{\int_{\Omega} #1 \ \mathrm{d} V}
\newcommand{\intda}[1]{\int_{\partial\omega} #1 \ \mathrm{d} a}
\newcommand{\intdv}[1]{\int_{\omega} #1 \ \mathrm{d} v}
\newcommand{\intdA}[1]{\int_{\partial\Omega} #1 \ \mathrm{d} A}
\newcommand{\conref}{^{(0)}} 
\newcommand{\mydV}{\,\mathrm{d} V}
\newcommand{\mydA}{\,\mathrm{d} A}
\newcommand{\mydt}{\,\mathrm{d} t}
\newcommand{\myi}{{\mathrm{i}}}
\newcommand{\myb}{{\mathrm{b}}}

%% draw a cube
\usetikzlibrary{quotes,arrows.meta}
\tikzset{
  annotated cuboid/.pic={
    \tikzset{%
      every edge quotes/.append style={midway, auto},
      /cuboid/.cd,
      #1
    }
    \draw [every edge/.append style={pic actions, densely dashed, opacity=.5}, pic actions]
    (0,0,0) coordinate (o) -- ++(-\cubescale*\cubex,0,0) coordinate (a) -- ++(0,-\cubescale*\cubey,0) coordinate (b) edge coordinate [pos=1] (g) ++(0,0,-\cubescale*\cubez)  -- ++(\cubescale*\cubex,0,0) coordinate (c) -- cycle
    (o) -- ++(0,0,-\cubescale*\cubez) coordinate (d) -- ++(0,-\cubescale*\cubey,0) coordinate (e) edge (g) -- (c) -- cycle
    (o) -- (a) -- ++(0,0,-\cubescale*\cubez) coordinate (f) edge (g) -- (d) -- cycle;
  },
  /cuboid/.search also={/tikz},
  /cuboid/.cd,
  width/.store in=\cubex,
  height/.store in=\cubey,
  depth/.store in=\cubez,
  units/.store in=\cubeunits,
  scale/.store in=\cubescale,
  width=10,
  height=10,
  depth=10,
  units=cm,
  scale=.1,
}

\theoremstyle{nonumberbreak}
\newtheorem{remarknn}{Remark}

\begin{comment}
\newcommand{\bb}[1]{\mathbb{#1}}
\newcommand{\vv}[1]{\pmb{#1}}
\newcommand{\bg}[1]{\pmb{#1}}
\newcommand{\nk}[1]{\mathrm{#1}}
\newcommand{\abl}[1]{\dot{#1}}
\newcommand{\pd}[2]{\frac{\partial{#1}}{\partial{#2}}}
\newcommand{\ablzeit}{\frac{\nk{d}}{\nk{d}t} }
\newcommand{\intok}[1]{\int\limits_{\Omega} #1 \, \nk{d}v }
\newcommand{\intdV}[1]{\int_{\Omega} #1 \ \mathrm{d} V}
\newcommand{\intda}[1]{\int_{\partial\omega} #1 \ \mathrm{d} a}
\newcommand{\intdv}[1]{\int_{\omega} #1 \ \mathrm{d} v}
\newcommand{\intdA}[1]{\int_{\partial\Omega} #1 \ \mathrm{d} A}
\newcommand{\conref}{^{(0)}}
%%\newcommand{\dV}[1]{\,\mathrm{dV}{#1}
%\newcommand{\norm[1]}{\left\lVert#1\right\rVert}
\end{comment}

% .. define more if needed
%
% double stroke
%
%\usepackage[bb=boondox]{mathalfa}

\begin{comment}
\newcommand{\bb}[1]{\mathbb{#1}}
\newcommand{\vv}[1]{\pmb{#1}}
\newcommand{\bg}[1]{\pmb{#1}}
\newcommand{\nk}[1]{\mathrm{#1}}
\newcommand{\abl}[1]{\dot{#1}}
\newcommand{\pd}[2]{\frac{\partial{#1}}{\partial{#2}}}
\newcommand{\ablzeit}{\frac{\nk{d}}{\nk{d}t} }
\newcommand{\intok}[1]{\int\limits_{\Omega} #1 \, \nk{d}v }
\newcommand{\intdV}[1]{\int_{\Omega} #1 \ \mathrm{d} V}
\newcommand{\intda}[1]{\int_{\partial\omega} #1 \ \mathrm{d} a}
\newcommand{\intdv}[1]{\int_{\omega} #1 \ \mathrm{d} v}
\newcommand{\intdA}[1]{\int_{\partial\Omega} #1 \ \mathrm{d} A}
\newcommand{\conref}{^{(0)}}
%%\newcommand{\dV}[1]{\,\mathrm{dV}{#1}
%\newcommand{\norm[1]}{\left\lVert#1\right\rVert}

\end{comment}

%\linenumbers
\title{
First- and Second-Order Phase Transformation Modeling Based on the Hamilton Principle: A Coupled Thermo-Mechanical Approach for Glass Additive Manufacturing
}
\author{}
\date{} 
\maketitle
{\large
\noindent{Tobias Rudolf}, Meisam Soleimani, Philipp Junker\\[0.5mm]
}
Leibniz University Hanover, Institute of Continuum Mechanics, Hanover, Germany\\[2mm]
{
Corresponding author:\\[0.5mm]
Tobias Rudolf, \color{LUHblue}{\Letter \hskip 1mm rudolf@ikm.uni-hannover.de} 
}
    
%\listoftodos
\section*{Abstract}
Additive manufacturing of glass inherently involves complex thermal histories characterized by extreme heating and rapid cooling rates. These extreme conditions directly govern the final microstructure and mechanical integrity of the printed material. This work presents a comprehensive multi-physics material model derived from the extended Hamilton principle, establishing a unified variational framework for coupled thermal, mechanical, and phase transformation processes at finite strains. The formulation integrates a rigorous thermodynamic description of first-order melting and second-order glass transitions with a kinematic split accounting for thermal expansion, phase specific density changes, and viscoelastic deformation. A temperature dependent viscosity model is employed to capture the kinetic freezing of the microstructure inherent to vitrification. The numerical implementation utilizes a monolithic Neighbored Element Method (NEM) for the solution of the heat equation, ensuring computational efficiency and stability. Numerical investigations at the material point level validate the model’s ability to reproduce Time-Temperature-Transformation (TTT) behavior under varying cooling rates. Furthermore, three dimensional Finite Element simulations in ANSYS$^{\copyright}$ of a laser-based deposition process demonstrate the accumulation of residual stresses and macroscopic warpage resulting from the interplay between phase transformation kinetics and viscous relaxation.

\section{Introduction}
The precise control of phase transformations during thermal processing is fundamental to tailoring the microstructure and properties of engineering materials. Processes such as additive manufacturing, laser welding, and glass forming subject materials to complex thermal histories characterized by rapid heating, localized melting, and spatially varying cooling rates. Under such conditions, the final material state is determined not by thermodynamic equilibrium alone, but by the intricate interplay between thermodynamic driving forces and kinetic limitations, giving rise to path-dependent phenomena such as vitrification and partial crystallization \cite{uhlmann1972kinetic,uhlmann1982kinetics,debenedetti2001supercooled}.

Phase transformations are traditionally classified according to the Ehrenfest scheme \cite{jaeger1998ehrenfest, stanley1971introduction}. \textit{First-order transformations}: such as melting and solidification, exhibit discontinuities in entropy and volume, manifesting as latent heat and abrupt volume changes, with kinetics governed by nucleation and growth mechanisms \cite{christian2002theory, porter2009phase}. \textit{Second-order transformations}, including the glass transition, exhibit continuous first derivatives but discontinuous second derivatives of the free energy, such as heat capacity. The glass transition represents a kinetic phenomenon wherein viscosity increases so dramatically that molecular rearrangement effectively ceases on experimental timescales \cite{angell1995formation, debenedetti2001supercooled}. The transition temperature $\theta^{\mathrm{g}}$ is therefore cooling-rate-dependent, reflecting frozen in configurational entropy rather than a true thermodynamic phase boundary \cite{richet1993entropy, ediger1996supercooled}.

Glass-forming materials exhibit a rich spectrum of behavior due to their ability to bypass crystallization if cooled sufficiently rapidly. Time-Temperature-Transformation (TTT) diagrams \cite{uhlmann1972kinetic, weinberg1991glass, musgraves2019springer} delineate the critical cooling rates required to avoid crystallization. Materials cooled faster than the critical rate achieve full vitrification, while slower cooling permits crystallization. Intermediate cooling rates, common in industrial processes yield, composite microstructures, comprising both crystalline and amorphous phases. Additionally, stress fields arising from thermal gradients or external loading can significantly influence transformation kinetics by modifying thermodynamic driving forces \cite{levitas2014phase}.

Computational modeling of these phenomena has been pursued through various frameworks. Phase-field models \cite{chen2002phase, steinbach2009phase, boettinger2002phase, ammar2009combining, mamivand2013phase} naturally capture complex morphologies but are computationally expensive due to interface resolution requirements. Sharp-interface approaches \cite{gibou2018review, karma2001quantitative} offer efficiency, but face challenges with topological changes. For engineering-scale simulations, phenomenological models treating phase fractions as internal state variables provide a pragmatic balance \cite{michaleris2014modeling, kollmannsberger2019computational, mercelis2006residual, denis2002prediction}. However, existing macroscale models often lack rigorous thermodynamic foundations and introduce phenomenological coupling terms without fundamental derivation. Furthermore, treatment of second-order transitions within these frameworks has received limited attention.

The extended Hamilton principle provides a unified variational framework for coupled multiphysics problems \cite{junker2021extended}. By augmenting the classical Lagrangian with functionals representing thermal energy, entropy production, and constraints, this approach yields governing equations through a single variational statement. This ensures: automatic satisfaction of thermodynamic consistency, systematic derivation of coupling terms, clear physical interpretation, and a natural framework for finite element discretization. While successfully applied to phase transformation \cite{soleimani2025continuum}, damage mechanics \cite{liu2025novel, kok2025variational}, and biomechanics \cite{klempt2024hamilton}, its application to glass-forming materials exhibiting both first- and second-order transitions remains largely unexplored.

This work develops a comprehensive multi-physics material model for glass-forming materials that captures coupled thermal conduction, mechanical deformation, viscous deformation, and phase transformations among crystalline, amorphous, and liquid states. Formulated within the extended Hamilton principle, the model features: unified treatment of first- and second-order transitions, temperature-dependent viscosity, rate-dependent kinetics governed by thermodynamic driving forces, finite-strain kinematics with multiplicative decomposition. The model is validated through material point simulations demonstrating TTT diagram behavior under varied cooling rates. To demonstrate the practical applicability of the proposed framework, the resulting coupled equations are numerically implemented in the commercial finite element software ANSYS$^{\copyright}$ via user-defined subroutines. 

\section{Continuum Thermodynamic Formulation}
This section presents a multi-physics material model for glass that captures the coupled effects of thermal, mechanical, and phase transformation processes under finite strains. The proposed framework integrates a finite-strain formulation to model mechanical responses with a thermodynamic framework that delineates phase transitions among crystalline, amorphous, and liquid states. The integration of thermal and mechanical domains is facilitated by variational principles and thermodynamic potentials.

\subsection{Fundamentals}
The extended Hamilton principle, as advanced by Junker et al. \cite{junker2021extended}, provides a unified variational framework for coupled multiphysics systems. Unlike the classical principle confined to conservative mechanical systems, the extended version accommodates non-conservative continua by incorporating thermomechanical coupling and dissipative microstructural evolution through augmentation with thermal work and energy dissipation terms derived from thermodynamic laws.
For the present model, the relevant state variables are: the displacement field $\bfu$ (mechanical configuration), the temperature field $\theta$ (thermal state), internal variables $\bflambda$ and $\bfC^{\mathrm{v}}$ (microstructural states representing phase volume fractions and viscous deformation, respectively), and the Lagrange multiplier $\gamma$ (enforcing physical constraints on $\bflambda$). The extended Hamilton functional is expressed by the integral
\begin{align}
    \calH&=\int_{t_0}^{t}\bigg[\int_\Omega\frac{1}{2}\rho^{(0)}||\Dot{\bfu}||^2\mydV-\int_\Omega\rho^{(0)}\overline{\Psi}\mydV+\int_\Omega\bfb^\star\cdot\bfu \mydV+\int_{\partial\Omega}\bft^\star\cdot\bfu\mydA+\int_\Omega h^{\star}\theta\mydV\\
    &-\int_\Omega\int\rho^{(0)}\Dot{\theta}\overline{s} \mydt \mydV-\int_\Omega\int\frac{1}{\theta}\bfQ\cdot\nabla_{\bfX}\theta\mydt\mydV-\int_\Omega\int\theta\Delta^{\mathrm{s}}\mydt\mydV+\int_\Omega\gamma g \mydV
    \bigg]\mydt\notag \,.
    \label{EQ: The extended Hamilton Functional My}
\end{align}
In this equation, $\rho^{(0)}$ denotes the reference density, $\overline{\Psi}(\bfu,\theta,\bflambda,\bfC^{\mathrm{v}})$ is the mass-specific Helmholtz free energy, $\bfb^\star$ and $\bft^\star$ are body forces and surface tractions, $h^{\star}$ is the external heat source, $\overline{s}$ is the entropy density, 
$\bfQ $ is the heat flux vector in the reference configuration, and $\Delta^{\mathrm{s}}(\bflambda,\bfC^{\mathrm{v}}$) is the entropy production accounting for internal dissipation. The governing equations are derived from the stationarity condition
\begin{equation}
\delta\calH=\delta_{\bfu}\calH+\delta_{\theta}\calH+\delta_{\bflambda}\calH+\delta_{\bfC^{\mathrm{v}}}\calH+\delta_{\gamma}\calH
=0 \quad \forall \delta_{\bfu}, \delta_{\theta}, \delta_{\bflambda},\delta_{\bfC^{\mathrm{v}}}, \delta_{\gamma}\,.
\end{equation}
The transition from the global variational principle to the local Euler-Lagrange equations relies on the fundamental lemma of variational calculus. A detailed derivation of these stationarity conditions is omitted here for brevity.  Interested readers are referred to the foundational work by Junker et al.\cite{junker2021extended}. Consequently, evaluating the variation with respect to the displacement field $\delta_{\bfu}\calH=0$ yields the balance of linear momentum 
\begin{align}
    -\rho\conref\Ddot{\bfu}+\nabla_{\bfX}\cdot\rho\conref\frac{\partial\overline{\Psi}}{\partial\bfF^\mathrm{T}}+\bfb^\star
    &=\boldsymbol{0}\quad \forall\in\Omega \\
    -\bfN\cdot\rho\conref\frac{\partial\overline{\Psi}}{\partial\bfF^\mathrm{T}}+\bft^\star&=\mathbf{0}\quad \forall\in\partial\Omega \notag
\end{align}
where $\bfF$ is the deformation gradient and $\bfN$ the unit vector in normal direction.

Similarly, the variation with respect to temperature, $\delta_{\theta}\mathcal{H}=0$, governs the thermal state of the continuum. By evaluating this stationarity condition, expanding the entropy rate via standard thermodynamic potential relations, and incorporating Fourier's law for the reference heat flux ($\mathbf{Q} = -\alpha^{\mathrm{cond}}\nabla_X\theta$), the strong form of the heat conduction equation is directly obtained as  
\begin{align}
\kappa\dot{\theta}+\alpha^{\mathrm{cond}} \nabla_{\bfX}^2 \theta-\left[\bfp^{\bflambda}-\theta\frac{\partial \bfp^{\bflambda}}{\partial\theta}\right]\cdot\Dot{\bflambda}
-\left[\bfp^{\bfC^{\mathrm{v}}}-\theta\frac{\partial \bfp^{\bfC^{\mathrm{v}}}}{\partial\theta}\right]:\Dot{\bfC^{\mathrm{v}}}-\theta\frac{\partial \bfP}{\partial\theta}:\Dot{\bfF}+h^{\star}=0\,,
\label{EQ: Heat EQ Strong}
\end{align}
where $\alpha^{\mathrm{cond}}$ is the thermal conductivity, and $\bfP$ the first Piola-Kirchhoff stress tensor. Furthermore, this equation naturally introduces the specific heat capacity $\kappa$ alongside the thermodynamic driving forces conjugate to the internal variables, defined as $\bfp^{\bflambda} $ and $\bfp^{\bfC^{\mathrm{v}}}$.

To determine the kinetic evolution of these microstructural state variables, we evaluate the variation of the extended Hamilton functional with respect to the phase fractions ($\delta_{\lambda}\mathcal{H}=0$). By modeling the internal energy dissipation through non-conservative thermodynamic forces derived from a scalar-valued dissipation potential $\Delta^{\mathrm{diss,\bflambda}}$, the thermodynamic stationarity condition takes the form
\begin{equation}
    \bfp^{\bflambda}+\frac{\partial\Delta^{\mathrm{diss,\bflambda}}}{\partial\Dot{\bflambda}}+\gamma\frac{\partial g}{\partial\bflambda}=\boldsymbol{0}\,.
    \label{EQ: Variation to lambda}
\end{equation}
The kinetic of the phase transformations is subsequently governed by the specific choice of the dissipation function. Here, the dissipation potential combines viscous and threshold-like behaviors, written as
\begin{equation}
    \Delta^{\mathrm{diss,\bflambda}}=\frac{1}{2}\Dot{\bflambda}\cdot\bfeta^{\bflambda}\cdot\Dot{\bflambda}+N^{\mathrm{thres}}||\Dot{\bflambda}||\,.
    \label{EQ: Diss lambda}
\end{equation}
To ensure that the phase volume fractions remain confined to a physically meaningful domain, a constraint function $g$ is introduced to restrict the sum of the fractions to unity at any given material point, taking the form
\begin{equation}
    g=\sum_i^3\lambda_i-1\,.
    \label{EQ: Consraint Lambda}
\end{equation}
By evaluating the derivative in \autoref{EQ: Variation to lambda} of the dissipation potential with respect to the phase fraction rate, the evolution equation is obtained as
\begin{equation}
    \bfp^{\bflambda}+\bfeta^{\bflambda}\cdot\Dot{\bflambda}+N^{\mathrm{thres}}\frac{\Dot{\bflambda}}{||\Dot{\bflambda}||}+\gamma\boldsymbol{1}^\mathrm{v}\ni \boldsymbol{0}\,.
    \label{EQ: Evolution Lambda}
\end{equation}
In this formulation, $\boldsymbol{1}^\mathrm{v}$ is the identity tensor and  $\bfeta^{\bflambda}$ denotes the  mobility tensor. We emphasize that each individual phase (crystalline, amorphous, and liquid) possesses its own unique, phase-specific viscosity parameter within this tensor. This distinction is crucial for accurately capturing the highly disparate kinetic rates inherent to crystallization versus the kinetic freezing during vitrification, while $N^{\mathrm{thres}}$ represents the energetic activation barrier preventing spontaneous phase transformation.

In a similar manner, evaluating the variation of the extended Hamilton functional with respect to the viscous internal variable $\delta_{\bfC^{\mathrm{v}}}\calH=0$, yields
\begin{equation}
    \bfp^{\bfC^{\mathrm{v}}}+\frac{\partial\Delta^{\mathrm{diss,\bfC^{\mathrm{v}}}}}{\partial\Dot{\bfC^{\mathrm{v}}}}=\boldsymbol{0}\,,
    \label{EQ: Variation to viscous defo}
\end{equation}
where the underlying macro-mechanical dissipation potential is governed by a viscosity function according to
\begin{equation}
\Delta^{\mathrm{diss},\bfC^{\mathrm{v}}} = \frac{1}{2}\eta^{\bfC^{\mathrm{v}}}\Dot{\bfC^{\mathrm{v}}}:\Dot{\bfC^{\mathrm{v}}} \,.
\label{EQ: Dissipation potential of viscous deformation}
\end{equation}
By evaluating the derivative in \autoref{EQ: Variation to viscous defo} of this macro-mechanical dissipation potential with respect to the rate of viscous deformation, the evolution equation is given by
\begin{equation}
    \bfp^{\bfC^{\mathrm{v}}}+\eta^{\bfC^{\mathrm{v}}}\Dot{\bfC^{\mathrm{v}}}=\boldsymbol{0}\,,
    \label{EQ: Evolution of viscous deformation}
\end{equation}
Here, $\eta^{\bfC^{\mathrm{v}}}$ represents the temperature-dependent mechanical viscosity regulating the material's resistance to irreversible flow. This formulation introduces an isotropic viscous response, which effectively captures the macroscale stress relaxation behavior of the glass melt while strictly maintaining thermodynamic consistency.

Consequently, performing the global variation with respect to the Lagrange multiplier, $\delta_{\gamma}\calH=0$, rigorously enforces this physical constraint throughout the coupled process
\begin{equation}
    g=0
\end{equation}

\begin{remarknn}
Unlike classical plasticity or standard viscoelastic flow formulations where the inelastic deformation gradient is strictly assumed to be isochoric, i.e., $\det[\bfF^{\mathrm{v}}] = 1$, the present multi-phase framework must accommodate volumetric alterations. Due to the conservation of mass and since the crystalline, amorphous, and liquid phases exhibit distinct reference densities, any change in the phase fractions $\bflambda$ inherently induces a macroscopic volume change. Therefore, the non-elastic portion of the deformation gradient is explicitly non-isochoric, meaning that $\det[\bfF^{\mathrm{v}} \cdot \bfF^\rho] \neq 1$ is a direct consequence to satisfy the fundamental principle of mass conservation.
\end{remarknn}

\subsection{Modeling the free Helmholtz potential}
The thermodynamic modeling of the material is based on the specification of the Helmholtz free energy, which serves as the fundamental potential for deriving the constitutive laws under non-isothermal conditions. Both the volumetric energy density $\Psi$ and the mass-specific energy density $\overline{\Psi}$ are related by the reference density $\rho^{(0)}$, given by
\begin{equation}
    \Psi=\rho^{(0)}\overline{\Psi} \,.
\end{equation}
The total free energy is formulated as a mixture of the solid phase, $\Psi^\mathrm{s}$, and a void or air phase, $\Psi^\mathrm{air}$. We define $\phi$ as the volume fraction of the solid phase, where $\phi=1$ represents the pure solid and $\phi=0$ represents the pure void. The contributions are weighted linear interpolation. The total energy is thus expressed as
\begin{equation}
\Psi=(1-\phi)\Psi^\mathrm{air}+\phi\Psi^\mathrm{s} \,.
\label{EQ: mixed free energy}
\end{equation}

\begin{figure}
    \centering
    \begin{tikzpicture}[scale=1.1]

% Definiere Pfeilstil
\tikzset{
    defarrow/.style={-{Stealth[length=3mm]}, thick, shorten >=2mm, shorten <=2mm},
    totalarrow/.style={-{Stealth[length=3.5mm]}, very thick, shorten >=2mm, shorten <=2mm}
}

% Referenzkonfiguration (oben links)
\begin{scope}[shift={(2.5,2)}]
    % Kartoffel - mittleres Volumen
    \draw[thick, fill=gray!40] plot[smooth cycle, tension=0.7] coordinates {
        (0,0.8) (0.6,0.5) (0.9,0) (0.6,-0.6) (0,-0.8) (-0.5,-0.5) (-0.7,0.1) (-0.4,0.6)
    };
    % Materialpunkt
    \fill (0.4,-0.4) circle (2pt) node[left=0.1ex] {$\bfX$};
    % Label
    \node at (0,0) {$\Omega_0$};
\end{scope}

% Aktuelle Konfiguration (oben rechts) - weitere FORMÄNDERUNG (elastisch)
\begin{scope}[shift={(7.5,2)}]
    % Kartoffel - ähnliches Volumen, zusätzliche elastische Verzerrung
    \draw[thick, fill=gray!40] plot[smooth cycle, tension=0.7] coordinates {
        (0.25,0.95) (0.85,0.48) (1.15,-0.18) (0.8,-0.8) (0.15,-1.05) (-0.45,-0.62) (-0.75,0.15) (-0.28,0.72)
    };
    % Materialpunkt
    \fill (0.8,-0.3) circle (2pt) node[left=2pt] {$\bfx_{\mathrm{el}}$};
    % Label
    \node at (0,0) {$\Omega_t$};
\end{scope}

% Nach F^ρ (links oben) - VOLUMENREDUKTION
\begin{scope}[shift={(2.7,-0.25)}]
    % Kartoffel - KLEINER (Kompression)
    \draw[thick, fill=gray!40] plot[smooth cycle, tension=0.7] coordinates {
        (0,0.55) (0.42,0.35) (0.6,0) (0.42,-0.42) (0,-0.58) (-0.36,-0.35) (-0.52,0.07) (-0.28,0.42)
    };
    % Materialpunkt
    \fill (0.35,-0.35) circle (2pt) node[left=0.1ex] {$\bfx_{\rho}$};
    % Label
    \node at (0,0) {$\Omega_{\rho}$};
\end{scope}

% Nach F^th (links unten) - VOLUMENEXPANSION
\begin{scope}[shift={(4.7,-1.4)}]
    % Kartoffel - GRÖSSER (thermische Expansion)
    \draw[thick, fill=gray!40] plot[smooth cycle, tension=0.7] coordinates {
        (0,0.95) (0.7,0.6) (1.05,0) (0.7,-0.7) (0,-0.95) (-0.58,-0.6) (-0.82,0.12) (-0.46,0.7)
    };

    % Materialpunkt
    \fill (0.4,-0.4) circle (2pt)node[left=0.1ex] {$\bfx_{\mathrm{th}}$};
    % Label
    \node at (0,0) {$\Omega_{\mathrm{th}}$};
\end{scope}

% Nach F^Cv (unten mitte) - FORMÄNDERUNG bei etwa gleichem Volumen
\begin{scope}[shift={(7.5,-1)}]
    % Kartoffel - ähnliches Volumen wie nach F^th, aber geschert
    \draw[thick, fill=gray!40] plot[smooth cycle, tension=0.7] coordinates {
        (0.15,0.95) (0.75,0.55) (1.1,-0.05) (0.7,-0.75) (0.05,-1.0) (-0.55,-0.65) (-0.85,0.08) (-0.35,0.7)
    };
    % Materialpunkt
    \fill (0.2,-0.5) circle (2pt) node[left=0.1ex] {$\bfx_{\bfC\mathrm{v}}$};
    % Label
    \node at (0,0) {$\Omega_{\bfC\mathrm{v}}$};
\end{scope}

% Pfeile und Beschriftungen
% F^ρ
\draw[defarrow] (2.2,1.35) to[bend right=15] node[left, pos=0.5, font=\small] {$\bfF^{\rho}$} (2.5,0.15);

% F^th
\draw[defarrow] (3,-0.65) to[bend right=15] node[left, pos=0.5, font=\small, yshift=-1ex] {$\bfF^{\mathrm{th}}$} (4.0,-1.45);

% F^Cv
\draw[defarrow] (5.65,-1.6) to[bend right=15] node[below, pos=0.5, font=\small] {$\bfF^{\bfC^{\mathrm{v}}}$} (6.9,-1.4);

% F^el
\draw[defarrow] (8.2,-0.4) to[bend right=15] node[right, pos=0.5, font=\small] {$\bfF^{\mathrm{el}}$} (8,1.05);

% Gesamtdeformation F (direkt von Referenz zu aktuell)
\draw[totalarrow, black, line width=1mm] (3.3,2.3) to[bend left=25] node[below, pos=0.5,yshift=4ex, font=\normalsize] {$\bfF = \bfF^{\mathrm{el}} \cdot \bfF^{\bfC^{\mathrm{v}}} \cdot \bfF^{\mathrm{th}} \cdot \bfF^{\rho}$} (6.7,2.3);

\end{tikzpicture}
    \caption{Visualization of the kinematic split.}
    \label{Fig: Visualization Defgrad}
\end{figure}
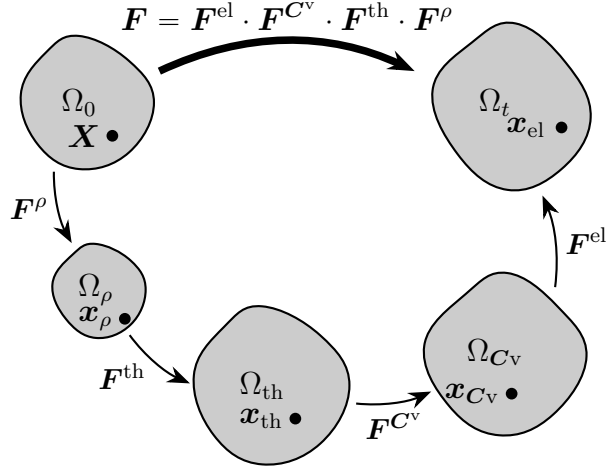

The primary contribution to the total energy comes from the solid phase. Its specific free energy, $\Psi^\mathrm{s}$, is further decomposed to account for different physical phenomena, taking the form
\begin{equation}
    \Psi^\mathrm{s}=\Psi^\mathrm{cal,tot}+\Psi^\mathrm{mech}+\Psi^\mathrm{pen} \,.
\end{equation}
Here, $\Psi^\mathrm{cal,tot}$ represents the caloric energy related to heat capacity, $\Psi^\mathrm{pen}$ is a penalty energy associated with phase transformations, and $\Psi^\mathrm{mech}$ is the mechanical strain energy which are determined from the Young's moduli $E_{i}$ and Poisson's ratios $\nu_{i}$ in the standard manner, yielding
\begin{align}
    \mu_i&=\frac{E_i}{2(1+\nu_i)}\\
    K_i&=\frac{E_i}{3(1-2\nu_i)} \,.
\end{align}
To describe the effective mechanical response of the multi-phase material, the phase-specific properties must be homogenized. Instead of applying a simple linear interpolation (Voigt bound), the effective shear modulus $\mu^\mathrm{tot}$ and the effective bulk modulus $K^\mathrm{tot}$ follow from the Reuss mixture rule \cite{saeb2016aspects}. They are defined by the harmonic mean weighted by the current volume phase fractions $\lambda_{i}$, yielding
\begin{align}
    \mu^{\mathrm{tot}}&=\left(\sum_i^3\mu_i^{-1}\lambda_i\right)^{-1}\\
    K^{\mathrm{tot}}&=\left(\sum_i^3K_i^{-1}\lambda_i\right)^{-1} \,.
\end{align}
Employing the Reuss mixture rule is a crucial constitutive modeling choice. It strictly ensures the convexity of the mechanical strain energy potential with respect to the phase fractions. This rigorously preserves the numerical stability of the multi-phase mixture.
With these homogenized effective moduli established, the mechanical behavior of the mixture under finite elastic deformations can be formulated. The corresponding specific strain energy is modeled using a compressible Neo-Hookean formulation, which reads
\begin{equation}
    \Psi^\mathrm{mech}=\frac{\mu^{\mathrm{tot}}}{2}\left(\overline{I}_1-3\right)+\frac{K^{\mathrm{tot}}}{2}\left(J^{\mathrm{el}}-1\right)^2 \,.
    \label{EQ: Psi mechanic}
\end{equation}
This isochoric-volumetric decoupled strain energy function relies on two primary invariants derived from the kinematic quantities. The first is the determinant of the elastic deformation gradient, given by
\begin{equation}
    J^{\mathrm{el}}=\det[\bfF^{\mathrm{el}}]\,,
\end{equation}
which represents the volume change due to elastic deformation. The other one is the first invariant of the isochoric (volume-preserving) part of the elastic left Cauchy-Green tensor $\bfb^{\mathrm{el}}$, given by
\begin{equation}
    \overline{I}_1=\mathrm{Tr}\left[(J^{\mathrm{el}})^{-\frac{2}{3}}\,\bfb^{\mathrm{el}}\right]\,.
    \label{EQ: First isochoric Invariant}
\end{equation}
The multiplicative decomposition of the deformation gradient reflects the physical sequence of density change ($\bfF^{\rho}$), thermal expansion ($\bfF^{\mathrm{th}}$), viscous deformation ($\bfF^{\bfC^{\mathrm{v}}}$), and finally elastic deformation ($\bfF^{\mathrm{el}}$), as shown in \autoref{Fig: Visualization Defgrad}. Following Simo and Miehe \cite{simo1992associative}, the elastic-first ordering ($ \bfF^{\mathrm{el}}\cdot\bfF^{\bfC^{\mathrm{v}}}\cdot\bfF^{\mathrm{th}}\cdot \bfF^{\rho}$) ensures that the reversible response operates on the current configuration, maintaining thermodynamic consistent. This structure allows for modular implementation of inelastic mechanisms \cite{sahraee2015homogenization} and separates reversible from irreversible processes. 
The elastic deformation gradient is isolated as
\begin{equation}
    \bfF^{\mathrm{el}}=\bfF\cdot\left(\bfF^{\rho}\right)^{-1}\cdot\left(\bfF^{\mathrm{th}}\right)^{-1}\cdot\left(\bfF^{\bfC^{\mathrm{v}}}\right)^{-1} \,,
\end{equation}
From this, we define the elastic left Cauchy-Green tensor $\bfb^{\mathrm{el}}$, which indicates the elastic deformation in the current configuration and is given by
\begin{equation}
    \bfb^{\mathrm{el}}=\bfF^{\mathrm{el}}\cdot\left(\bfF^{\mathrm{el}}\right)^\intercal \,.
\end{equation}
Expanding this definition yields 
\begin{equation}
        \bfb^{\mathrm{el}}=\left[\bfF\cdot\left(\bfF^{\rho}\right)^{-1}\cdot\left(\bfF^{\mathrm{th}}\right)^{-1}\cdot\left(\bfF^{\bfC^{\mathrm{v}}}\right)^{-1}\right]\cdot\left[\bfF\cdot\left(\bfF^{\rho}\right)^{-1}\cdot\left(\bfF^{\mathrm{th}}\right)^{-1}\cdot\left(\bfF^{\bfC^{\mathrm{v}}}\right)^{-1}\right]^\intercal 
\end{equation} 
and introducing the viscous internal variable $\bfC^{\mathrm{v}}$, the relationship can be written more explicitly. The explicit formulation of the viscous internal variable reads
\begin{equation}
\left(\bfC^{\mathrm{v}}\right)^{-1}=\left(\bfF^{\bfC^{\mathrm{v}}}\right)^{-1}\cdot\left(\bfF^{\bfC^{\mathrm{v}}}\right)^{-\intercal}\,.
\end{equation}
Substituting this into the definition of $\bfb^{\mathrm{el}}$ yields its expanded form as
\begin{equation}
\bfb^{\mathrm{el}}=\bfF\cdot\left(\bfF^{\rho}\right)^{-1}\cdot\left(\bfF^{\mathrm{th}}\right)^{-1}\cdot\left(\bfC^{\mathrm{v}}\right)^{-1}\cdot\left(\bfF^{\mathrm{th}}\right)^{-\intercal}\cdot\left(\bfF^{\rho}\right)^{-\intercal}\bfF^\intercal \,.
\end{equation}
The thermal deformation gradient, $\mathbf{F}^{\mathrm{th}}$, models the isotropic expansion and contraction of the material due to temperature changes. It is defined via
\begin{equation}
    \bfF^{\mathrm{th}}=\exp{\left[\alpha^{\mathrm{exp,tot}}(\theta-\theta^{\mathrm{ref}})\right]}\bfI \,.
\end{equation}
This deformation is governed by the total coefficient of linear thermal expansion, $\alpha^{\mathrm{exp,tot}}$, and the difference between the current temperature $\theta$ and a stress-free reference temperature $\theta^{\mathrm{ref}}$. The total thermal expansion coefficient is calculated as a weighted average of the coefficients for the individual phases $\alpha^{\mathrm{exp}}_i$ using the current phase fractions $\lambda_i$ 
\begin{equation}
\alpha^{\mathrm{exp,tot}}=\sum_i^3\alpha^{\mathrm{exp}}_i\lambda_i \,.
\end{equation}

Similarly, the phase-change deformation gradient $\bfF^{\rho}$ models the isotropic volumetric strain resulting from density changes during phase transformations. This term is defined by the ratio of the effective reference density to the current density as
\begin{equation}
    \bfF^{\rho}=\left(J^{\rho}\right)^{\frac{1}{3}}\bfI=\left(\frac{\rho^{\mathrm{ref}}}{\rho}\right)^{\frac{1}{3}}\bfI \,.
\end{equation}
The cube root in this expression relates the volumetric density change to the corresponding linear stretch. The effective reference density $\rho^{\mathrm{ref}}$ is a weighted average of the reference densities of the individual phases, $\rho\conref_i$, based on the volume phase fractions
\begin{equation}
\rho^{\mathrm{ref}}=\sum_i^3\rho\conref_i\lambda_i\conref \,.
\end{equation}
The current density, $\rho$, is itself a function of the current volume fraction $\bflambda$, i.e., 
\begin{equation}
    \rho=\sum_i^3\rho\conref_i\cdot\lambda_i\,.
\end{equation}
To strictly confine the phase fractions to their physical domain $0 \leq\lambda_i \leq 1$, a barrier potential $\Psi^{\mathrm{pen}}$ is introduced. This potential is negligible within the domain but grows asymptotically near the boundaries, effectively enforcing constraints without explicit optimization algorithms. It is defined as
\begin{equation}
\Psi^{\mathrm{pen}}=N^{\mathrm{pen}}\sum_i^3\frac{1}{\lambda_i^2(1-\lambda_i)^2}\,.
\end{equation}
The regularization parameter $N^{\mathrm{pen}}$ balances numerical stability against constraint violation.

The caloric part of the free energy $\Psi^{\mathrm{cal,tot}}$ accounts for the energy stored in the material as a function of temperature, and its formulation is derived directly from the definition of the specific heat capacity. The fundamental thermodynamic relationship connecting the specific free energy and the specific heat capacity at constant pressure is
\begin{equation}
c \coloneq - \theta \frac{\partial^2 \overline{\Psi}}{\partial \theta^2} \,.
\end{equation}
For each individual phase $i$ of the solid material, this relationship is expressed as
\begin{equation}
    c_{\mathrm{i}}^{\mathrm{S}}\rho^{(0),\mathrm{S}}_i=-\theta\rho^{(0),\mathrm{S}}_i\frac{\partial ^2\overline{\Psi^{\mathrm{cal}}}_i}{\partial\theta^2} \,.
\end{equation}
It should be noted that in the derivation of the specific heat capacity, the contribution of the mechanical energy potential $\Psi^{\mathrm{mech}}_i$ is omitted. This simplification is based on the well-founded assumption that for the class of thermo-mechanical problems considered here, the caloric energy storage significantly outweighs the coupling effects arising from thermo-elastic or viscous dissipation in its influence on the heat capacity. Neglecting the mechanical contribution facilitates a more direct analytical derivation of the caloric potentials without compromising the model's accuracy in capturing the primary phase transformation kinetics \cite{lion2000constitutive, simo1992associative}. The caloric free energy for each phase must be obtained by performing a double integration of the specific heat capacity, which takes the form
\begin{equation}
    \overline{\Psi}^\mathrm{cal}_i=-\iint\frac{c_\mathrm{i}^{\mathrm{S}}}{\theta}\mathrm{d}\theta\mathrm{d}\theta \,.
\end{equation}
To accurately capture the complex, temperature-dependent behavior of the specific heat capacity particularly across the glass transition and melting points, the overall caloric free energy is constructed piecewise. In the following, the specific formulations for the liquid, crystalline, and amorphous states are systematically derived.

The derivation begins by constructing a continuous free energy function $\overline{\Psi}^\mathrm{cal}_l(\theta)$ that describes the material in its liquid state and its extension into the supercooled liquid region below the melting point. This function serves as the baseline from which the distinct solid phases (crystalline and amorphous) will branch. This is based on a piecewise model of the specific heat capacity $c_{l}(\theta)$. The specific heat capacity is defined across three distinct temperature regions:
\begin{itemize}
    \item \textbf{Solid state for $\theta\leq\theta^{\mathrm{g}}$:} The specific heat is a constant value $c_a$.  The integrated free energy has the form
    \begin{equation}
        \overline{\Psi}^{\mathrm{cal}}_{l,a}=-c_a\theta\left(\log[\theta]-1\right)+a_1\theta+a_2
    \end{equation}
    \item \textbf{Transition region for $\theta^{\mathrm{g}}<\theta<\theta^{\mathrm{m}}$:} The specific heat capacity transitions smoothly from $c_a$ to $c_l$. This transition is modeled using a cubic Hermite polynomial 
    \begin{equation}
        c_t=c_a+(c_l-c_a)\left[-2\left(\frac{\theta-\theta^{\mathrm{g}}}{\theta^{\mathrm{m}}-\theta^{\mathrm{g}}}\right)^3+3\left(\frac{\theta-\theta^{\mathrm{g}}}{\theta^{\mathrm{m}}-\theta^{\mathrm{g}}}\right)^2\right]
    \end{equation}
    to ensure the function and its first derivative are continuous at both the glass transition temperature $\theta^{\mathrm{g}}$ and the melting temperature $\theta^{\mathrm{m}}$. The resulting free energy is computed as
    \begin{align}
        \overline{\Psi}_{l,t}^{\mathrm{cal}}& =-\iint\frac{c_t}{\theta}\mathrm{d}\theta\mathrm{d}\theta= c_4 + c_3 \theta + \frac{1}{6 (\theta^{\mathrm{g}} - \theta^{\mathrm{m}})^3} \\
        &\Bigg[ c_a \theta \Big( \theta^3 - 3(\theta^{\mathrm{g}} + \theta^{\mathrm{m}})\theta^2 + 18 \theta^{\mathrm{g}}\theta^{\mathrm{m}}\theta + 6(\theta^{\mathrm{m}})^2(3\theta^{\mathrm{g}} - \theta^{\mathrm{m}}) \Big)\notag\\
        &- c_l \theta \Big( \theta^3 - 3(\theta^{\mathrm{g}} + \theta^{\mathrm{m}})\theta^2 + 18 \theta^{\mathrm{g}}\theta^{\mathrm{m}}\theta - 6(\theta^{\mathrm{g}})^2(\theta^{\mathrm{g}} - 3\theta^{\mathrm{m}}) \Big) \notag\\
        &+ 6 \theta \Big( c_a (\theta^{\mathrm{m}})^2 (\theta^{\mathrm{m}} - 3\theta^{\mathrm{g}}) - c_l (\theta^{\mathrm{g}})^2 (\theta^{\mathrm{g}} - 3\theta^{\mathrm{m}}) \Big) \ln(\theta) \Bigg]\notag
    \end{align}  
    \item \textbf{Liquid state for $\theta\geq\theta^{\mathrm{m}}$:} For the stable liquid phase, the specific heat assumes a different constant value $c_{l}$. Integrating this constant yields the free energy representation for the stable liquid state as
    \begin{equation}
        \overline{\Psi}^{\mathrm{cal}}_{l,l}=-c_l\theta\left(\log[\theta]-1\right)+a_5\theta+a_6
    \end{equation}
\end{itemize}
The free energy $\overline{\Psi}^\mathrm{cal}_l(\theta)$ is derived by integrating the heat capacity model and determining the six integration constants (two for each region) by enforcing physical boundary and continuity conditions.
\begin{itemize}
\item \textbf{Reference State for $\theta\geq\theta^{\mathrm{m}}$:} Since only energy differences are physically relevant, we can define the free energy and its derivative (negative entropy) to be zero at the reference temperature $\theta^{\mathrm{ref}}$, which imposes 
    \begin{align}
            \overline{\Psi}^\mathrm{cal}_{l,l}(\theta^{\mathrm{ref}})=0\quad , \quad
            \frac{\partial\overline{\Psi}^\mathrm{cal}_{l,l}}{\partial\theta}\Bigg|_{\theta^{\mathrm{ref}}}=0 
    \end{align}
\item \textbf{Continuity at $\theta^{\mathrm{g}}<\theta<\theta^{\mathrm{m}}$:} The integration constants for the transition region are determined by enforcing $C^1$-continuity (continuity of the function and its first derivative) with the liquid region at the melting temperature $\theta^{\mathrm{m}}$ as 
    \begin{align}
    \overline{\Psi}^\mathrm{cal}_{l,t}(\theta^{\mathrm{m}})=\overline{\Psi}^\mathrm{cal}_{l,l}(\theta^{\mathrm{m}}) \quad , \quad   \frac{\partial\overline{\Psi}^\mathrm{cal}_{l,t}}{\partial\theta}\Bigg|_{\theta^{\mathrm{m}}}=\frac{\partial\overline{\Psi}^\mathrm{cal}_{l,l}}{\partial\theta}\Bigg|_{\theta^{\mathrm{m}}}
    \end{align}
\item \textbf{Continuity at $\theta^{\mathrm{g}}$:} The final two constants for the supercooled region are found by enforcing $C^1$-continuity with the transition region at the glass transition temperature $\theta^{\mathrm{m}}$, i.e.,
\begin{align}
        \overline{\Psi}^\mathrm{cal}_{l,s}(\theta^{\mathrm{g}})=\overline{\Psi}^\mathrm{cal}_{l,t}(\theta^{\mathrm{g}})\quad , \quad
     \frac{\partial\overline{\Psi}^\mathrm{cal}_{l,s}}{\partial\theta}\Bigg|_{\theta^{\mathrm{g}}}=\frac{\partial\overline{\Psi}^\mathrm{cal}_{l,t}}{\partial\theta}\Bigg|_{\theta^{\mathrm{g}}}
\end{align}
\end{itemize}
This procedure yields a single, smooth, and continuous piecewise function for $\overline{\Psi}^\mathrm{cal}_l(\theta)$. The free energy of the crystalline phase $\overline{\Psi}^\mathrm{cal}_c(\theta)$ is derived by linking it to the liquid phase at the melting point $\theta^{\mathrm{m}}$ accounting for the latent heat of fusion $\Delta H$. Latent heat is fundamental to the kinetics of crystallization. It is the thermal energy released during the formation of the crystal structure from the liquid phase. For the crystal to continue growing, this heat must be conducted away from the liquid-crystal interface into the surroundings. If the rate of heat removal is lower than the rate of its production, the local temperature at the interface will rise, which in turn decreases the growth rate \cite{musgraves2019springer}. In our thermodynamic model, this release of energy is mathematically captured by relating the latent heat $\Delta H$ to the difference in entropy between the liquid and crystalline phases at the melting temperature. 

The two integration constants for $\overline{\Psi}^\mathrm{cal}_c(\theta)$ (which has a constant specific heat $c_c$) are found by enforcing two physical conditions at $\theta=\theta^{\mathrm{m}}$:
\begin{itemize}
    \item The free energies of the crystalline and liquid phases are equal, requiring“ 
    \begin{equation}
        \overline{\Psi}^\mathrm{cal}_c(\theta^{\mathrm{m}})=\overline{\Psi}^\mathrm{cal}_l(\theta^{\mathrm{m}}) \,.
    \end{equation}
    \item The entropy difference between the phases creates a jump in the first derivative of the free energy equal to the latent heat as 
    \begin{equation}
\frac{\partial\overline{\Psi}^\mathrm{cal}_c}{\partial\theta}\Bigg|_{\theta^{\mathrm{m}}}    -\frac{\partial\overline{\Psi}^\mathrm{cal}_l}{\partial\theta}\Bigg|_{\theta^{\mathrm{m}}}=\frac{\Delta H}{\theta^{\mathrm{m}}}
\label{EQ: Free Energie Crystalline mit latente heat}
    \end{equation}
\end{itemize}
The free energy of the amorphous phase is initially introduced as an intermediate variable, denoted by $\overline{\Psi}^\mathrm{cal^\star}_a$. This term, which represents the 'frozen-in' solid structure, branches from the supercooled liquid function at the glass transition temperature $\theta=\theta^{\mathrm{g}}$.
To mathematically ensure that the glass transition is modeled correctly as a second-order phase transformation with a continuous entropy, $C^{1}$-continuity of the free energy is strictly enforced at this transition point. This requires satisfying two conditions:
\begin{itemize}
    \item The free energies of the amorphous and supercooled liquid phases are equal at the glass transition temperature, leading to
    \begin{equation}
        \overline{\Psi}^\mathrm{cal^\star}_a(\theta^{\mathrm{g}})=\overline{\Psi}^\mathrm{cal}_l(\theta^{\mathrm{g}}) 
    \end{equation}
    \item The entropies of the two phases are also equal, meaning their derivatives are continuous as 
    \begin{equation}
\frac{\partial\overline{\Psi}^\mathrm{cal^\star}_a}{\partial\theta}\Bigg|_{\theta^{\mathrm{g}}}  =\frac{\partial\overline{\Psi}^\mathrm{cal}_l}{\partial\theta}\Bigg|_{\theta^{\mathrm{g}}}
    \end{equation}
\end{itemize}
To provide a distinct energy potential for the amorphous state while satisfying these conditions, a regularization term is added. This term introduces a slight energy difference away from the transition point, which aids numerical stability without creating a non-physical entropy jump \cite{soleimani2025continuum}. The modified caloric free energy for the amorphous phase is defined as
\begin{equation}
    \overline{\Psi}^{\mathrm{cal}}_a = \overline{\Psi}^{\mathrm{cal^\star}}_a +     \epsilon^{\mathrm{a}}\left(\theta-\theta^{\mathrm{g}}\right)^3
    \label{EQ: Free Energie amorph}
\end{equation}
where $\epsilon^{\mathrm{a}}$ is a small regularization parameter. This ensures the necessary distinction between the amorphous and supercooled liquid states in a thermodynamically consistent manner.
\\
The resulting free energy curves for each of the three phases are plotted as a function of temperature in the \autoref{fig: Caloric energy} below. The diagram illustrates the physical characteristics captured by the model. The liquid curve (\autoref{Diag Line: Liquid}) serves as the continuous baseline from which the solid phases branch. At the melting point, the crystalline curve (\autoref{Diag Line: Crystalline}) intersects the liquid curve. The magnified view shows a distinct "kink" at this point, which represents the $C^0$-continuity (equal energy values) but $C^1$-discontinuity (a jump in the first derivative). This jump in the slope corresponds to the change in entropy due to the latent heat of fusion. At the glass transition temperature, the amorphous curve (\autoref{Diag Line: Amorphous}) branches from the liquid curve. The change in slope here is much less pronounced, reflecting the fact that the energy offset $\epsilon^{\mathrm{a}}$ is significantly smaller than $\Delta H$. At any given temperature, the phase with the lowest free energy is the most thermodynamically stable. The diagram clearly shows how the crystalline phase is the stable state below the melting temperature.

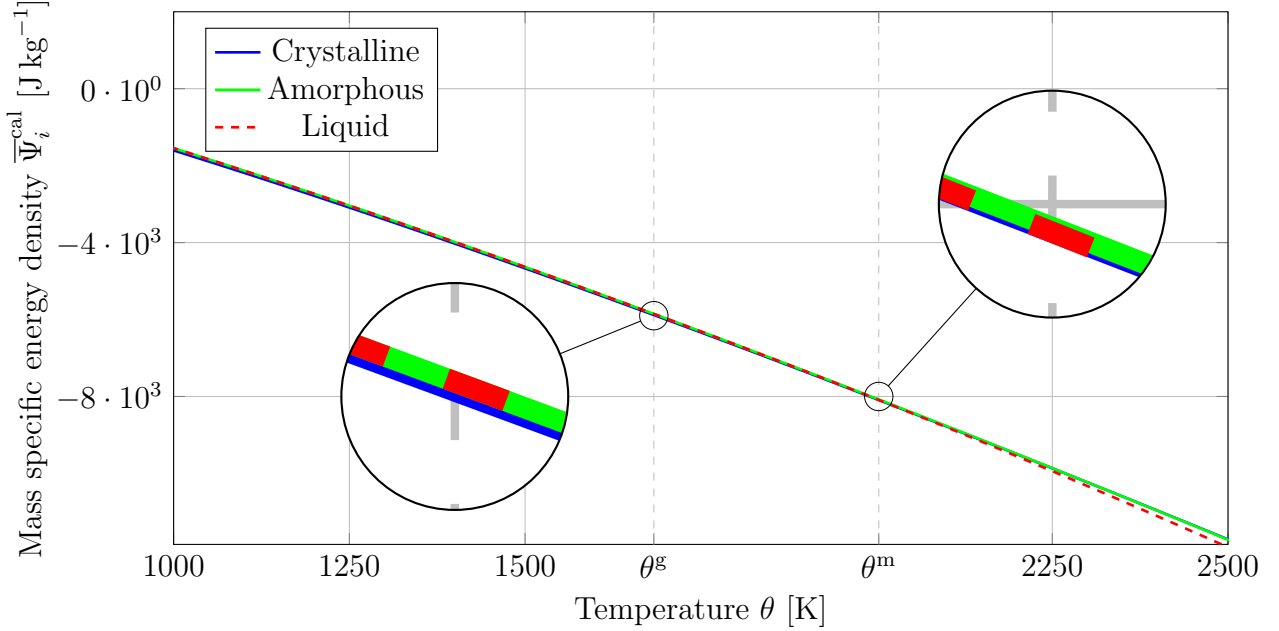
\begin{figure}[h]
    \centering
    \begin{tikzpicture}[spy using outlines=
    {circle, magnification=8, connect spies}]
\begin{axis}[
width=0.9\textwidth,height=0.50\textwidth,
xlabel= {Temperature $\theta$ [\SI{}{\kelvin}] },
ylabel={Mass specific energy density $\overline{\Psi}^\mathrm{cal}_i$ [\SI{}{\joule\per\kilogram}] },
y label style={at={(-0.05,0.5)}},
xtick={0,250,500,750,1000,1250,1500,2250,2500,2750,3000},
ymax=2E3,
xmin=1000,xmax=2500,
ytick={0,-4000,-8000,-12000,-16000},
grid=major,
xticklabel style={
 /pgf/number format/fixed,
 /pgf/number format/precision=0,
 /pgf/number format/fixed zerofill,
 /pgf/number format/1000 sep={}
},
yticklabel style={
 /pgf/number format/sci,
 %/pgf/number format/sci generic={mantissa sep=\times,exponent={10^{#1}}},
 /pgf/number format/precision=4,
 %/pgf/number format/fixed zerofill,
},
scaled ticks=false,
enlarge y limits=false,
enlarge x limits=false,
%legend style={at={(0.1,-0.37)},
% anchor=south west, legend columns=-1},
legend pos=north west,
extra x ticks={1683.15, 2003.15},
        extra x tick labels={$\theta^{\mathrm{g}}$,$\theta^{\mathrm{m}}$},
        extra x tick style={
            grid=major,
            grid style={dashed}, % Gilt für alle extra ticks, man kann es auch pro tick setzen
        },
]
\addplot [color=blue,line width=1pt] table [x=temp,y=psi_c,col sep=semicolon]{Diagrams/Caloric_Energy/CalEnergy.txt};
\label{Diag Line: Crystalline}

\addplot [color=green,line width=1] table [x=temp,y=psi_a_dc-0-1,col sep=semicolon]{Diagrams/Caloric_Energy/CalEnergy.txt};
\label{Diag Line: Amorphous}

\addplot [color=red,line width=1pt,dashed] table [x=temp,y=psi_l,col sep=semicolon]{Diagrams/Caloric_Energy/CalEnergy.txt};
\label{Diag Line: Liquid}

% --- VERTICAL LINES ADDED HERE ---
% Line for glass transition temperature
%\draw[black, thick, dotted] (axis cs:1683.15, \pgfkeysvalueof{/pgfplots/ymin}) -- (axis cs:1683, \pgfkeysvalueof{/pgfplots/ymax})
%    node[pos=0.8,xshift=7pt, rotate=00, anchor=south, font=\small] {$\theta^{\mathrm{g}}$};

% Line for melting temperature
%\draw[black, thick, dotted] (axis cs:2003.15, \pgfkeysvalueof{/pgfplots/ymin}) -- (axis cs:2003.15, \pgfkeysvalueof{/pgfplots/ymax})
%    node[pos=0.8,xshift=7pt, rotate=0, anchor=south, font=\small] {$\theta^{\mathrm{m}}$};
% ---------------------------------

%Coordinates for spy
 \coordinate (melt)   at (axis cs: 2003.15,-8*10^3);
 \coordinate (melt2) at (axis cs: 2250,-3*10^3);
 \coordinate (trans) at (axis cs: 1683.15,-5.9*10^3);
\coordinate (trans2) at (axis cs: 1400,-8*10^3);
 
\legend{Crystalline, Amorphous, Liquid};
\end{axis};
    \spy [black, size=3cm] on (melt)   in node[fill=white] at (melt2);
    \spy [black, size=3cm] on (trans)   in node[fill=white] at (trans2);
\end{tikzpicture}
    \caption{Caloric energy of the crystalline, amorphous and liquid phase.}
    \label{fig: Caloric energy}
\end{figure}
The individual caloric free energy potentials for the crystalline phase ($\overline{\Psi}_1=\overline{\Psi}_c$), the amorphous phase ($\overline{\Psi}_2=\overline{\Psi}_a$), and the liquid phase ($\overline{\Psi}_3=\overline{\Psi}_l$) are now fully defined. The total caloric free energy of the material, is the weighted average of these individual potentials, using the phase fractions as the weighting factors as
\begin{equation}
\overline{\Psi}^{\mathrm{cal,tot}}=\sum_{i=1}^3 \lambda_i \overline{\Psi}^\mathrm{cal}_i(\theta)\,.
\end{equation}

While the energy contribution of the air or void phase $\Psi^{\mathrm{air}}$ is typically negligible compared to the solid phases, a small but finite stiffness is assigned to ensure numerical stability and prevent the material model from collapsing under severe compression. To maintain thermodynamic consistency within the variational framework, the air phase is described using the same Neo-Hookean constitutive law as employed for the solid phases. The specific free energy for the air phase is therefore given by, 
\begin{equation}
\Psi^\mathrm{air} = \frac{\mu_i}{2}\left(\overline{I}_1^{\mathrm{air}}-3\right)+\frac{K_i}{2}\left(J^{\mathrm{air}}-1\right)^2
\end{equation}
where $\mu_i$ and $K_i$ denote the shear and bulk moduli of the void phase, respectively. The first invariant $\overline{I}_1^{\mathrm{air}}$ and the Jacobian $J^{\mathrm{air}}$ are calculated directly from the total deformation gradient $\bfF$. Assuming purely elastic deformation of the void phase, i.e., $\bfF^{\mathrm{air}}$ = $\bfF$, the elastic left Cauchy-Green tensor for the air phase becomes
\begin{equation}
    \bfb^{\mathrm{air}}=\bfF\cdot\bfF^\intercal\,.
\end{equation}
This simplified kinematic description is justified by the absence of viscoelastic deformation mechanisms in the void phase.

\section{Numerical Treatment}
This chapter details the numerical implementation of the derived multi-physics material model. The integration of the highly non-linear, coupled evolution equations requires robust solution for the internal variables, followed by the spatial solution of the thermal field using the Neighbored Element Method (NEM).

\subsection{Numerical Treatment of the Internal Variables}
The evolution of the phase fractions $\lambda$ is governed by the thermodynamic stationarity condition previously established in \autoref{EQ: Evolution Lambda}. Furthermore, to remain physically meaningful, this evolution must strictly obey the holonomic constraint $g=0$ introduced in \autoref{EQ: Consraint Lambda}, ensuring the sum of all phase fractions exactly equals unity at any given time. Enforcing such holonomic constraints analytically does not necessarily guarantee that those constraints are also preserved numerically upon spatiotemporal discretization. Traditionally, preserving constraint manifolds during time integration, such as isochoric plastic flow or phase volume conservation, often necessitates computationally expensive numerical schemes like exponential mapping.

To avoid the computational complexity of exponential maps, we employ the simpler, constraint-conforming Backward-Euler discretization approach proposed by Bode et al.\cite{bode2024constraint}. First, the unconstrained thermodynamic residual is defined by
\begin{equation}
    \bfe^{\bflambda}\coloneq \bfp^{\bflambda}+\frac{\partial\Delta^{\mathrm{diss,\bflambda}}}{\partial\Dot{\bflambda}}=\frac{\partial\left(\Psi^{\mathrm{cal,tot}}+\Psi^{\mathrm{pen}}+\Psi^\mathrm{mech,tot}\right)}{\partial\bflambda}+\bfeta^{\bflambda}\cdot\Dot{\bflambda}+N^{\mathrm{thres}}\frac{\Dot{\bflambda}}{||\Dot{\bflambda}||}\,.    
\end{equation}
To project this unconstrained residual onto the permissible tangent space of the constraint surface, a projection operator $\dsP^{\bflambda}$ is constructed, reading
\begin{equation}
    \dsP^{\bflambda}=\bfI-\frac{\frac{\partial g}{\partial\bflambda}\otimes\frac{\partial g}{\partial\bflambda}}{\frac{\partial g}{\partial\bflambda}\cdot\frac{\partial g}{\partial\bflambda}}\,.
\end{equation}
Applying this projection operator maps the evolution equation into a constraint-conforming subspace, which inherently creates a linear dependency in the evolution equations. The effective, constraint-conforming residual equation for the local Newton-Raphson iteration takes the form
\begin{equation}
    \Tilde{\bfe}^{\bflambda}=\dsP^{\bflambda}\cdot\bfe^{\bflambda} \,.
        \label{EQ: Evolution Equation Lambda Projected}
\end{equation}
Similarly, the evolution of the viscous internal variable $\bfC^{\mathrm{v}}$ is evaluated based on the thermodynamic balance between the driving force and the dissipative resistance derived in \autoref{EQ: Evolution of viscous deformation}. To ensure numerical stability and preserve the physical symmetry of the internal viscous metric during the time integration, the conjugate thermodynamic driving force $\bfp^{\bfC^{\mathrm{v}}}$ is explicitly symmetrized. The symmetric driving force is obtained via
\begin{equation}
    \bfp^{\bfC^{\mathrm{v}},\mathrm{sym}}=\frac{1}{2}\left[\bfp^{\bfC^{\mathrm{v}}}+\left(\bfp^{\bfC^{\mathrm{v}}}\right)^{\intercal}\right]\,.
\end{equation}
By substituting this symmetric force into the stationarity condition, the local residual for the viscous deformation reduces to
\begin{equation}  \bfe^{\bfC^\mathrm{v}}\coloneq\bfp^{\bfC^{\mathrm{v}},\mathrm{sym}}+\frac{\partial\Delta^{\mathrm{diss},\bfC^{\mathrm{v}}}}{\partial\Dot{\bfC^{\mathrm{v}}}}= \frac{\partial\Psi^\mathrm{mech,tot}}{\partial\bflambda}+\eta^{\bfC^{\mathrm{v}}}\Dot{\bfC^{\mathrm{v}}}\,.
    \label{EQ: Evolution Equation Cv symmetrized}
\end{equation}
Having established the individual residuals, the complete local system can be assembled. Because the projection operator $\dsP^{\bflambda}$ creates a linear dependency among the components of $\Tilde{\bfe}^{\bflambda}$, a direct numerical solution of the full set of projected equations would result in a singular Jacobian matrix. To formulate a well-posed local system and resolve this dependency, one of the linearly dependent equations must be replaced by the constraint in order to determine the remaining unknowns. By explicitly substituting one component of the phase evolution equations with the exact algebraic constraint $g=0$, a standard Backward-Euler discretization can be applied without suffering a loss of accuracy due to unfulfilled constraints.

Consequently, taking advantage of the symmetry of the viscous tensor $\bfC^{\mathrm{v}}$ and applying the explicit constraint replacement, the fully assembled local residual vector $\bfR$ comprises nine independent unknowns and is simultaneously solved at each integration point via a local Newton-Raphson scheme
\begin{equation}
    \bfR=
    \left(\begin{matrix}
        \Tilde{e}^{\bflambda}_1
         &\Tilde{e}^{\bflambda}_2
         &g
         &e^{\bfC^{\mathrm{v}}}_{11}
         &e^{\bfC^{\mathrm{v}}}_{22}
         &e^{\bfC^{\mathrm{v}}}_{33}
         &e^{\bfC^{\mathrm{v}}}_{12}
         &e^{\bfC^{\mathrm{v}}}_{23}
         &e^{\bfC^{\mathrm{v}}}_{13}  
    \end{matrix}
    \right)^\intercal
    =    \boldsymbol{0} \,.
\label{EQ: Material Point Lambda Evolution}
\end{equation}

\subsection{The Neighbored Element Method (NEM)}
\label{Sec: NEM}
The central concept of the NEM is to decouple the mechanical problem from the thermal problem using a staggered operator-split scheme. Within each time step, the mechanical equilibrium is first solved using implicit FEM. Subsequently, the evolution equations for temperature (including Laplacian terms) are solved in their strong form. As demonstrated by Rudolf et al. \cite{rudolf2025computational} and \cite{von2026fast}, this approach reduces computational costs and memory requirements by orders of magnitude compared to monolithic FEM solvers.

Following the approach of Blaszczyk et al. \cite{blaszczyk2022application}, the NEM approximates spatial derivatives using a generalized Finite Difference Method based on Taylor series expansion combined with Weighted Least Squares (WLS). This formulation allows accurate treatment of arbitrary meshes, addressing a known limitation of simpler finite difference schemes \cite{vogel2020adaptive}. For each interior point $\theta_0 \forall \bfX \in \Omega$, spatial derivatives are computed from a local linear system involving its $n$ nearest neighbors, yielding the linear system

\begin{align}
\underbrace{
\begin{bmatrix}
\frac{1}{2} \Delta x_1^2 & \frac{1}{2} \Delta y_1^2 &  \frac{1}{2} \Delta z_1^2 & \Delta x_1 \Delta y_1 & \Delta x_1 \Delta z_1 & \Delta y_1 \Delta z_1 & \Delta x_1 & \Delta y_1 & \Delta z_1 \\
\vdots & \vdots & \vdots & \vdots & \vdots  & \vdots  & \vdots  & \vdots  & \vdots \\
\frac{1}{2} \Delta x_n^2 & \frac{1}{2} \Delta y_n^2 &  \frac{1}{2} \Delta z_n^2 & \Delta x_n \Delta y_n & \Delta x_n \Delta z_n & \Delta y_n \Delta z_n & \Delta x_n & \Delta y_n & \Delta z_n  
\end{bmatrix}}_{\mathbf{A}}
\underbrace{
\begin{bmatrix}
\theta_{xx} \\
\theta_{yy} \\
\theta_{zz} \\
\theta_{xy} \\
\theta_{xz} \\
\theta_{yz} \\
\theta_{x} \\
\theta_{y} \\
\theta_{z}
\end{bmatrix}}_{\mathbf{m}}
=
\underbrace{
\begin{bmatrix}
\theta_1 - \theta_0 \\
\vdots \\
\theta_n - \theta_0
\end{bmatrix}}_{\mathbf{b}}
\end{align}
In this context, the solution vector $\mathbf{m}$ is employed to denote the approximated derivatives of the temperature field. Specifically, its components represent the first partial derivatives (e.g., $\theta_{x}$) and second partial derivatives (e.g., $\theta_{xx}$). The right-hand side vector $\mathbf{b}$ contains the temperature differentials $\theta_{i}-\theta_{0}$ between the neighboring points (subscript $i=1...n$) and the pivot point (subscript $0$). The matrix $\mathbf{A}$ is constructed from the spatial distances $ \Delta x_i, \Delta y_i, \Delta z_i $, which are the components of the vector $\Delta \bfX=\bfX_i-\bfX_0$ pointing from the neighbors  to the pivot point in the reference configuration. The solution to this system can be obtained through the implementation of a weighted least-squares approach if $n$ is larger than the number of spatial derivatives
\begin{align}
    \min_{\mathbf{m}} \| \mathbf{W_D}^{\frac{1}{2}} \cdot (\mathbf{A} \cdot \mathbf{m} - \mathbf{b}) \|_2^2 = \min_{\mathbf{m}} ((\mathbf{A} \cdot \mathbf{m} - \mathbf{b})^T \cdot \mathbf{W_D} \cdot (\mathbf{A} \cdot \mathbf{m} - \mathbf{b}))\,.
\end{align}
A diagonal weight matrix $\mathbf{W_D}$ has been developed to assign relative importance to each neighbor. The solution vector $\bfm$ can be used to directly extract the Laplacian coefficients required for the local form of the heat equation, taking the form
\begin{align}
    \nabla^2 \theta  \approx \sum_{i=1}^3 m_i=\sum_{i=1}^3 \sum_{j=1}^n D_{ij} \cdot b_j \quad \text{with} \quad\mathbf{D}=:(\mathbf{A}^T \cdot \mathbf{W_D} \cdot \mathbf{A})^{-1} \cdot \mathbf{A}^T \cdot \mathbf{W_D} \,.
\end{align}

This results in an explicit formula for the derivatives, which can be expressed via a local operator matrix.

For each boundary point $\theta_0\forall\bfX\in\partial\Omega$, the local system must be adapted to concurrently solve for the unknown temperature value at the boundary itself. In contrast to the interior formulation, the approximation at the boundary relies on a first-order Taylor series expansion. To fully determine this system, the physical boundary conditions must be directly embedded into the local matrix structure. Consequently, the linear system is extended to incorporate the prescribed flux, yielding

\begin{align}
\underbrace{
\begin{bmatrix}
\Delta x_1 & \Delta y_1 & \Delta z_1 &1\\
\vdots  & \vdots  & \vdots &\vdots\\
 \Delta x_n & \Delta y_n & \Delta z_n & 1\\
N_x&N_y&N_z&0
\end{bmatrix}}_{\mathbf{\Tilde{A}}}
\underbrace{
\begin{bmatrix}
\theta_{x} \\
\theta_{y} \\
\theta_{z} \\
\theta_{0}
\end{bmatrix}}_{\mathbf{\Tilde{m}}}
=
\underbrace{
\begin{bmatrix}
\theta_1  \\
\vdots \\
\theta_n \\
H^{\mathrm{q}}
\end{bmatrix}}_{\mathbf{\Tilde{b}}}
\end{align}
while simultaneously enforcing the Robin boundary condition ($-\alpha^{\mathrm{cond}}\nabla_{\bfX}\theta_0\cdot\bfN=H^{\mathrm{q}}$) with the outward pointing normal vector $\bfN=(N_x, N_y, N_z)$. This yields a similar expression with a modified local operator matrix 

\begin{align}
    \theta_0  \approx \sum_{j=1}^{n} \Tilde{D}_{4,j} \cdot \Tilde{b}_j +\Tilde{D}_{4,n+1}\cdot H^{\mathrm{q}}\quad \text{with} \quad\mathbf{\Tilde{D}}=:(\mathbf{\Tilde{A}}^T \cdot \mathbf{\Tilde{W_D}} \cdot \mathbf{\Tilde{A}})^{-1} \cdot \mathbf{\Tilde{A}}^T \cdot \mathbf{\Tilde{W_D}} \,.
\end{align}
In this work, a monolithic modification to the staggered NEM scheme \cite{blaszczyk2022application} is proposed. By assembling a single global operator matrix, we eliminate the time-lag errors inherent in staggered solutions and improve the stability and convergence rate of the numerical scheme, particularly for problems with strong coupling. Instead of solving locally, we first assemble the local operator matrices for all points into a single global operator matrix, $\bfG \in \dsR^{k \times k}$ with $k \in \left(\Omega+\partial\Omega\right)$. This matrix relates the vector of all temperature values, $\boldsymbol{\theta}^{\myi} \in \dsR^p$ with $p \in \Omega$, to the vector of their Laplacians, and boundary values $\boldsymbol{\theta}^{\myb} \in \dsR^q$  with $q \in \partial\Omega$. This global system can be partitioned into blocks corresponding to the inner points (superscript $\myi$) and the boundary points (superscript $\myb$), taking the form
\begin{equation}
\begin{bmatrix} \nabla_{\bfX}^2\boldsymbol{\theta}^{\myi} \\ \boldsymbol{\theta}^{\myb} \end{bmatrix} = \bfG \cdot \boldsymbol{\theta} = \begin{bmatrix} \bfG^{\myi\myi} & \bfG^{\myi\myb} \\ \bfG^{\myb\myi} & \bfG^{\myb\myb} \end{bmatrix} \begin{bmatrix} \boldsymbol{\theta}^{\myi} \\ \boldsymbol{\theta}^{\myb} \end{bmatrix} +\begin{bmatrix}
    \boldsymbol{0} \\ \bfG^{\mathrm{\bfH}}\cdot\bfH^{\mathrm{q}} 
\end{bmatrix}
\,.
\label{EQ: NEM local BC}
\end{equation}
Here, the vector $ \bfH^{\mathrm{q}}$ aggregates the source terms from the Neumann boundary conditions for all boundary points. Each entry in the diagonal matrix $\bfG^{\mathrm{\bfH}} \in \dsR^{q \times q}$ corresponds to the term $\Tilde{D}_{10,n}$ for the respective point. In \cite{blaszczyk2022application}, the local operators were applied in a staggered manner. First, the boundary point values are updated based on the inner point values from the previous state. Then, these updated boundary values are used to calculate the Laplacians, $\nabla^2_{\bfX}\theta^{\myi}$, for the inner points. 
\begin{enumerate}
    \item {Coupling the boundary points according to
    \begin{equation}
        \boldsymbol{\theta}^{\myb} = \bfG^{\myb\myi}\cdot \boldsymbol{\theta}^{\myi}+\bfG^{\myb\myb}\cdot\boldsymbol{\theta}^{\myb}+\bfG^{\mathrm{\bfH}}\cdot\bfH^{\mathrm{q}}
        \label{EQ: NEM Staggered for BC}
    \end{equation}
    }
    \item {Coupling the inner point Laplacians via
    \begin{equation}
        \nabla_{\mathbf{X}}^2\boldsymbol{\theta}^{\myi}=\bfG^{\myi\myi}\cdot\boldsymbol{\theta}^{\myi}+\bfG^{\myi\myb}\cdot\boldsymbol{\theta}^{\myb}
        \label{EQ: NEM Staggered for Inner}
    \end{equation}
    }
\end{enumerate}
The source term $\bfH^{\mathrm{q}}$ can represent various physical phenomena. For the thermal problem considered here, it models heat convection across the boundary surface $\partial\Omega$ and is defined via
\begin{equation}
    \bfH^{\mathrm{q}}=-\frac{\alpha^{\mathrm{conv}}}{\alpha^{\mathrm{cond}}}\cdot\left(\boldsymbol{\theta}^{\myb}-\theta^{\mathrm{ref}}\boldsymbol{1}^\mathrm{v}\right) \,.
    \label{EQ: Convection Law}
\end{equation}
where $\alpha^{\mathrm{conv}}$ is the heat transfer coefficient, $\alpha^{\mathrm{cond}}$ the heat conduction coefficent, and $\boldsymbol{1}^\mathrm{v}$ is an identity vector of appropriate size. 

\subsubsection{Time integration of the Heat Equation inside the FEM Framework}

The spatial discretization is realized by embedding the monolithic Laplacian operator, defined in \autoref{EQ: NEM local BC}, directly into the strong form of the heat equation, yielding
\begin{align}
\kappa\dot{\boldsymbol{\theta}}^{\myi}+\alpha^{\mathrm{cond}}\left[\bfG^{\myi\myi}\cdot\boldsymbol{\theta}^{\myi}+\bfG^{\myi\myb}\cdot\boldsymbol{\theta}^{\myb}\right]
-\left[\bfp^{\bflambda}-\boldsymbol{\theta}^{\myi}\cdot\frac{\partial \bfp^{\bflambda}}{\partial\boldsymbol{\theta}^{\myi}}\right]\cdot\Dot{\bflambda} \notag \\ 
-\left[\bfp^{\bfC^{\mathrm{v}}}-\boldsymbol{\theta}^{\myi}\cdot\frac{\partial \bfp^{\bfC^{\mathrm{v}}}}{\partial\boldsymbol{\theta}^{\myi}}\right]:\Dot{\bfC^{\mathrm{v}}}-\boldsymbol{\theta}^{\myi}\cdot\frac{\partial \bfP}{\partial\boldsymbol{\theta}^{\myi}}:\Dot{\bfF}+h^{\star}&=\boldsymbol{0} \quad \forall \bfX \in\Omega \,, \\
\boldsymbol{\theta}^{\myb} - \bfG^{\myb\myi}\cdot \boldsymbol{\theta}^{\myi}-\bfG^{\myb\myb}\cdot\boldsymbol{\theta}^{\myb}+\bfG^{\mathrm{\bfH}}\cdot\bfH^{\mathrm{q}}&=\boldsymbol{0} \quad \forall \bfX \in\partial\Omega \,. \notag
\end{align}

To balance computational efficiency with numerical robustness, a semi-implicit time-integration scheme is adopted. The linear, high-stiffness terms, specifically the rate of caloric energy change and the Fourier heat conduction, are treated implicitly using a Backward Euler discretization. This ensures unconditional stability regarding the thermal diffusion. Conversely, the highly non-linear coupling terms, including the latent heat of transformation and the viscous dissipation work, are evaluated explicitly. They are based on the converged state at the previous time increment $t$. This operator-split avoids the requirement for a global Newton-Raphson iteration of the multi-physics system while maintaining a high degree of accuracy for typical manufacturing cooling rates. Aggregating these explicit coupling contributions into an effective thermal source vector $\bfQ^{\mathrm{coup}}_t$ takes the form
\begin{equation}
\bfQ^{\mathrm{coup}}_t = -\left[\bfp^{\bflambda}_t-\boldsymbol{\theta}^{\myi}_t\cdot\frac{\partial \bfp^{\bflambda}}{\partial\boldsymbol{\theta}^{\myi}}\Bigg|_{t}\right]\cdot\Dot{\bflambda}_t
- \left[\bfp^{\bfC^{\mathrm{v}}}_t-\boldsymbol{\theta}^{\myi}_t\cdot\frac{\partial \bfp^{\bfC^{\mathrm{v}}}}{\partial\boldsymbol{\theta}^{\myi}}\Bigg|_{t}\right]:\Dot{\bfC^{\mathrm{v}}}_t - \boldsymbol{\theta}^{\myi}_t\cdot\frac{\partial \bfP}{\partial\boldsymbol{\theta}^{\myi}}\Bigg|_{t}:\Dot{\bfF}_t - h^{\star}_t \,.
\end{equation}

Equating the implicitly treated diffusion terms to this explicit coupling source yields the final linear system of equations for the updated temperature field $\boldsymbol{\theta}^{t+\Delta t}$, expressed by
\begin{equation}
    \kappa\frac{\boldsymbol{\theta}^{t+\Delta t}-\boldsymbol{\theta}^t}{\Delta t}+\alpha^{\mathrm{cond}}\left[\bfG^{\myi\myi}\cdot\boldsymbol{\theta}^{\myi, t+\Delta t}+\bfG^{\myi\myb}\cdot\boldsymbol{\theta}^{\myb, t+\Delta t}\right] =- \bfQ^{\mathrm{coup}}_t \,.
\end{equation}
This assembled linear equation system can subsequently be solved with high computational efficiency using standard linear solvers.

\section{Material, Process and Numerical Parameters}
This chapter details the material, numerical, and process parameters that define the constitutive model for the material under investigation. The fundamental thermomechanical behavior of the material is described by the parameters listed in \autoref{tab:Material Parameter}. It is important to note that these values can vary significantly across different literature sources. Therefore, the parameters used here represent averaged or commonly accepted values from the literature to provide a representative model. These values, compiled from various literature sources, represent the material's response in its crystalline ($c$), amorphous ($a$), and liquid ($l$) phases. Parameters marked with an $^{\star}$ were not taken from literature but were chosen specifically to ensure numerical stability within the model, particularly in modelling the liquid phase. The air or void volume fraction is represented by subscript $air$ in the model. To capture convective heat transfer at the domain boundaries, the air phase is assigned a high thermal conductivity. This elevated conductivity effectively simulates the rapid heat transport via convection that occurs at the boundary elements, ensuring that thermal energy is quickly redistributed to the edges of the computational domain.

\begin{table}[H]
    \centering
    \begin{tabular}{|p{5cm}|p{2cm}|p{4cm}|p{2cm}|p{2.5cm}|}
        \hline
        \textbf{Name} & \textbf{Sign} & 
        \textbf{Value} & \textbf{Unit}&\textbf{Reference} \\
        \hline
        \rowcolor[HTML]{EFEFEF} 
        Latent heat& $\Delta H$  & $1.5 \cdot 10^5 $& \SI{}{\joule\per\kilogram}&\cite{shackelford2001materials,richet1986thermochemical} \\
        Specific heat capacity (c/a/l/air)&$c_c$/$c_a$/$c_l$\newline$c_{\mathrm{air}}$  & $740$/$740$/$1427$\newline$1021$& \SI{}{\joule \per \kilogram \per\kelvin}&\cite{bansal1986handbook, richet1993entropy, NISTJanafSiO2,kohlhaas1965warmeleitfahigkeit} \\
         \rowcolor[HTML]{EFEFEF} 
        Young's modulus (c/a/l/air)&$E_c$/$E_a$
        \newline$E_l$/$E_{\mathrm{air}}$&$7.3\cdot 10^{10}$/$7.3\cdot 10^{10}$/\newline$7.3\cdot 10^{7\star}$/$1\cdot 10^{7\star}$& \SI{}{\pascal}&\cite{musgraves2019springer,shackelford2001materials, bansal1986handbook, richet1986thermochemical, robu2024Datasheet}\\
        Reference temperature & $\theta^{\mathrm{ref}}$& $295.15$& \SI{}{\kelvin}&\\
        \rowcolor[HTML]{EFEFEF}
         Glass transition temperature & $\theta^{\mathrm{g}}$& $1523.15$/$1683.15^{\star}$& \SI{}{\kelvin}&\cite{stebbins1984heat,bansal1986handbook,richet1986thermochemical,schaeffer2020herstellung}\\
        Melting Temperature &$\theta^{\mathrm{m}}$& $2003.15$ & \SI{}{\kelvin}&\cite{tooley1985handbook,uhlmann1980glass,seward2005high,shackelford2001materials, schaeffer2020herstellung}\\
        \rowcolor[HTML]{EFEFEF}
        Poisson's ratio (c/a/l/air) &$\nu_c$/$\nu_a$/$\nu_l$\newline$\nu_{\mathrm{air}}$& $0.17$/$0.17$/$0.25$\newline$0.05^{\star}$&-&\cite{musgraves2019springer,robu2024Datasheet,shackelford2001materials}\\
        Coefficient of Thermal expansion (c/a/l)&$\alpha^{\mathrm{exp}}$&$5.7\cdot10^{-7}$&\SI{}{\per \kelvin}&\cite{bansal1986handbook,musgraves2019springer,shackelford2001materials,robu2024Datasheet}\\
        \rowcolor[HTML]{EFEFEF}
        Density (c/a/l/V) &$\rho_c$/$\rho_a$/$\rho_l$\newline$\rho_{\mathrm{air}}$&$2650$/$2200$/$2000$\newline$1.16$&\SI{}{\kilogram \per  \cubed\metre}&\cite{bansal1986handbook,musgraves2019springer,robu2024Datasheet,shackelford2001materials, robu2024Datasheet,Howie1996silica,baumgartner2020reappraising}\\
        Coefficient of thermal conduction (c/a/l/air) &$\alpha^{\mathrm{cond}}_c$\newline
        $\alpha^{\mathrm{cond}}_a$\newline
        $\alpha^{\mathrm{cond}}_l$\newline
        $\alpha^{\mathrm{cond}}_{\mathrm{air}}$\newline
        &$1.1$\newline
        $1.1$\newline
        $1.1$\newline
        $0.02852$&\SI{}{\watt\per\metre\per\kelvin}&\cite{bansal1986handbook,schaeffer2020herstellung,musgraves2019springer,shackelford2001materials}\\
    \hline
    \end{tabular}
    \caption{Material parameter}
    \label{tab:Material Parameter}
\end{table}
The glass transition temperature $\theta^{\mathrm{g}}$ is fundamentally a kinetic phenomenon rather than a fixed thermodynamic equilibrium point. Its observed value is intrinsically dependent on the cooling rate. The literature value of $1523.15 \SI{}{\kelvin}$ (\autoref{tab:Material Parameter}) corresponds to standard quasi-static measurement conditions (e.g., differential scanning calorimetry), where cooling rates are typically low. However, the laser-based additive manufacturing process subjects the material to extreme cooling rates. Under such rapid quenching conditions, the structural relaxation time of the supercooled liquid exceeds the available process time at significantly higher temperatures, causing the microstructure to \textit{freeze} earlier than in equilibrium experiments. To account for this kinetic shift, $\theta^{\mathrm{g}}$ is treated as a process-dependent parameter in this framework. The adjusted value of $\theta^{\mathrm{g}}=1683.15 \SI{}{\kelvin}$ representing an upward shift of $160\SI{}{\kelvin}$ is therefore necessary to accurately capture the transition from the viscous liquid to the non-equilibrium glassy state under the specific high-rate conditions of the simulation.

A critical aspect of the model is the temperature-dependent viscosity of the crystalline phase, which is modeled as a continuous function of temperature reading
\begin{equation}
        \eta_c^{\lambda}=\SI{10}{\pascal\second}+\frac{\eta^{\lambda,\mathrm{max}}-\SI{10}{\pascal\second}}{2}\cdot\left(1-\tanh{\left[\frac{\theta-\left(\theta^{\mathrm{m}}-\SI{10}{\kelvin}\right)}{\SI{1}{\kelvin}}\right]}\right) \,.
        \label{EQ: eta cryst}
\end{equation}
This function creates a steep, step-like transition from a very high viscosity (solid state) to a low viscosity (liquid state) centered around the melting temperature. The hyperbolic tangent function was chosen over a standard sigmoid function due to its advantageous numerical properties. Specifically, the derivative of the tanh function is wider and smoother, which leads to better-behaved gradients during the numerical solution process and thus enhances the overall stability of the simulation. 
For the amorphous and liquid phases, a constant viscosity is applied, taking the form
\begin{equation}
    \eta_a^{\lambda}=\eta_l^{\lambda}=1\SI{}{\pascal\second} \,.
\label{EQ: eta amorph liquid}
\end{equation} 

\begin{figure}[H]
    \centering
    \begin{tikzpicture}[spy using outlines=
    {circle, magnification=8, connect spies}]
\begin{axis}[
    width=0.9\textwidth,
    height=0.5\textwidth,
    xlabel={Temperature $\theta$ [\SI{}{\kelvin}]},
    ylabel={Viscosity $\eta$ [\SI{}{\pascal\second}]},
    grid=major,
    ymode=log,
    domain=0:3000,
    legend pos=south west,
    legend cell align={left}, 
    xtick={0,250,500,750,1000,1250,1500,2250,2500,2750,3000},
    ymin=1e-1,
    ymax=1e21,
    enlarge y limits=false, 
    enlarge x limits=false, 
    % --- Definition der Konstanten aus der Tabelle ---
    declare function={
        % Parameter für eta_a (amorph/liquid)
        Na_A = 1e-1;
        Na_B = 2500;
        Na_T0 = 1173.15;        
        % Parameter für eta_c (kristallin)
        Nc_A = 10.0;
        Nc_B = 10.0;
        Nc_T0 = 1.0;        
        % Gemeinsame Parameter
        eta_max = 1e20;        
        % ACHTUNG: Angenommener Wert für die Schmelztemperatur!
        % Bitte durch den korrekten Wert ersetzen.
        theta_m = 2003.15; 
    }, % <--- DAS FEHLERHAFTE KOMMA WURDE HIER ENTFERNT
    xticklabel style={
        /pgf/number format/fixed,
        /pgf/number format/precision=0,
        /pgf/number format/fixed zerofill,
        /pgf/number format/1000 sep={}
    },
    yticklabel style={
        /pgf/number format/sci,
        /pgf/number format/precision=4,
    },
extra x ticks={1683.15, 2003.15},
        extra x tick labels={$\theta^{\mathrm{g}}$,$\theta^{\mathrm{m}}$},
        extra x tick style={
            grid=major,
            grid style={dashed}, % Gilt für alle extra ticks, man kann es auch pro tick setzen
        },
]

% --- Plot für die kristalline Phase (eta_c) ---
\addplot[
    blue,
    line width=1pt, 
    samples=250
    % Hier ist keine besondere Option nötig, da die Kurve im Bereich bleibt
] 
{ Nc_A + (eta_max - Nc_A) * 0.5 * (1 - tanh((x - (theta_m - Nc_B)) / Nc_T0)) };

\addplot[
    black,
    line width=1pt, 
    samples=250
    % Hier ist keine besondere Option nötig, da die Kurve im Bereich bleibt
] 
{ (x <= 1) ? eta_max : min(max(10^(-22+(30300/(x)))*10^6,1),eta_max };
\label{Line: Viscosity mechanical}

% --- Vertikale Linien (unverändert) ---
% Linie für die Glasübergangstemperatur (angenommener Wert)
%\draw[black, thick, dotted] (axis cs:1683.15, 1e-1) -- (axis cs:1683.15, 1e21)
 %   node[pos=0.85,xshift=7pt, rotate=00,yshift=00pt, anchor=south, font=\small] {$\theta^{\mathrm{g}}$};

% Linie für die Schmelztemperatur
%\draw[black, thick, dotted] (axis cs:2003.15, 1e-1) -- (axis cs:2003.15, 1e21)
%    node[pos=0.85,xshift=7pt, rotate=0, anchor=south, font=\small] {$\theta^{\mathrm{m}}$};

%Coordinates for spy
 \coordinate (melt)   at (axis cs: 1997,7e19);
 \coordinate (melt2) at (axis cs: 2450,1e16);
  \coordinate (melt3)   at (axis cs: 2003,13);
 \coordinate (melt4) at (axis cs: 1000,1e1);
\legend{$\eta^{\lambda}_c$,$\eta^{\bfC^{\mathrm{v}}}$};
\end{axis};
%    \spy [black, size=2.5cm] on (melt)   in node[fill=white] at (melt2);
%    \spy [black, size=2.5cm] on (melt3)   in node[fill=white] at (melt4);

\end{tikzpicture}
    \caption{The phase transformation viscosity for the crystalline, amorphous, liquid volume fraction and the mechanical viscosity over the temperature domain.}
    \label{fig: eta lambda}
\end{figure}
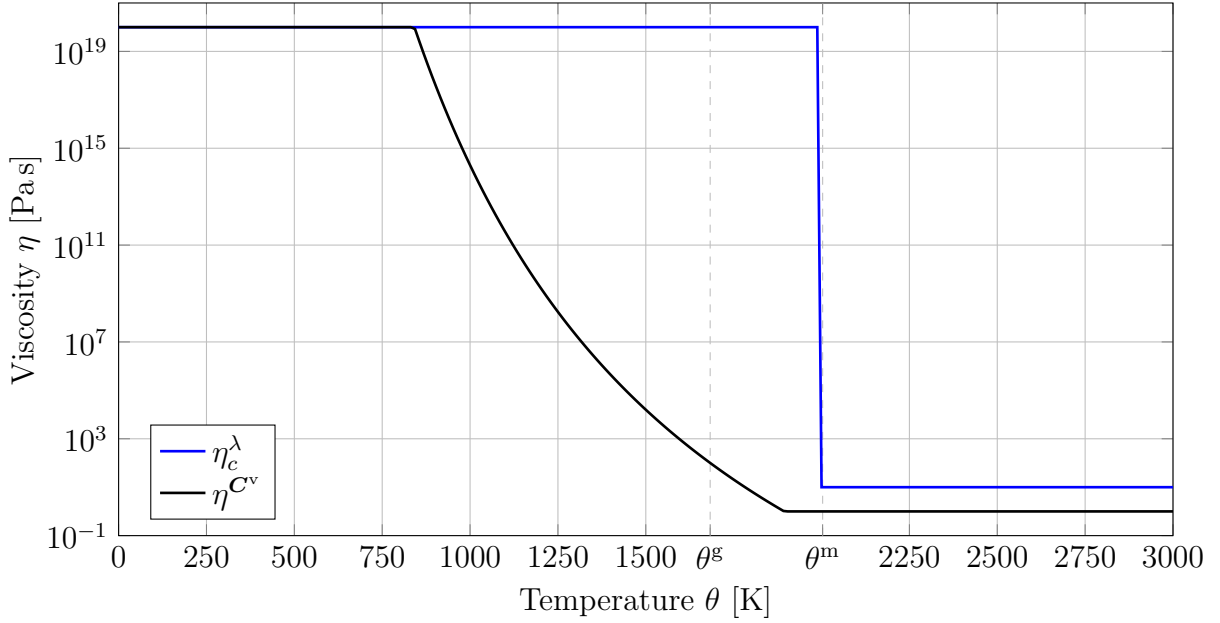

In addition to the tensor of the phase transformation viscosity $\bfeta^{\bflambda}$, the model incorporates a unified temperature-dependent mechanical viscosity $\eta^{\bfC^{\mathrm{v}}}$ to describe the macroscopic viscous deformation of the material. This viscosity directly determines the dissipative forces in the evolution equation (\autoref{EQ: Evolution of viscous deformation}) for the viscous strain $\bfC^{\mathrm{v}}$. Based on established data for glass-forming materials, the mechanical viscosity takes the form
\begin{equation}
    \eta^{\bfC^{\mathrm{v,dev}}}=
    \begin{cases}
        \eta^{\bfC^{\mathrm{v,max}}} \SI{}{\pascal\second}& \text{if }\theta \leq \SI{600}{\kelvin}\\
        \min\left[\max\left(\exp\left[-22+\frac{\SI{30000}{\kelvin}}{\theta}\right],\eta^{\bfC^{\mathrm{v,min}}}\right),\eta^{\bfC^{\mathrm{v,max}}}\right] \SI{}{\pascal\second}&\text{if }\theta > \SI{600}{\kelvin}\,.\\
    \end{cases}
\end{equation}
This formulation captures the characteristic relaxation behavior of the material and is known as Vogel-Fulcher-Tammann equation \cite{nascimento2007data,ikeda2013understanding,mauro2009viscosity}. As illustrated in \autoref{fig: eta lambda} (\autoref{Line: Viscosity mechanical}), $\eta^{\bfC^{\mathrm{v}}}$ maintains a physically realistic response across the temperature domain \cite{shackelford2001materials,scholze1988glas,shelby2005introduction,bansal1986handbook}, bounded by $\eta^{\bfC^{\mathrm{v,max}}}$ and $\eta^{\bfC^{\mathrm{v,min}}}$ to prevent numerical instabilities associated with the singularity of the exponential term at low and high temperatures, respectively. Consequently, these bounds prevent numerical divergence caused by the extreme viscosity contrasts between the liquid and solid phases ($\approx 10^{5}\SI{}{\pascal\second}$ to $\approx 10^{20}\SI{}{\pascal\second}$), thereby facilitating a stable simulation of the entire phase transformation cycle.

The distinct behavior of the crystalline transformation viscosity and the mechanical viscosity are illustrated in \autoref{fig: eta lambda}. The crystalline phase $\eta_c^{\lambda}$ maintains an extremely high viscosity until the melting temperature, indicating its solid nature. The diagram reveals that this transition is not modeled as a discontinuity, but rather as a steep, continuous gradient between the viscosity limits. 

To ensure the stability and accuracy of the numerical implementation, a set of specific parameters is required, as detailed in \autoref{tab:Numerical Parameter}. These are not physical properties of the material, but are essential for the computational solution of the governing equations.
\begin{table}[H]
    \centering
    \begin{tabular}{|p{5cm}|p{2cm}|p{4cm}|p{2cm}|}
        \hline
        \textbf{Name} & \textbf{Sign} & 
        \textbf{Value} & \textbf{Unit} \\
        \hline
        \rowcolor[HTML]{EFEFEF} 
    \hline
    Regularization parameter &$\epsilon^{\mathrm{a}}$ &$0.5$&\SI{}{\joule\per\kilogram}\\
    Penalty parameter &$N^{\mathrm{pen}}$ &$10^{-8}$&\SI{}{\joule\per\metre\cubed}\\
    \rowcolor[HTML]{EFEFEF}
    Threshold &$N^{\mathrm{thres}}$&$10^{-2}$&\SI{}{\pascal}\\
    Maximum viscosity limit for volume fraction &$\eta^{\lambda,\mathrm{max}}$&$10^{20}$&\SI{}{\pascal\second}\\
    \rowcolor[HTML]{EFEFEF}
    Minimum viscosity limit for mechanical viscosity &$\eta^{\bfC^{\mathrm{v,min}}}$&$1$&\SI{}{\pascal\second}\\
    Maximum viscosity limit for mechanical viscosity &$\eta^{\bfC^{\mathrm{v,max}}}$&$10^{20}$&\SI{}{\pascal\second}\\
    \rowcolor[HTML]{EFEFEF}
    Viscosity of the amorphous phase &$\eta_a^{\lambda}$&$1$&\SI{}{\pascal\second}\\
    Viscosity of the amorphous phase &$\eta_l^{\lambda}$&$1$&\SI{}{\pascal\second}\\
    \hline
    \end{tabular}
    \caption{Numerical parameter}
    \label{tab:Numerical Parameter}
\end{table}
An external heat source drives the model, with process parameters specified in \autoref{tab:Process Parameter}. These parameters define a Gaussian laser beam, which is commonly used in processes such as selective laser melting or welding \cite{jibben20233d,prisco2021shape,eagar1983temperature,zhang2018simulation}. The spatial distribution of the volumetric heat flux $h^{\star}$ is mathematically expressed by
\begin{equation}
    h^{\star} = \frac{L^{\mathrm{A}}}{\pi} \exp\left(-2 \frac{\Delta x^2 + \Delta y^2}{r^2} - \frac{\Delta z^2}{\left(r^{\mathrm{d}}\right)^2}\right)\,.
    \label{EQ: Gauß laser beam}
\end{equation}
Here, $\Delta x$, $\Delta y$, and $\Delta z$ represent the spatial distance from the center of the laser beam. The parameter $r$ defines the beam radius in the radial direction, and $r^{\mathrm{d}}$ controls the effective penetration depth of the laser energy. The term $L^{\mathrm{A}}$ is defined via
\begin{equation}
    L^{\mathrm{A}} = \frac{2 P}{r^2r^{\mathrm{d}} \sqrt{\pi}}
    \label{EQ: Laser amplitude}
\end{equation}
and serves as a normalization factor that ensures the total energy input rate exactly corresponds to the specified laser power $P$.
\begin{table}[H]
    \centering
    \begin{tabular}{|p{5cm}|p{2cm}|p{4cm}|p{2cm}|}
        \hline
        \textbf{Name} & \textbf{Sign} & 
        \textbf{Value} & \textbf{Unit} \\
        \hline
    \rowcolor[HTML]{EFEFEF} 
    \hline
    Laser power&$P$ &$0.35$&\SI{}{\watt}\\
    Laser radius &$r$&$0.18$&\SI{}{\milli\metre}\\
    \rowcolor[HTML]{EFEFEF} 
    Laser depth &$r^{\mathrm{d}}$&$0.02$&\SI{}{\milli\metre}\\
    Laser speed &$v$&$0.5$&\SI{}{\milli\metre\per\second}\\
    \rowcolor[HTML]{EFEFEF} 
    Convection coefficient &$\alpha^{\mathrm{conv}}$&$10$&\SI{}{\watt\per\metre\squared\per\kelvin}\\
    \hline
    \end{tabular}
    \caption{Process parameter}
    \label{tab:Process Parameter}
\end{table}

\section{Numerical Results}
This chapter presents the numerical findings obtained from the simulations, highlighting the model's ability to capture the complex path-dependency of glass-forming materials. The analysis begins with material point investigations to characterize the intrinsic kinetic behavior of the phase transformations. By prescribing specific thermal cycles, we evaluate the model’s performance across a broad kinetic spectrum, ranging from rapid quenching to quasi-static equilibrium. Following this fundamental validation, the results of the three-dimensional finite element simulation are presented to discuss the accumulation of residual stresses in additive manufacturing.

\subsection{Material Point}
\label{Sec: Material Point Results}
The initial set of numerical experiments focuses on a single, zero-dimensional material point to verify the fundamental thermodynamic coupling between temperature, phase fractions, and internal dissipation. To rigorously isolate the intrinsic response driven by phase transformations, external mechanical loading and thermal expansion are neglected by constraining both the total deformation gradient $\bfF$ and the thermal deformation gradient $\bfF^{\mathrm{th}}$ to the identity matrix $\bfI$. Under these specific boundary conditions, the kinematic state is governed solely by the interplay between phase-change-induced density shifts $\bfF^{\rho}$ and the viscous internal variable $\bfC^{\mathrm{v}}$. Consequently, any volumetric mismatch resulting from the evolving phase fractions generates internal stresses, which are dissipated exclusively through the material's internal viscous relaxation mechanism. This configuration ensures a precise assessment of the model's ability to capture the kinetic freezing of the microstructure during vitrification.

To evaluate this path-dependent behavior, each simulation prescribes a thermal cycle wherein $\SI{8000}{\joule}$ of energy is supplied during heating and subsequently removed during cooling, with the cycle reversing at the simulation midpoint. Varying the cycle duration allows investigation of the model's ability to capture the critical influence of cooling rate on the final phase composition, a defining characteristic of glass-forming materials \cite{ediger1996supercooled,angell1995formation}. This rate-dependent behavior is fundamentally governed by the transformation kinetics, a phenomenon extensively documented in Time-Temperature-Transformation (TTT) \cite{musgraves2019springer, weinberg1990critical, uhlmann1982kinetics}. These diagrams delineate the critical cooling rates required to suppress crystallization and achieve vitrification. To bypass the \textit{nose} of the crystallization curve and freeze the liquid structure into an amorphous solid, the material must be cooled faster than a specific critical rate. Conversely, slower cooling provides sufficient time for atomic rearrangement, allowing crystallization to proceed toward thermodynamic equilibrium.

The following sections systematically explore this spectrum across three regimes: \autoref{Sec: High Cooling Rate} examines rapid cooling exceeding the critical rate, resulting in complete vitrification;  \autoref{Sec: Slow Cooling Rate} investigates slow cooling well below the critical rate, allowing full crystallization and \autoref{Sec: Intermediate Cooling Rate} focuses on intermediate cooling rates producing mixed-phase microstructures
    
\subsubsection{High Cooling Rate: Vitrification  and Hysteresis}
\label{Sec: High Cooling Rate}
\begin{figure}
    \centering
    \begin{tikzpicture}
        \begin{axis}[
            width=0.8\textwidth,
            height=0.4\textwidth,
            xlabel={Time $t$ [$\SI{}{\second}$]},
            ylabel={Volume fraction $\lambda$ [-]},
            ytick={0,0.5,1},
            grid=major,
            ymax=1.2,
            enlarge x limits=false,
            enlarge y limits=false,
            ]
            % Ersten Datensatz zeichnen (linke Y-Achse)
            \addplot[ blue,thick] table [x index=0,y index=2,col sep=semicolon] {Diagrams/Phase_trans/Datas/fast_time_temp_cryst_amorph_liq.txt};;
            \label{Line: fast Crystalline}
            \addplot[ green,thick] table [x index=0,y index=3,col sep=semicolon] {Diagrams/Phase_trans/Datas/fast_time_temp_cryst_amorph_liq.txt};
            \label{Line: fast Amorph}
            \addplot[ red,thick] table [x index=0,y index=4,col sep=semicolon] {Diagrams/Phase_trans/Datas/fast_time_temp_cryst_amorph_liq.txt};
            \label{Line: fast Liquid}
        \end{axis}
        
        % Zweite Achse
        \begin{axis}[
            width=0.8\textwidth,
            height=0.4\textwidth,
            xlabel={},
            ylabel={Temperature $\theta$ [$\SI{}{\kelvin}$]},
            ylabel near ticks,
            ymin=300,
            ytick={\pgfkeysvalueof{/pgfplots/ymin}, \pgfkeysvalueof{/pgfplots/ymax}},
            axis y line*=right,
            axis x line=none,
            legend style={at={(0.05,-0.25)},anchor=north west,legend columns=-1
            },
            enlarge x limits=false,
            yticklabel style={/pgf/number format/1000 sep=,/pgf/number format/precision=0},
        extra y ticks={1683.15, 2003.15},
        extra y tick labels={$\theta^{\mathrm{g}}$,$\theta^{\mathrm{m}}$},
        extra y tick style={
            grid=major,
            grid style={dashed}, % Gilt für alle extra ticks, man kann es auch pro tick setzen
        },
       % ymajorgrids,
       % grid style={dashed},
            ]
            \addlegendimage{/pgfplots/refstyle=Line: fast Crystalline}\addlegendentry{Crystalline}
            \addlegendimage{/pgfplots/refstyle=Line: fast Amorph}\addlegendentry{Amorphous}
            \addlegendimage{/pgfplots/refstyle=Line: fast Liquid}\addlegendentry{Liquid}
            \addplot[black,thick
            ] table [x index=0,y index=1, col sep=semicolon
            ] {Diagrams/Phase_trans/Datas/fast_time_temp_cryst_amorph_liq.txt};
            \addlegendentry{Temperature}
        
%            \draw [gray, dashed]({axis cs:0, 2126}) -- ({axis cs:40, 2126});
            
        % Schattierter Bereich für die latente Wärme
        \addplot[gray,opacity=0.3, fill=gray, fill opacity=0.3] coordinates {
            (16.8, 300) (16.8, 2009.1) (18.3, 2009.1) (18.3, 300)
        } \closedcycle;
        % Beschriftung für die latente Wärme
%        \node at (axis cs:12.25, 1150) [anchor=center, fill=white, draw=black, rounded corners, inner sep=2pt] {Latente Wärme};            \legend{Crystalline,Amorphous,Liquid,Temperature}
        \end{axis}
    \end{tikzpicture}
    \caption{Phase and temperature evolution during a rapid heating and cooling cycle.}
    \label{fig: fast cooling}
\end{figure}
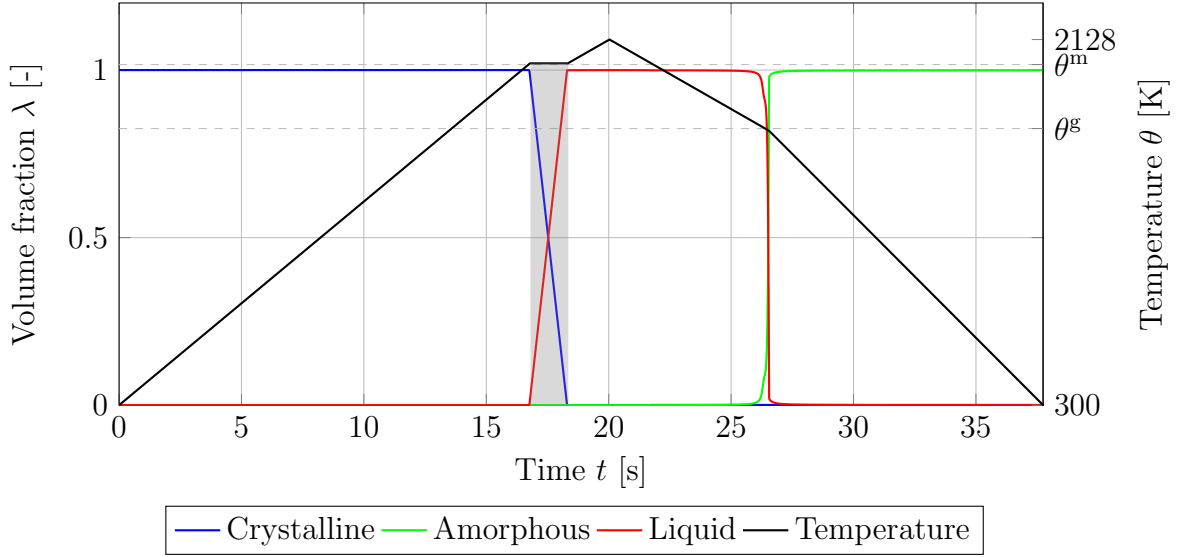

The initial set of numerical experiments investigates the material's intrinsic kinetic behavior under a high cooling rate of approximately $-2.5\cdot10^5\SI{}{\kelvin\per\hour}$ between $\theta^{\mathrm{m}}$ and $\theta^{\mathrm{g}}$, a critical condition for glass formation. \autoref{fig: fast cooling} illustrates the evolution of the phase fractions and temperature throughout this cycle. During the heating phase, the material behaves as expected: upon reaching the melting temperature $\theta^{\mathrm{m}}$, the crystalline phase completely transforms into the liquid phase, as evidenced by the sharp transition in the phase fraction curves.

The subsequent rapid cooling phase, however, demonstrates a key prediction of the model: complete suppression of crystallization. The supercooled liquid lacks sufficient time to overcome the kinetic barriers governing nucleation and growth of the ordered crystalline lattice. Consequently, upon reaching the glass transition temperature $\theta^{\mathrm{g}}$, the system vitrifies, transforming into a fully amorphous solid. This outcome demonstrates the model's ability to capture vitrification, a path-dependent process wherein kinetic limitations prevent the system from achieving its thermodynamically most stable state.
The gray region \protect\tikz[baseline]{\protect\filldraw(0,0)[color=gray,opacity=0.3] rectangle(0.3,0.3) ;} visible during the crystalline-to-liquid transformation in \autoref{fig: fast cooling} corresponds to the latent heat release associated with this first-order phase transition. Conversely, the amorphous-to-liquid transition manifests a lack of such a plateau, as it signifies a second-order phase transition devoid of latent heat. The variation in the grade of the temperature progression is attributable to the modification of material properties, which is accompanied by a phase fraction change. This fundamental distinction manifests in the transformation kinetics: the first-order crystallization is accompanied by discontinuous changes in entropy and volume, while the glass transition proceeds continuously.

\begin{figure}
    \centering
    \begin{tikzpicture}[spy using outlines=
    {circle, magnification=4, connect spies}]
        \begin{axis}[
            width=0.8\textwidth,
            height=0.4\textwidth,
            xlabel={Time $t$ [$\SI{}{\second}$]},
            ylabel={Volume fraction $\lambda$ [-]},
            ytick={0,0.5,1},
            grid=major,
            xmin=0,
            ymax=1.2,
            enlarge x limits=false,
            enlarge y limits=false,
            ]
            % Ersten Datensatz zeichnen (linke Y-Achse)
            \addplot[ blue,thick] table [x index=0,y index=2,col sep=semicolon] {Diagrams/Phase_trans/Datas/fast_time_temp_cryst_amorph_liq_reduced_Cycle.txt};
            \label{Line: fast cycle Crystalline}
            \addplot[ green,thick] table [x index=0,y index=3,col sep=semicolon] {Diagrams/Phase_trans/Datas/fast_time_temp_cryst_amorph_liq_reduced_Cycle.txt};
            \label{Line: fast cycle Amorph}
            \addplot[ red,thick] table [x index=0,y index=4,col sep=semicolon] {Diagrams/Phase_trans/Datas/fast_time_temp_cryst_amorph_liq_reduced_Cycle.txt};
            \label{Line: fast cycle Liquid}
         \coordinate (p1)   at (axis cs: 12.5,0.5);
 \coordinate (p2) at (axis cs: 7,0.35);
  \coordinate (p3)   at (axis cs: 17.5,0.5);
 \coordinate (p4) at (axis cs: 23,0.35);
        \end{axis}
       
        % Zweite Achse
        \begin{axis}[
            width=0.8\textwidth,
            height=0.4\textwidth,
            xlabel={},
            ylabel={Temperature $\theta$ [$\SI{}{\kelvin}$]},
            ylabel near ticks,
           ymin=300,
            ytick={\pgfkeysvalueof{/pgfplots/ymin}, \pgfkeysvalueof{/pgfplots/ymax}},
            axis y line*=right,
            axis x line=none,
            xmin=0,
            legend style={at={(0.05,-0.25)},anchor=north west,legend columns=-1
            },
            enlarge x limits=false,
            yticklabel style={/pgf/number format/1000 sep=,/pgf/number format/precision=0},
        extra y ticks={1683.15, 2003.15},
        extra y tick labels={$\theta^{\mathrm{g}}$,$\theta^{\mathrm{m}}$},
        extra y tick style={
            grid=major,
            grid style={dashed}, % Gilt für alle extra ticks, man kann es auch pro tick setzen
        },
            ]
            \addlegendimage{/pgfplots/refstyle=Line: fast cycle Crystalline}\addlegendentry{Line: fast cycle Crystalline}
            \addlegendimage{/pgfplots/refstyle=Line: fast cycle Amorph}\addlegendentry{Line: fast cycle Amorph}
            \addlegendimage{/pgfplots/refstyle=Line: fast cycle Liquid}\addlegendentry{Line: fast cycle Liquid}
            \addplot[black,thick
            ] table [x index=0,y index=1, col sep=semicolon
            ] {Diagrams/Phase_trans/Datas/fast_time_temp_cryst_amorph_liq_reduced_Cycle.txt};
%            \draw [gray, dashed]({axis cs:0, 2267.5}) -- ({axis cs:80, 2267.5});
            \legend{Crystalline,Amorphous,Liquid,Temperature}
        \end{axis}
%    \spy [black, size=2.8cm] on (p1)   in node[fill=white] at (p2);
%    \spy [black, size=2.8cm] on (p3)   in node[fill=white] at (p4);
\end{tikzpicture}
    \caption{Phase and temperature evolution during two consecutive rapid heating and cooling cycles.}
    \label{fig: fast cycle cooling}
\end{figure}
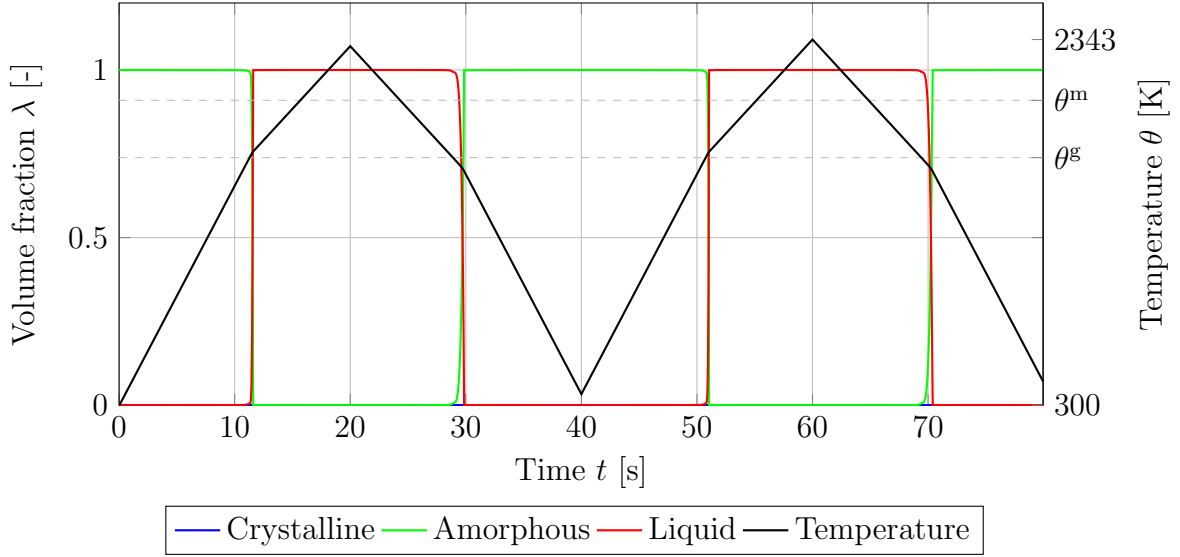

To examine the stability and characteristics of the glass transition itself, the material point, in an amorphous state, is subjected to further rapid thermal cycles, as depicted in \autoref{fig: fast cycle cooling}. Unlike the initial cycle, which began from a crystalline state to demonstrate the primary melting and subsequent vitrification, these consecutive cycles intentionally begin from the amorphous state. This approach isolates the second-order glass transition, demonstrating that under continued rapid thermal cycling, the material remains kinetically trapped in the amorphous-liquid regime without reverting to an ordered crystalline structure. The results confirm the model's robustness: the system demonstrates consistent and reversible transformation between the amorphous and liquid states. Furthermore, the magnified insets reveal a hysteretic behavior, where the heating and cooling paths of the transformation do not coincide.

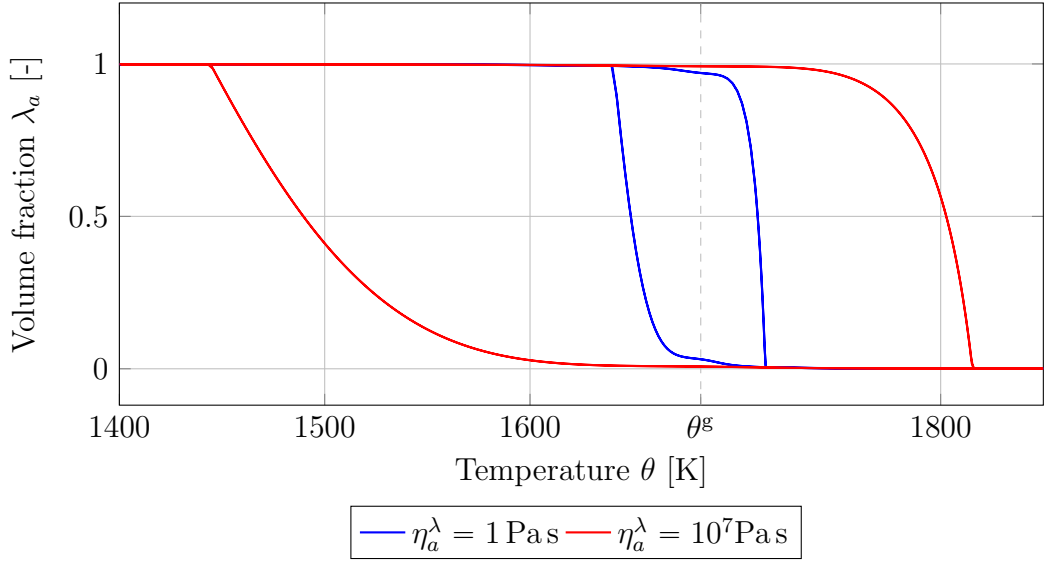
\begin{figure}
    \centering
    \begin{tikzpicture}
        \begin{axis}[
            width=0.8\textwidth,
            height=0.4\textwidth,
            xlabel={Temperature $\theta$ [$\SI{}{\kelvin}$]},
            ylabel={Volume fraction $\lambda_a$ [-]},
            ytick={0,0.5,1},
            grid=major,
            xmin=1400,
            xmax=1850,
            xtick={1400,1500,1600,1800},
            ymax=1.2,
            enlarge x limits=false,
            xticklabel style={
 /pgf/number format/fixed,
 /pgf/number format/precision=0,
 /pgf/number format/fixed zerofill,
 /pgf/number format/1000 sep={}
},
legend style={at={(0.25,-0.25)},anchor=north west,legend columns=-1
}, 
% colormap/bluered,
%    colorbar,
%    colorbar style={
%    title={Time $t$ [$\SI{}{\second}$] },  % Beschriftung der Farblegende mit Prozentzeichen
%    title style={rotate=90, xshift=-80pt,yshift=-45pt},        % Titel vertikal ausrichten
    %,
%    },
extra x ticks={1683.15},
        extra x tick labels={$\theta^{\mathrm{g}}$},
        extra x tick style={
            grid=major,
            grid style={dashed}, % Gilt für alle extra ticks, man kann es auch pro tick setzen
        },
            ]
\addplot[ blue,thick] table[x index=1,y index=3,col sep=semicolon] {Diagrams/Phase_trans/Datas/fast_hyster_time_temp_cryst_amorph_liq_reduced.txt};
\label{Line: Amorph Hyste eta=1}

\addplot[ red,thick] table[x index=1,y index=3,col sep=semicolon] {Diagrams/Phase_trans/Datas/fast_hyster_time_temp_cryst_amorph_liq_reduced_Eta_1E-1.txt};
\label{Line: Amorph Hyste eta=10^7}
%           \addplot[scatter
%        ,scatter src=explicit
%        ,only marks
%        ,mark size=1.5pt
%        ,point meta=explicit % Dritte Spalte für Farbkodierung
%        ] 
%table[x index=1,y index=3, meta index=0,col sep=semicolon] {Diagrams/Phase_trans/Datas/fast_hyster_time_temp_cryst_amorph_liq_reduced.txt};
\legend{$\eta_a^{\lambda}=\SI{1}{\pascal\second}$,$\eta_a^{\lambda}=10^{7}\SI{}{\pascal\second}$}
        \end{axis}
    \end{tikzpicture}
    \caption{Hysteresis loop of the amorphous phase during the rapid heating and cooling cycle.}
    \label{fig: fast hyste cooling}
\end{figure}

\autoref{fig: fast hyste cooling} provides a more explicit analysis of this hysteretic phenomenon by plotting the amorphous volume fraction as a function of temperature. To highlight the influence of transformation kinetics, the material response is evaluated for two different amorphous viscosity values ($\eta_{a}^{\lambda} = \SI{1}{\pascal\second}$ \autoref{Line: Amorph Hyste eta=1}  and $\eta_{a}^{\lambda} =10^7 \SI{}{\pascal\second}$ \autoref{Line: Amorph Hyste eta=10^7}). The diagram  illustrates the gradual nature of the glass transition: rather than a hard thermodynamic boundary, the phase change occurs over a smooth, continuous transition region. Furthermore, the distinct widths of the two transformation regions directly reflect the impact of the viscosity parameters. The higher viscosity imposes a strong resistance to microstructural rearrangement, leading to a broader transformation region with significant kinetic lag. In contrast, the lowered viscosity allows the system to response more rapidly to thermal changes, thereby narrowing the hysteresis loop. In both cases, the path of the transformation from liquid to amorphous during cooling (left side of the loop) is clearly shifted to lower temperatures compared to the path of the amorphous-to-liquid transformation during heating (right side of the loop). The origin of this hysteresis is embedded in the constitutive formulation of the model.

The evolution of the phase fraction is described by the rate-dependent \autoref{EQ: Evolution Lambda}, where the rate of change $\Dot{\bflambda}$ is proportional to the thermodynamic driving force but inversely proportional to the temperature-dependent mobility tensor $\bfeta^{\bflambda}$ (\autoref{EQ: Diss lambda}). 
This rate- and temperature-dependent kinetic constraint is the fundamental mechanism responsible for the simulated hysteretic behavior, directly analogous to experimental observations in glass-forming materials. It is crucial to emphasize that this hysteresis visible in the volume fraction persists even though the mobility tensor is modeled as constant and identical for both the amorphous and liquid phases. Consequently, the loop does not result from a variation in material resistance, but rather from the kinetic lag induced by the rapid process speed. In the vicinity of the equilibrium temperature $\theta^{\mathrm{g}}$, the thermodynamic driving force $\bfp^{\bflambda}$ is vanishingly small. Under the imposed rapid cooling, the temperature must drop significantly below $\theta^{\mathrm{g}}$ to generate a driving force sufficient to produce a transformation rate that can keep pace with the thermal boundary conditions. The width of the hysteresis loop is, therefore, a direct quantifier of the system's deviation from thermodynamic equilibrium. The model thus correctly captures the non-equilibrium nature of the glass transition, validating its applicability to vitrification phenomena.

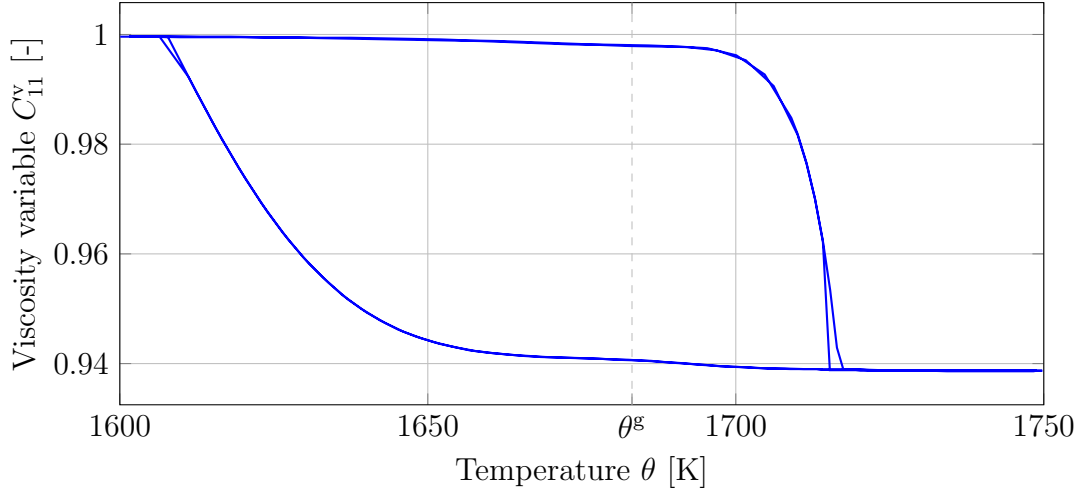
\begin{figure}
    \centering
    \begin{tikzpicture}
        \begin{axis}[
            width=0.8\textwidth,
            height=0.4\textwidth,
            xlabel={Temperature $\theta$ [$\SI{}{\kelvin}$]},
            ylabel={Viscosity variable $C^{\mathrm{v}}_{11}$ [-]},
            xtick={1600,1650,1700,1750},
            grid=major,
            xmin=1600,
            xmax=1750,
            enlarge x limits=false,
            xticklabel style={
 /pgf/number format/fixed,
 /pgf/number format/precision=0,
 /pgf/number format/fixed zerofill,
 /pgf/number format/1000 sep={}
},
extra x ticks={1683.15},
        extra x tick labels={$\theta^{\mathrm{g}}$},
        extra x tick style={
            grid=major,
            grid style={dashed}, % Gilt für alle extra ticks, man kann es auch pro tick setzen
        },
            ]
\addplot[ blue,thick] table[x index=1,y index=2,col sep=semicolon]{Diagrams/Phase_trans/Datas/fast_RCv11_time_temp_reduced.txt};
        \end{axis}
    \end{tikzpicture}
    \caption{Hysteresis loop of the viscosity variable $C^{\mathrm{v}}_{11}$ during the rapid heating and cooling cycle.}
    \label{fig: fast Hyst Cv}
\end{figure}

Furthermore, the internal viscous variable (\autoref{fig: fast Hyst Cv}) necessarily follows this hysteretic path. Since the total deformation is constrained to the identity ($\bfF=\bfI$) in this material point setting, the volumetric mismatch induced by the phase transformation ($\bfF^{\rho}$) generates significant internal stress. To restore thermodynamic equilibrium, these stresses must be relaxed through viscous evolution, thereby coupling the viscosity variable directly to the phase transformation kinetics.

\subsubsection{Slow cooling rate}
\label{Sec: Slow Cooling Rate}
This section examines the material's response under quasi-static conditions, where the system is allowed to approach its thermodynamic equilibrium. A very slow cooling rate of approximately $\SI{-0.0001}{\kelvin\per\hour}$ is applied between the melting and glass transition temperatures. 

\begin{figure}[H]
    \centering
    \begin{tikzpicture}
        \begin{axis}[
            width=0.8\textwidth,
            height=0.4\textwidth,
            xlabel={Time $t$ [$\SI{}{\second}$]},
            ylabel={Volume fraction $\lambda$ [-]},
            ytick={0,0.5,1},
            grid=major,
            xmin=0,
            ymax=1.2,
            enlarge x limits=false,
            enlarge y limits=false,
            scaled x ticks=true,
            every x tick scale label/.style={
          at={(xticklabel cs:1)}, % ganz rechts bei den xtick-Labels
          anchor=base west,
          xshift=-30pt,yshift=-8pt,
          execute at begin node={\let\cdot\times}
        },
            ]
            % Ersten Datensatz zeichnen (linke Y-Achse)
            \addplot[ blue,thick] table [x index=0,y index=2,col sep=semicolon] {Diagrams/Phase_trans/Datas/slow_time_temp_cryst_amorph_liq_reduced_Cycle.txt};
            \label{Line: slow cycle Crystalline}
            \addplot[ green,thick] table [x index=0,y index=3,col sep=semicolon] {Diagrams/Phase_trans/Datas/slow_time_temp_cryst_amorph_liq_reduced_Cycle.txt};
            \label{Line: slow cycle Amorph}
            \addplot[ red,thick] table [x index=0,y index=4,col sep=semicolon] {Diagrams/Phase_trans/Datas/slow_time_temp_cryst_amorph_liq_reduced_Cycle.txt};
            \label{Line: slow cycle Liquid}
        \end{axis}
        
        % Zweite Achse
        \begin{axis}[
            width=0.8\textwidth,
            height=0.4\textwidth,
            xlabel={},
            ylabel={Temperature $\theta$ [$\SI{}{\kelvin}$]},
            ylabel near ticks,
            ymin=300,
            ytick={\pgfkeysvalueof{/pgfplots/ymin}, \pgfkeysvalueof{/pgfplots/ymax}},
            axis y line*=right,
            axis x line=none,
            xmin=0,
            legend style={at={(0.05,-0.25)},anchor=north west,legend columns=-1
            },
               enlarge x limits=false,
            yticklabel style={/pgf/number format/1000 sep=,/pgf/number format/precision=0},
        extra y ticks={1683.15, 2003.15},
        extra y tick labels={$\theta^{\mathrm{g}}$,$\theta^{\mathrm{m}}$},
        extra y tick style={
            grid=major,
            grid style={dashed}, % Gilt für alle extra ticks, man kann es auch pro tick setzen
        },
        scaled x ticks=true,
                every x tick scale label/.style={
                    at={(xticklabel cs:1)}, % ganz rechts bei den xtick-Labels
                    anchor=base west,
                    xshift=-25pt,yshift=-8pt,
                    % Auch hier für die zweite Achse eintragen:
                    execute at begin node={\let\cdot\times}
                },
            ]
            \addlegendimage{/pgfplots/refstyle=Line: slow cycle Crystalline}\addlegendentry{Line: slow cycle Crystalline}
            \addlegendimage{/pgfplots/refstyle=Line: slow cycle Amorph}\addlegendentry{Line: slow cycle Amorph}
            \addlegendimage{/pgfplots/refstyle=Line: slow cycle Liquid}\addlegendentry{Line: slow cycle Liquid}
            \addplot[black,thick
            ] table [x index=0,y index=1, col sep=semicolon
            ] {Diagrams/Phase_trans/Datas/slow_time_temp_cryst_amorph_liq_reduced_Cycle.txt};
              %\draw [gray, dashed] ({axis cs:0, 2128.9}) -- ({axis cs:16e10, 2128.9});
            \legend{Crystalline,Amorphous,Liquid,Temperature}
        \end{axis}
    \end{tikzpicture}
    \caption{Phase and temperature evolution during two consecutive slow heating and cooling cycles.}
    \label{fig: slow cycle cooling}
\end{figure}
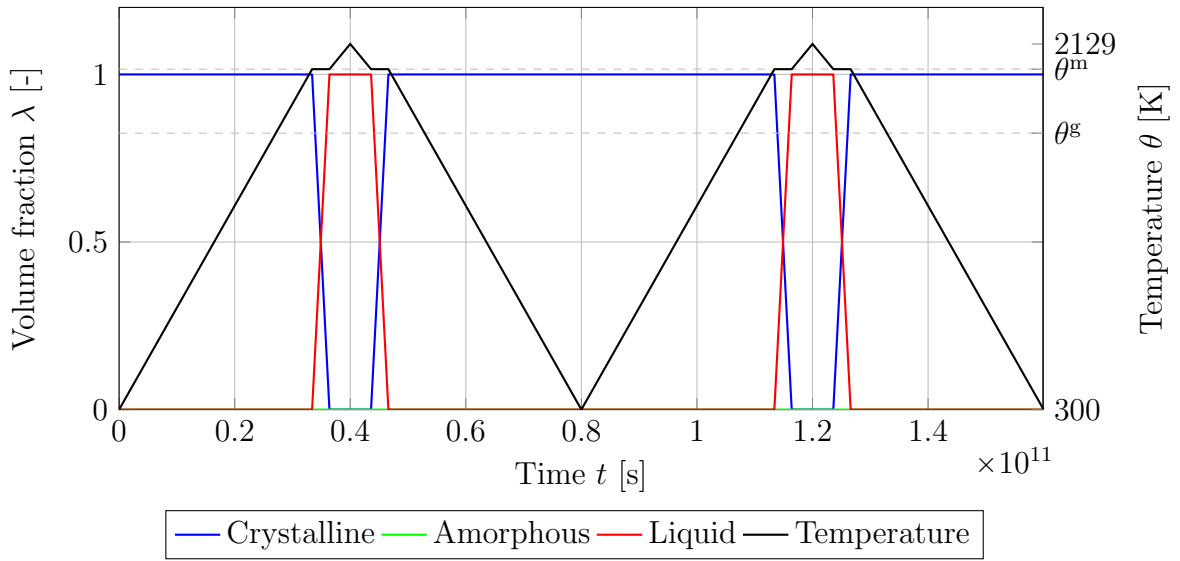

The evolution of phase fractions and temperature over two consecutive thermal cycles is depicted in \autoref{fig: slow cycle cooling}. During the heating phase, the material behaves as previously observed, with the crystalline phase transforming entirely into a liquid upon reaching the melting temperature. During the subsequent slow cooling period, the system is afforded sufficient time to overcome any kinetic barriers to nucleation and growth. Consequently, as the temperature drops back to the melting point, the liquid phase transforms back into the crystalline phase. This process continues to completion, and the material returns to a fully crystalline state. The formation of the amorphous phase is entirely suppressed,  when kinetic constraints are removed. The model correctly predicts that the system will evolve towards its state of lowest free energy, the crystalline solid.

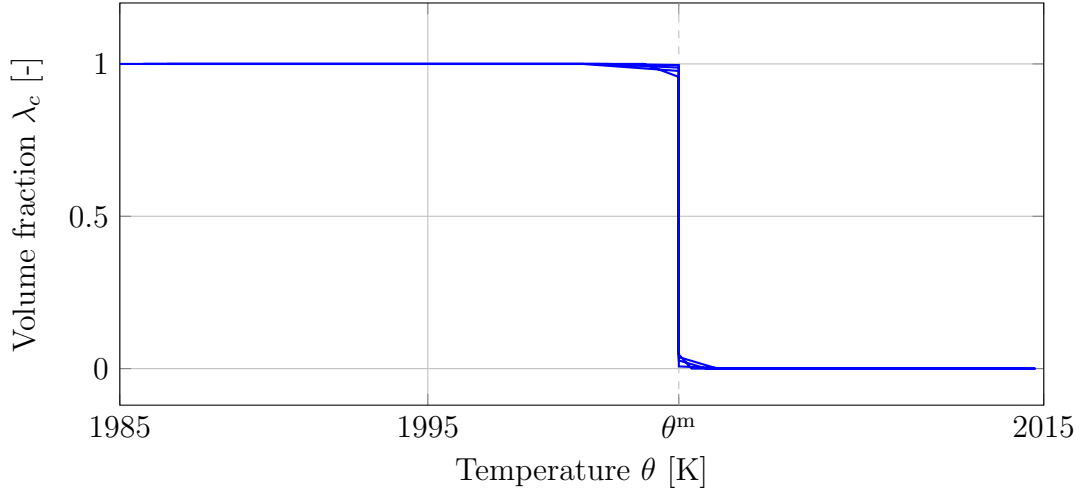
\begin{figure}[H]
    \centering
    \begin{tikzpicture}
        \begin{axis}[
            width=0.8\textwidth,
            height=0.4\textwidth,
            xlabel={Temperature $\theta$ [$\SI{}{\kelvin}$]},
            ylabel={Volume fraction $\lambda_c$ [-]},
            ytick={0,0.5,1},
            grid=major,
            xmin=1985,
            xmax=2015,
            xtick={1985,1995,2015},
            ymax=1.2,
            enlarge x limits=false,
            xticklabel style={
 /pgf/number format/fixed,
 /pgf/number format/precision=0,
 /pgf/number format/fixed zerofill,
 /pgf/number format/1000 sep={}
},
extra x ticks={2003.15},
        extra x tick labels={$\theta^{\mathrm{m}}$},
        extra x tick style={
            grid=major,
            grid style={dashed}, % Gilt für alle extra ticks, man kann es auch pro tick setzen
        },
            ]
\addplot[ blue,thick] table[x index=1,y index=2,col sep=semicolon]  {Diagrams/Phase_trans/Datas/slow_hyster_time_temp_cryst_reduced.txt};
        \end{axis}
    \end{tikzpicture}
    \caption{Hysteresis behavior of the crystalline phase during the slow heating and cooling cycle.}
    \label{fig: slow hyste cooling}
\end{figure}

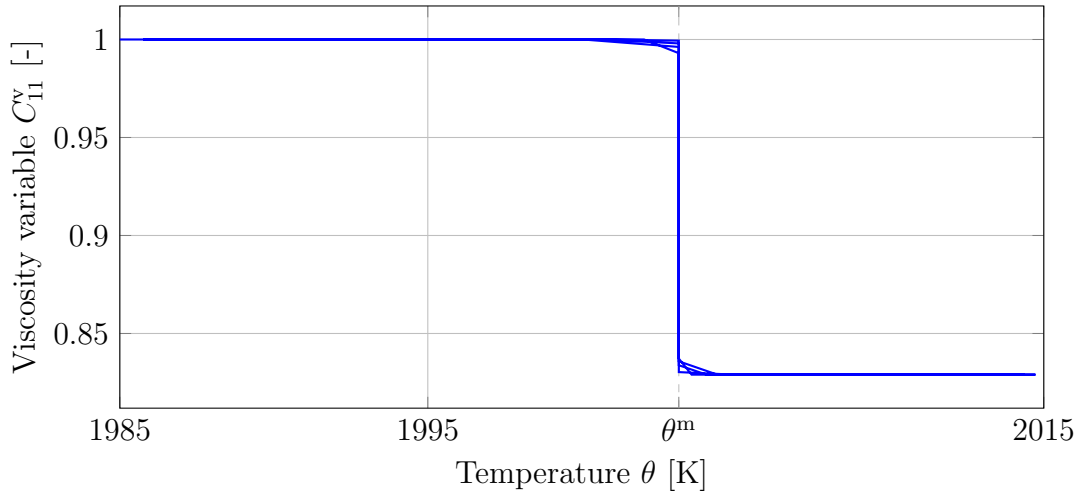
\begin{figure}[H]
    \centering
    \begin{tikzpicture}
        \begin{axis}[
            width=0.8\textwidth,
            height=0.4\textwidth,
            xlabel={Temperature $\theta$ [$\SI{}{\kelvin}$]},
            ylabel={Viscosity variable $C^{\mathrm{v}}_{11}$ [-]},
            grid=major,
            xmin=1985,
            xmax=2015,
            xtick={1985,1995,2015},
            grid=major,
            enlarge x limits=false,
            xticklabel style={
 /pgf/number format/fixed,
 /pgf/number format/precision=0,
 /pgf/number format/fixed zerofill,
 /pgf/number format/1000 sep={}
},
extra x ticks={2003.15},
        extra x tick labels={$\theta^{\mathrm{m}}$},
        extra x tick style={
            grid=major,
            grid style={dashed}, % Gilt für alle extra ticks, man kann es auch pro tick setzen
        },
        ]
\addplot[ blue,thick] table[x index=1,y index=2,col sep=semicolon]{Diagrams/Phase_trans/Datas/slow_RCv11_time_temp_reduced.txt};
        \end{axis}
    \end{tikzpicture}
    \caption{Hysteresis loop of the viscosity variable $C^{\mathrm{v}}_{11}$ during the slow heating and cooling cycle.}
    \label{fig: slow hyste Cv}
\end{figure}

\autoref{fig: slow cycle cooling} confirms that this behavior is stable and perfectly reversible over multiple cycles. The phase evolution trajectories for heating and cooling are identical, indicating that the system operates under quasi-equilibrium conditions throughout the entire process.
The defining characteristic of this equilibrium behavior is illustrated in \autoref{fig: slow hyste cooling} which plots the crystalline volume fraction as a function of temperature. The plot reveals a complete absence of hysteresis. The transformation from solid to liquid during heating and from liquid to solid during cooling occur at precisely the same equilibrium melting temperature $\theta^{\mathrm{m}}$, manifesting as a vertical phase transition. This signifies that the transformation is not kinetically limited. The available time is sufficiently long that even an infinitesimal thermodynamic driving force is capable of driving the phase change.
Finally, the evolution of the internal viscosity variable $C^{\mathrm{v}}_{11}$ is explicitly plotted in \autoref{fig: slow hyste Cv}. As expected for this quasi-static regime, it exhibits the exact same reversible, non-hysteretic behavior as the phase fractions. Because the cooling rate is extremely low, the volumetric mismatch induced by the phase transformation is accommodated immediately by viscous relaxation. Unlike the rapid cooling scenario, no transient internal stresses accumulate. This allows the viscous variable to follow the thermodynamic equilibrium path of the volume fractions perfectly and without any kinetic lag, resulting in a sharp, vertical transition at the melting temperature.

This result confirms the thermodynamic consistency of the model which correctly captures the full spectrum of behavior from complete vitrification under rapid cooling (\autoref{fig: fast hyste cooling}) to full, reversible crystallization under quasi-static conditions (\autoref{fig: slow hyste cooling}). The striking difference between these limiting cases, wide hysteresis loops versus vertical phase transitions, validates the model's ability to represent the transition from kinetically controlled to thermodynamically controlled regimes.

\subsubsection{Intermediate Cooling Rate: Partial Crystallization and Vitrification Formation} 
\label{Sec: Intermediate Cooling Rate}
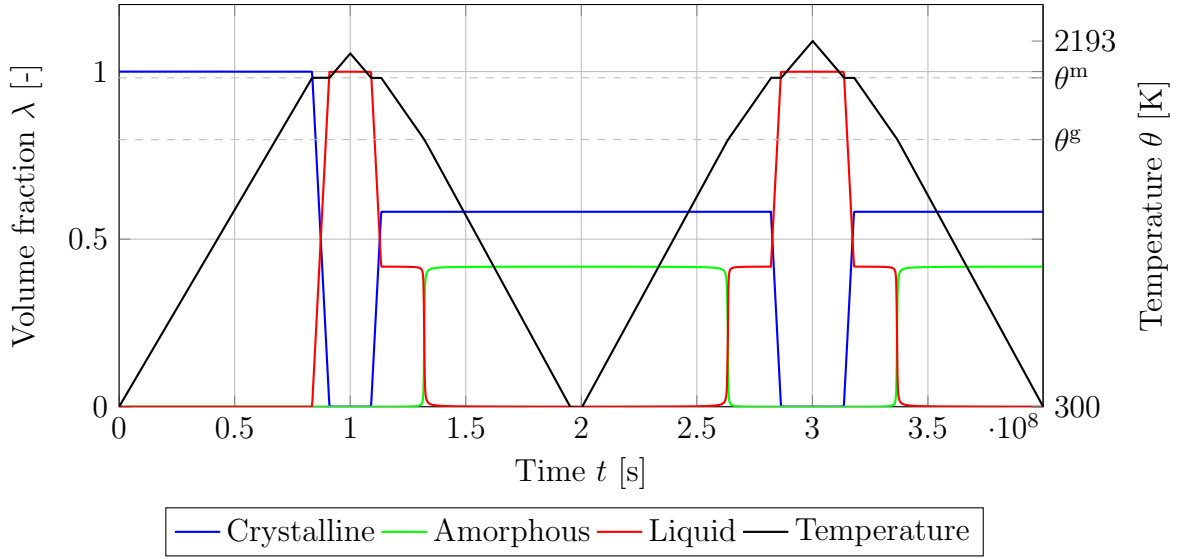
\begin{figure}[H]
    \centering
    \begin{tikzpicture}
        \begin{axis}[
            width=0.8\textwidth,
            height=0.4\textwidth,
            xlabel={Time $t$ [$\SI{}{\second}$]},
            ylabel={Volume fraction $\lambda$ [-]},
            ytick={0,0.5,1},
            xtick={0,0.5e8,1e8,1.5e8,2e8,2.5e8,3e8,3.5e8},
            grid=major,
            xmin=0,
            ymax=1.2,
            enlarge x limits=false,
            enlarge y limits=false,
            scaled x ticks=true,
        % Der korrekte Schlüssel, der in einem sauberen Projekt
            % mit TeX Live 2024 funktionieren MUSS.
            every x tick scale label/.style={
          at={(xticklabel cs:1)}, % ganz rechts bei den xtick-Labels
          anchor=base west,
          xshift=-25pt,yshift=0.75ex
        },
            ]
            % Ersten Datensatz zeichnen (linke Y-Achse)
            \addplot[ blue,thick] table [x index=0,y index=2,col sep=semicolon] {Diagrams/Phase_trans/Datas/middle_time_temp_cryst_amorph_liq_reduced_Cycle.txt};
            \label{Line: middle cycle Crystalline}
            \addplot[ green,thick] table [x index=0,y index=3,col sep=semicolon] {Diagrams/Phase_trans/Datas/middle_time_temp_cryst_amorph_liq_reduced_Cycle.txt};
            \label{Line: middle cycle Amorph}
            \addplot[ red,thick] table [x index=0,y index=4,col sep=semicolon] {Diagrams/Phase_trans/Datas/middle_time_temp_cryst_amorph_liq_reduced_Cycle.txt};
            \label{Line: middle cycle Liquid}
        \end{axis}
        
        % Zweite Achse
        \begin{axis}[
            width=0.8\textwidth,
            height=0.4\textwidth,
            xlabel={},
            ylabel={Temperature $\theta$ [$\SI{}{\kelvin}$]},
            ylabel near ticks,
            ymin=300,
            ytick={\pgfkeysvalueof{/pgfplots/ymin}, \pgfkeysvalueof{/pgfplots/ymax}},
            axis y line*=right,
            axis x line=none,
            xmin=0,
            legend style={at={(0.05,-0.25)},anchor=north west,legend columns=-1
            },
               enlarge x limits=false,
            yticklabel style={/pgf/number format/1000 sep=,/pgf/number format/precision=0},
        extra y ticks={1683.15, 2003.15},
        extra y tick labels={$\theta^{\mathrm{g}}$,$\theta^{\mathrm{m}}$},
        extra y tick style={
            grid=major,
            grid style={dashed}, % Gilt für alle extra ticks, man kann es auch pro tick setzen
        },
            ]
            \addlegendimage{/pgfplots/refstyle=Line: middle cycle Crystalline}\addlegendentry{Line: middle cycle Crystalline}
            \addlegendimage{/pgfplots/refstyle=Line: middle cycle Amorph}\addlegendentry{Line: middle cycle Amorph}
            \addlegendimage{/pgfplots/refstyle=Line: middle cycle Liquid}\addlegendentry{Line: middle cycle Liquid}
            \addplot[black,thick
            ] table [x index=0,y index=1, col sep=semicolon
            ] {Diagrams/Phase_trans/Datas/middle_time_temp_cryst_amorph_liq_reduced_Cycle.txt};
                %\draw [gray, dashed] ({axis cs:0, 2108.41}) -- ({axis cs:4e8, 2108.41});
            \legend{Crystalline,Amorphous,Liquid,Temperature}
        \end{axis}
    \end{tikzpicture}
    \caption{Phase and temperature evolution during two consecutive intermediate heating and cooling cycles.}
    \label{fig: middle cycle cooling}
\end{figure}

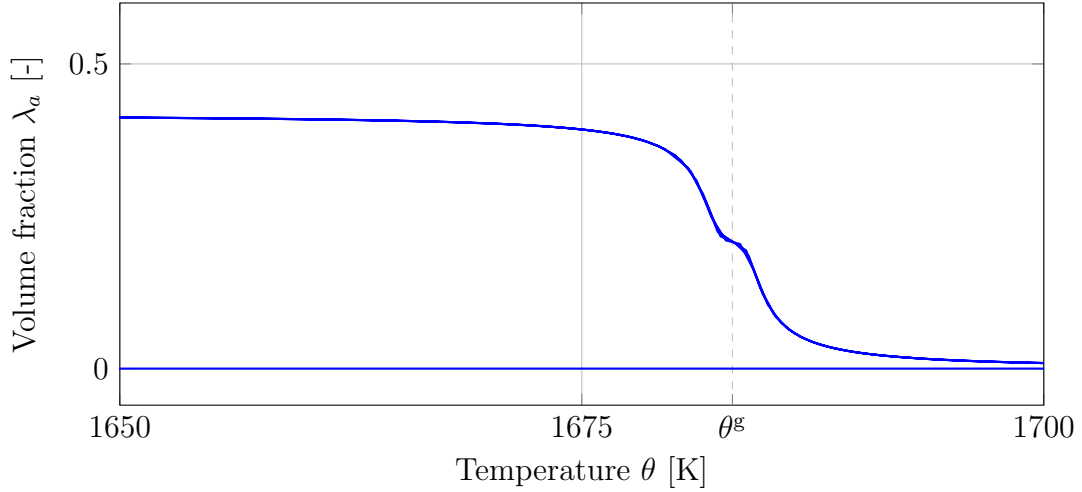
\begin{figure}[H]
    \centering
    \begin{tikzpicture}
        \begin{axis}[
            width=0.8\textwidth,
            height=0.4\textwidth,
            xlabel={Temperature $\theta$ [$\SI{}{\kelvin}$]},
            ylabel={Volume fraction $\lambda_a$ [-]},
            ytick={0,0.5},
            xtick={1650,1675,1700,1725,1750},
            grid=major,
            xmin=1650,
            xmax=1700,
            ymax=0.6,
            enlarge x limits=false,
            xticklabel style={
 /pgf/number format/fixed,
 /pgf/number format/precision=0,
 /pgf/number format/fixed zerofill,
 /pgf/number format/1000 sep={}
},
extra x ticks={1683.15},
        extra x tick labels={$\theta^{\mathrm{g}}$},
        extra x tick style={
            grid=major,
            grid style={dashed}, % Gilt für alle extra ticks, man kann es auch pro tick setzen
        },
            ]
\addplot[ blue,thick] table[x index=1,y index=2,col sep=semicolon]  {Diagrams/Phase_trans/Datas/middle_Hyst_time_temp_amorph_reduced.txt};
        \end{axis}
    \end{tikzpicture}
    \caption{Hysteresis loop of the amorphous phase during the intermediate heating and cooling cycle.}
    \label{fig: middle hyste amorph cooling}
\end{figure}

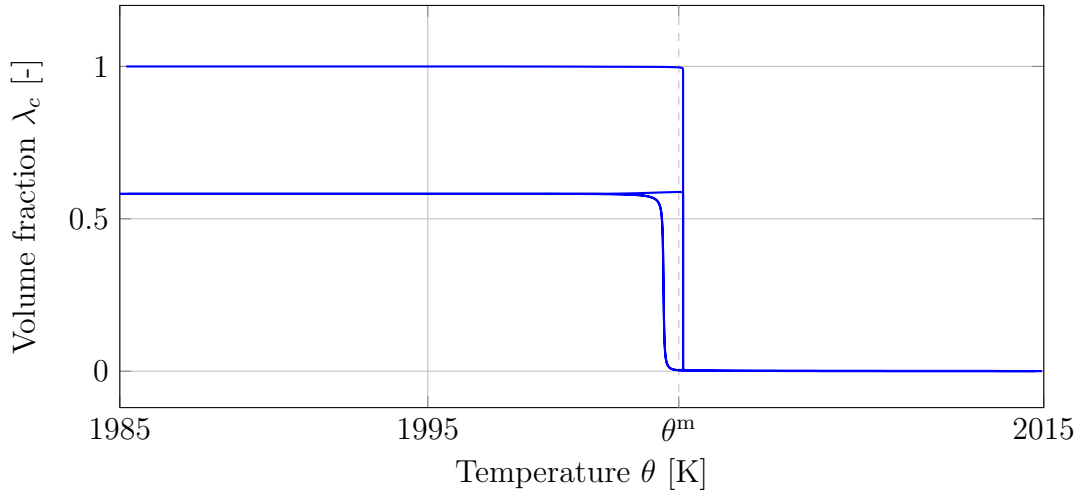
\begin{figure}[H]
    \centering
    \begin{tikzpicture}
        \begin{axis}[
            width=0.8\textwidth,
            height=0.4\textwidth,
            xlabel={Temperature $\theta$ [$\SI{}{\kelvin}$]},
            ylabel={Volume fraction $\lambda_c$ [-]},
            ytick={0,0.5,1},
            xtick={1985,1995,2015},
            grid=major,
            xmin=1985,
            xmax=2015,
            ymax=1.2,
            enlarge x limits=false,
            xticklabel style={
 /pgf/number format/fixed,
 /pgf/number format/precision=0,
 /pgf/number format/fixed zerofill,
 /pgf/number format/1000 sep={}
},
extra x ticks={2003.15},
        extra x tick labels={$\theta^{\mathrm{m}}$},
        extra x tick style={
            grid=major,
            grid style={dashed}, % Gilt für alle extra ticks, man kann es auch pro tick setzen
        },
            ]
\addplot[ blue,thick] table[x index=1,y index=2,col sep=semicolon]  {Diagrams/Phase_trans/Datas/middle_Hyst_time_temp_cryst_reduced.txt};
        \end{axis}
    \end{tikzpicture}
    \caption{Hysteresis loop of the crystalline phase during the intermediate heating and cooling cycle.}
    \label{fig: middle hyste crystalline cooling}
\end{figure}

This section investigates the material's response within the critical kinetic regime lying between the extremes of rapid vitrification and slow, full crystallization. Applying a moderate cooling rate of approximately  $-5\SI{}{\kelvin\per\hour}$is applied between the melting and glass transition temperatures.
The behavior over consecutive thermal cycles is presented in \autoref{fig: middle cycle cooling}. The heating phase exhibits a distinctive two-step melting process: the amorphous fraction transforms into a liquid at $\theta^{\mathrm{g}}$, followed by the melting of the crystalline fraction at $\theta^{\mathrm{m}}$. Upon subsequent cooling, the model consistently reproduces the partial crystallization and subsequent vitrification of the remaining liquid. The transformation is kinetically arrested before the entire volume can arrange into an ordered lattice, yielding a stable composite microstructure comprising approximately $58\%$ crystalline and $42\%$ amorphous phases. This reproducibility confirms that the intermediate cooling rate produces a stable material state governed by the precise interplay between thermodynamic driving forces and kinetic constraints.
\autoref{fig: middle hyste amorph cooling} and \autoref{fig: middle hyste crystalline cooling} provide explicit hysteretic analyses by plotting the phase fractions as functions of temperature, and these plots reveal the most significant finding of this section: the hysteresis loops in the mixed-phase material are substantially narrower than those observed in the single-phase materials under comparable cooling rates.

In the rapid cooling scenario (\autoref{fig: fast hyste cooling}), the pure amorphous phase exhibits a wide hysteresis loop spanning approximately $\SI{105}{\kelvin}$ (from $\SI{1610}{\kelvin}$ during cooling to $\SI{1715}{\kelvin}$ during heating). In remarkable contrast, the amorphous fraction within the mixed-phase material (\autoref{fig: middle hyste amorph cooling}) displays a hysteresis loop with a significantly smaller width. This reduction in hysteresis width is a direct consequence of the rate-dependent nature of the glass transition. The hysteresis width serves as a measure of the system's deviation from thermodynamic equilibrium. In the fast cooling case, the extreme cooling rate ($-2.5\cdot10^5\SI{}{\kelvin\per\hour}$) drives the material into a highly non-equilibrium state, creating a large kinetic lag between the temperature change and the phase response. Conversely, the intermediate cooling rate ($-5\SI{}{\kelvin\per\hour}$), while fast enough to induce partial vitrification, is orders of magnitude slower. This allows the system to operate much closer to thermodynamic equilibrium, enabling the phase evolution to track the temperature change with significantly less kinetic lag.

Similarly, the crystalline phase in the mixed-phase material (\autoref{fig: middle hyste crystalline cooling}) exhibits a narrow but measurable hysteresis loop near the melting temperature $\theta^{\mathrm{m}}$, whereas the pure crystalline phase under equilibrium conditions (\autoref{fig: slow hyste cooling}) shows virtually no hysteresis, a vertical transition at the melting point. The crystalline hysteresis in \autoref{fig: middle hyste crystalline cooling}, while much smaller than the amorphous hysteresis, represents a kinetic effect absent under equilibrium conditions.

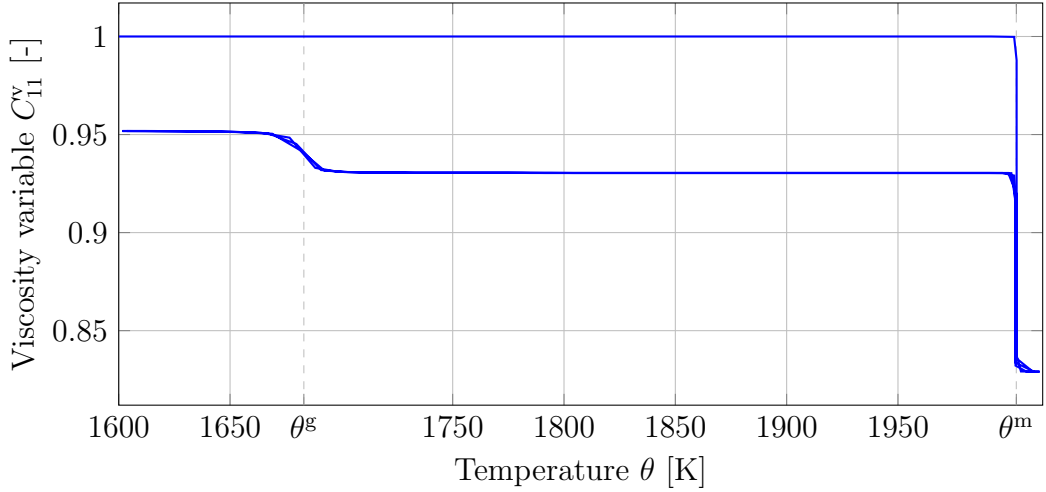
\begin{figure}[H]
    \centering
    \begin{tikzpicture}
        \begin{axis}[
            width=0.8\textwidth,
            height=0.4\textwidth,
            xlabel={Temperature $\theta$ [$\SI{}{\kelvin}$]},
            ylabel={Viscosity variable $C^{\mathrm{v}}_{11}$ [-]},
            xtick={1600,1650,1750,1800,1850,1900,1950},
            grid=major,
            xmin=1600,
            xmax=2015,
            enlarge x limits=false,
            xticklabel style={
 /pgf/number format/fixed,
 /pgf/number format/precision=0,
 /pgf/number format/fixed zerofill,
 /pgf/number format/1000 sep={}
},
extra x ticks={1683.15,2003.15},
        extra x tick labels={$\theta^{\mathrm{g}}$,$\theta^{\mathrm{m}}$},
        extra x tick style={
            grid=major,
            grid style={dashed}, % Gilt für alle extra ticks, man kann es auch pro tick setzen
        },
            ]
\addplot[ blue,thick] table[x index=1,y index=2,col sep=semicolon]  {Diagrams/Phase_trans/Datas/middle_RCv11_time_temp_reduced.txt};
        \end{axis}
    \end{tikzpicture}
    \caption{Hysteresis loop of the viscosity variable $C^{\mathrm{v}}_{11}$ during the intermediate heating and cooling cycle.}
    \label{fig: middle hyste Cv}
\end{figure}

Finally, the viscosity variable (\autoref{fig: middle hyste Cv}) essentially acts as an indicator of the macroscopic material state, bridging the behavior of the two phases. It follows the composite hysteresis path, reflecting the combined stress relaxation mechanisms of the partial crystallization and the glass transition of the remaining matrix.

\subsection{FEM Results for the Rectangular Prism within $\mathrm{ANSYS}^{\copyright}$}
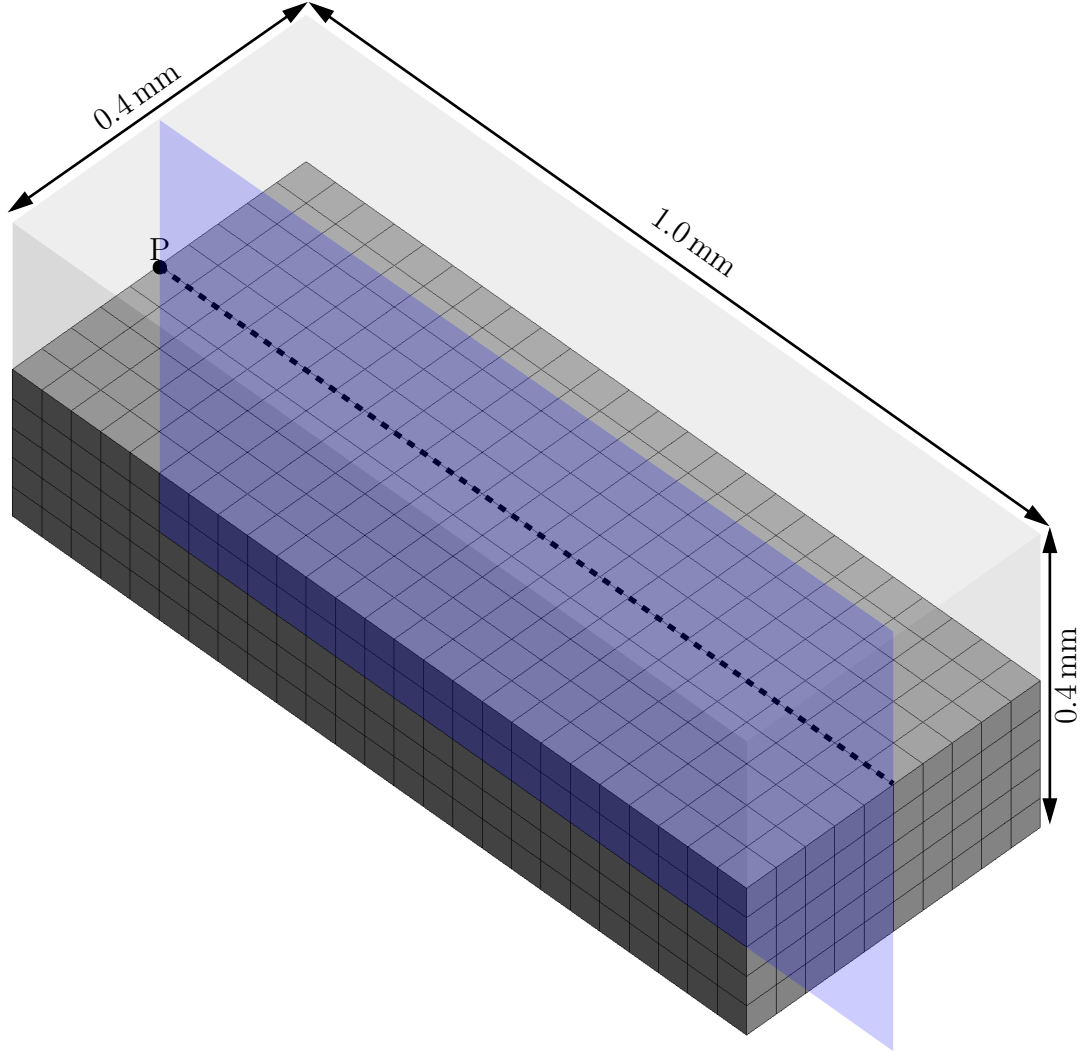
\begin{figure}[h]
    \centering
    \begin{tikzpicture}
        % Lädt das Bild und definiert ein relatives Koordinatensystem
        \node[anchor=south west,inner sep=0] (image) at (0,0) {\includegraphics[width=0.8\textwidth]{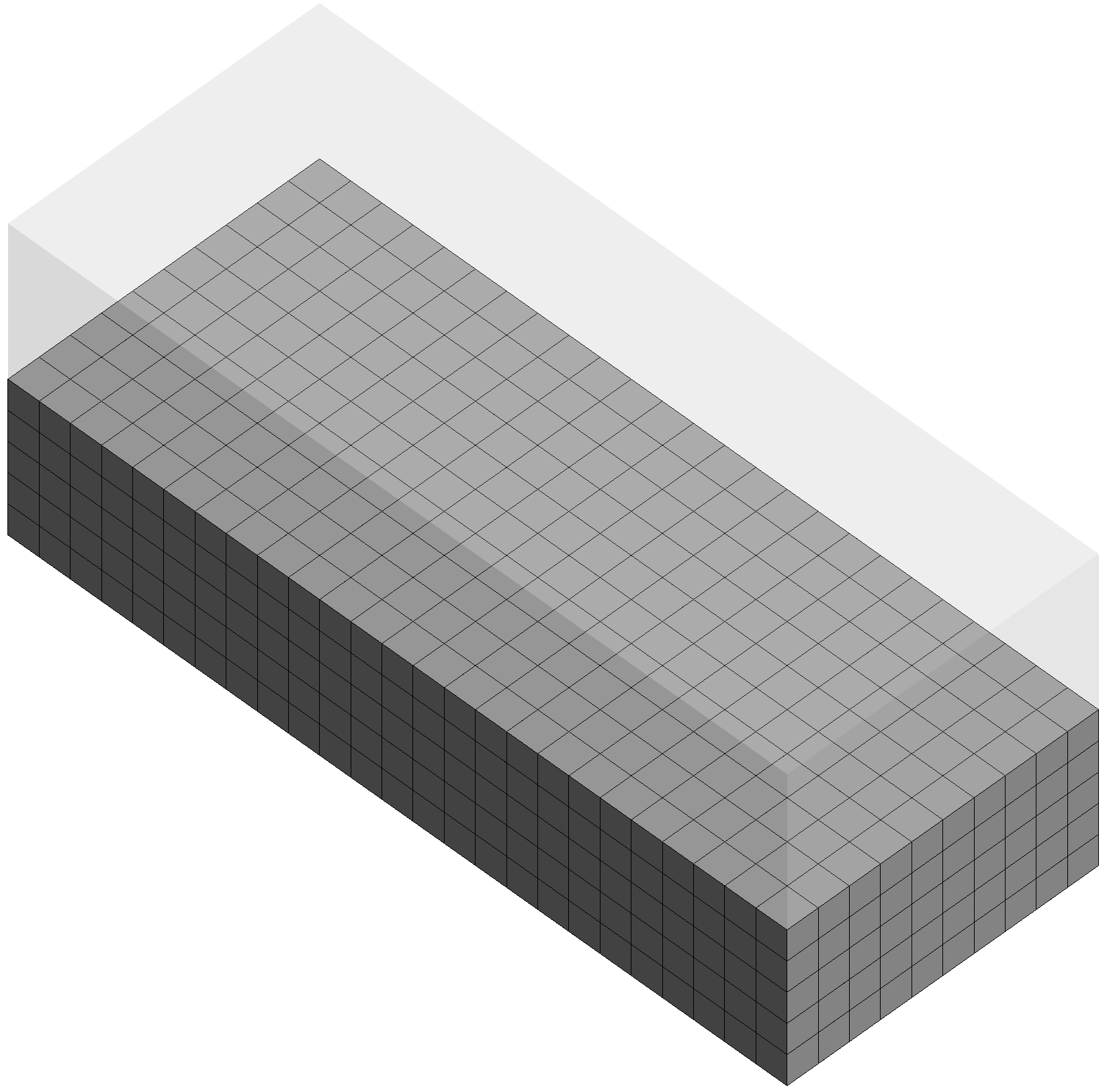}};
        \begin{scope}[x={(image.south east)},y={(image.north west)}]
            
%\draw[help lines,xstep=.05,ystep=.05] (0,0) grid (1,1);
%\foreach \x in {0,1,...,9} { \node [anchor=north] at (\x/10,0) {0.\x}; }
%\foreach \y in {0,1,...,9} { \node [anchor=north] at (0,\y/10) {0.\y}; }

%\draw[line width=2pt,color=green] (0.444,0.26) -- (0.79,0.54);

            %% Messpfeile für die neuen Abmaße
            
            % Maß für die LÄNGE (1.0) unten
            \draw[line width=1pt,color=black,arrows={Stealth[inset=0pt, length=10pt, angle'=30]-Stealth[inset=0pt, length=10pt, angle'=30]}] 
            (0.29, 1.01) -- (1, 0.5) node[midway, above=2pt, sloped] {\SI{1.0}{\milli \metre}};

            % Maß für die HÖHE (0.4) rechts
            \draw[line width=1pt,color=black,arrows={Stealth[inset=0pt, length=10pt, angle'=30]-Stealth[inset=0pt, length=10pt, angle'=30]}] 
            (1, 0.21) -- (1, 0.5) node[midway, xshift=6pt, rotate=90] {\SI{0.4}{\milli \metre}};

            % Maß für die BREITE (0.4) oben links
            \draw[line width=1pt,color=black,arrows={Stealth[inset=0pt, length=10pt, angle'=30]-Stealth[inset=0pt, length=10pt, angle'=30]}] 
            (0.002, 0.805) -- (0.29,1.01) node[midway, above, xshift=-4pt, sloped] {\SI{0.4}{\milli \metre}};

         \filldraw (0.148,0.752) circle (2.5pt);
        \node at (0.148,0.77) {P};
        \draw[line width=2pt,color=black,dashed] (0.148,0.752) -- (0.85,0.25);
        %\blue plane
        \fill[blue, opacity=0.2] (0.148,0.495) -- (0.148,0.895) -- (0.85,0.398) -- (0.85,-0.01) -- cycle;

        \end{scope}
    \end{tikzpicture}
    \caption{Geometry, mesh and dimensions of the rectangular prism model.}
    \label{Fig: Geometry}
\end{figure}

This section presents the simulation results of a three-dimensional boundary value problem described. The simulation shows the additive manufacturing process of glass, utilizing the full coupling between thermal transport, phase transformation kinetics, and mechanical deformation. To implement this macroscale problem, the custom constitutive laws derived in the previous chapters were integrated into the commercial finite element software $\mathrm{ANSYS}^{\copyright}$, utilizing the numerical solution strategies detailed in \autoref{Sec: NEM}. The model geometry consists of a rectangular prism with dimensions of \SI{1}{\milli \metre}$ \times $\SI{0.4}{\milli \metre}$ \times $\SI{0.4}{\milli \metre}, as illustrated in \autoref{Fig: Geometry}. The domain is discretized with a structured mesh of $12,000$  first-order hexahedral elements, maintaining a uniform element size of \SI{0.02}{\milli \metre}. Crucially, the computational mesh encompasses both the solid elements representing the progressively deposited material and the surrounding air elements that define the void space into which the glass track is built. To simulate a rigid connection to an underlying manufacturing platform, the bottom face of the substrate is fully fixed, completely constraining all translational degrees of freedom ($\bfu=\boldsymbol{0}\forall\in\partial\Omega(z=0)$).

The progressive material deposition is simulated along a single linear track, following the path indicated by the dashed line \protect\tikz[baseline]{\protect\draw[line width=2pt,color=black,dashed] (0,.8ex)--++(0.5,0) ;} indicated by P. One cross-sectional plane, delineated by blue \protect\tikz[baseline]{\protect\filldraw(0,0)[color=blue,opacity=0.2] rectangle(0.3,0.3) ;} in the \autoref{Fig: Geometry}, are designated for subsequent post-processing and result visualization. The coupling of the user-defined multi-physics material model with the $\mathrm{ANSYS}^{\copyright}$ solver is achieved through two dedicated user subroutines : the USERMAT subroutine manages the mechanical response, calculating the residual stress $\bfsigma$ and the consistent tangent stiffness matrix. While the USSFIN subroutine evaluates the localized evolution of temperature $\theta$, internal viscous variable $\bfC^{\mathrm{v}}$ and the phase volume fractions $\bflambda$ at each integration point. This coupled system is solved efficiently using the monolithic operator-split of the Neighbored Element Method (NEM) \cite{rudolf2025computational,vogel2020adaptive}. In the following sections, the resulting fields are systematically analyzed in terms of their thermal history, spatial microstructural evolution, and the accumulation of residual stresses and macroscopic distortions.

\subsubsection{Thermal History and Material Deposition}

\begin{figure}
    \centering

  %% --- Spaltenüberschriften ---
  \begin{subfigure}{0.20\textwidth}
    \quad % Leerraum über der Zeit-Spalte
  \end{subfigure}
  %\hspace{0.1\textwidth}
  \begin{subfigure}{0.30\textwidth}
    \centering\textbf{Printed material $\phi$}
  \end{subfigure}
  \hspace{0.1\textwidth}
  \begin{subfigure}{0.30\textwidth}
    \centering\textbf{Body view}
  \end{subfigure}
  \hfill
  \vspace{0.2em}
    
%%%% t=0.39s%%%%%%%%%%%%%%%%%%%%%%%%
    \begin{subfigure}{0.20\textwidth}
        \centering
        \adjustbox{raise ={1ex}{\height}}{\rotatebox{90}{\textbf{$t=\SI{0.39}{\second}$}}}
    \end{subfigure}
    %\hspace{0.1\textwidth}
    \begin{subfigure}{0.30\textwidth}
    \centering
    \includegraphics[width=1\textwidth]{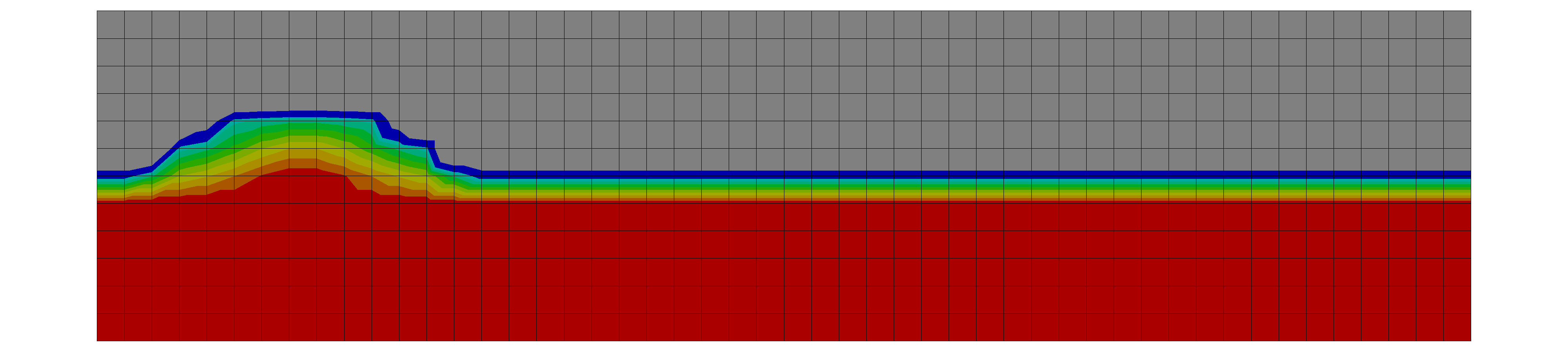} 
    \end{subfigure}
     \hspace{0.1\textwidth}
    \begin{subfigure}{0.30\textwidth}
    \centering
    \adjustbox{raise ={-2ex}}{
    \includegraphics[width=1\textwidth]{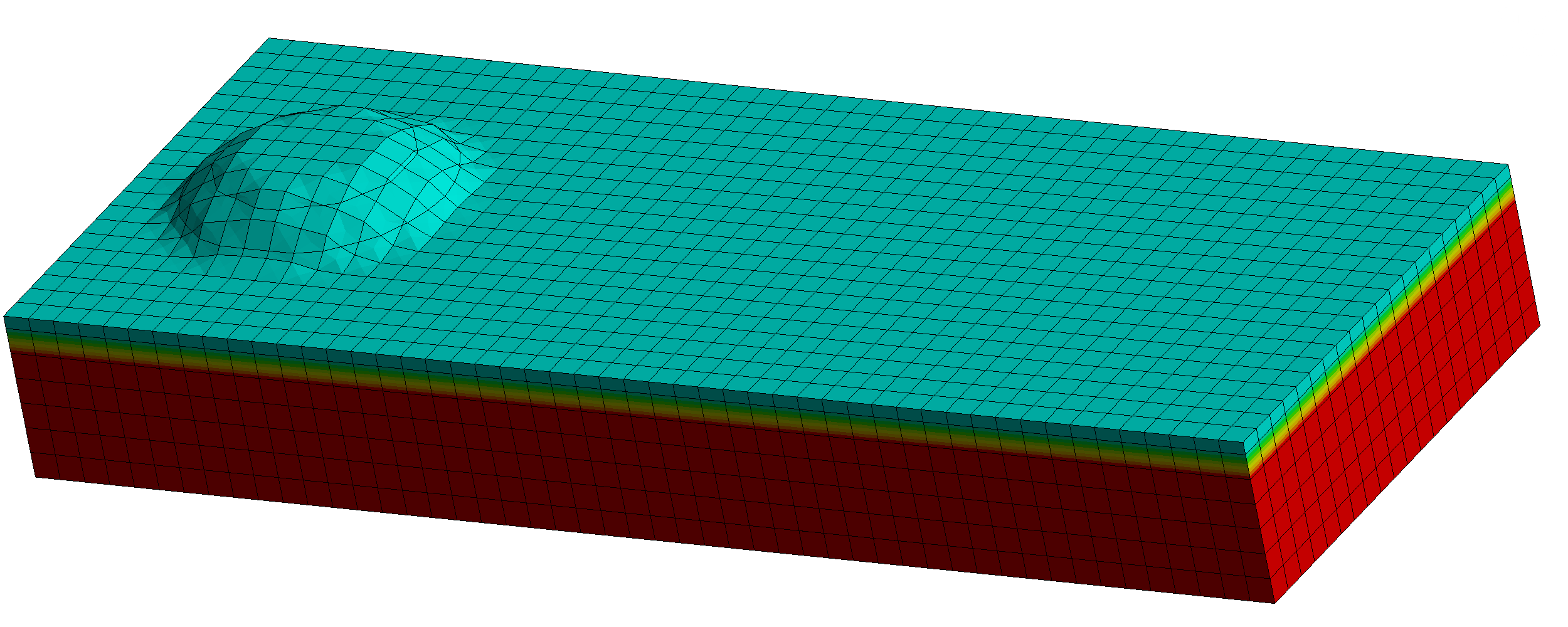}}
    \end{subfigure}
    \hfill
    \vspace{0.2em}
 
%%%% t=0.5s%%%%%%%%%%%%%%%%%%%%%%%%
    \begin{subfigure}{0.2\textwidth}
        \centering
        \adjustbox{raise ={1ex}{\height}}{\rotatebox{90}{\textbf{$t=\SI{0.5}{\second}$}}}
    \end{subfigure}
    \begin{subfigure}{0.30\textwidth}
    \includegraphics[width=1\textwidth]{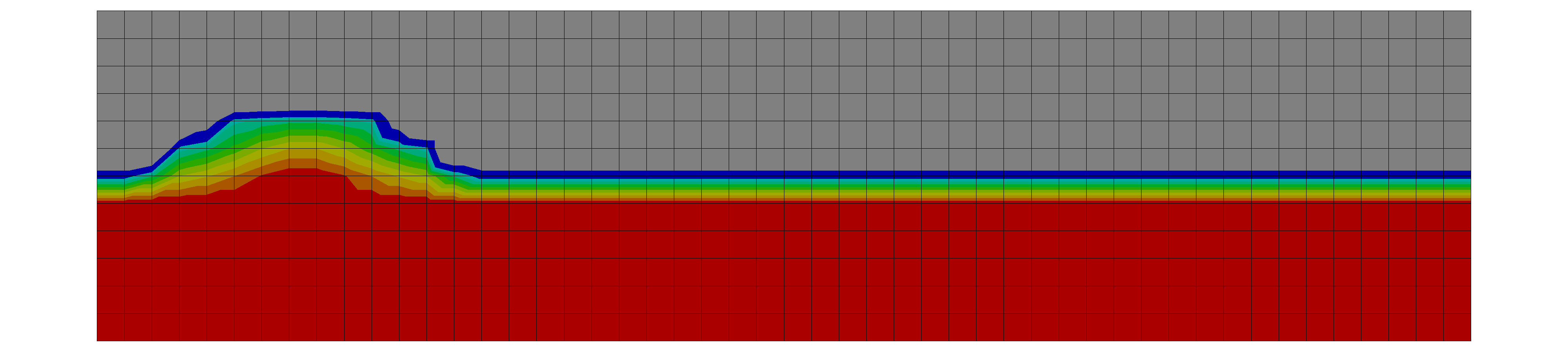} 
    \end{subfigure}
    \hspace{0.1\textwidth}
    \begin{subfigure}{0.30\textwidth}
    \centering
    \adjustbox{raise ={-2ex}}{
    \includegraphics[width=1\textwidth]{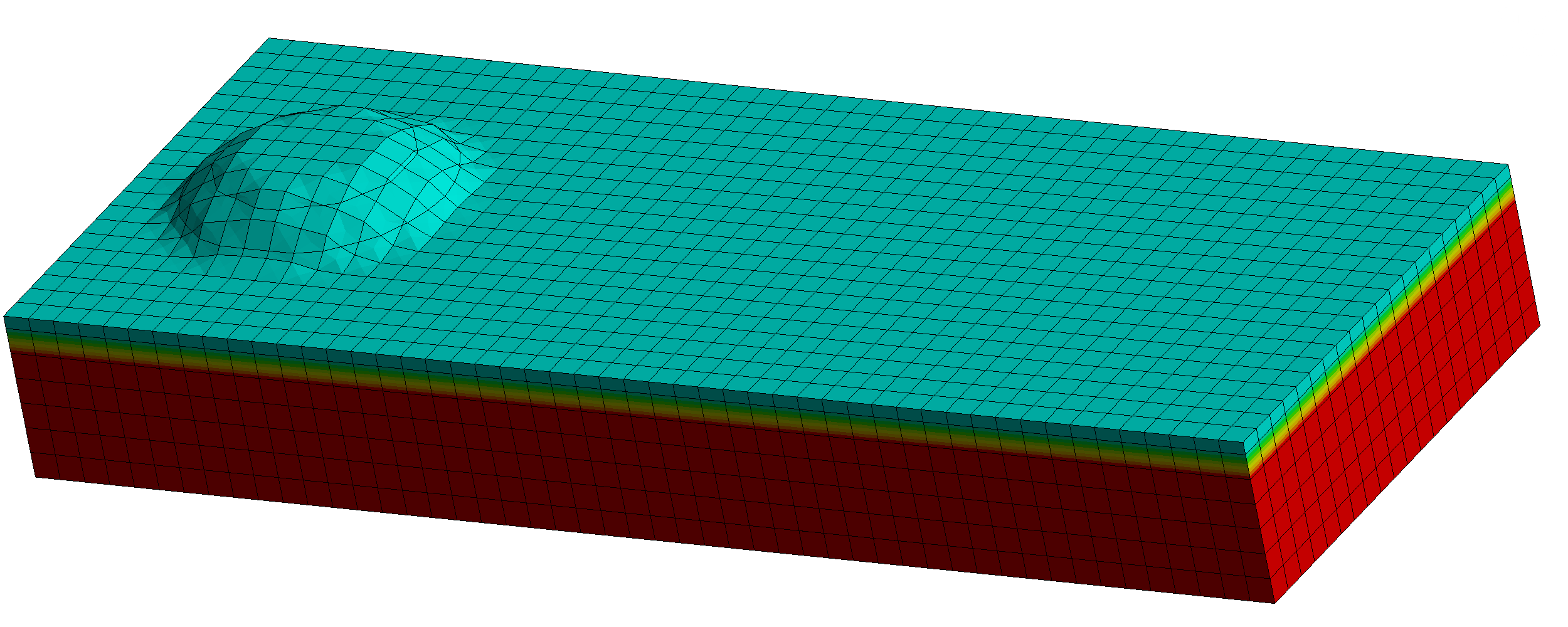}}
    \end{subfigure}
    \hfill
    \vspace{0.2em}
    
%%%% t=1.3s%%%%%%%%%%%%%%%%%%%%%%%%
    \begin{subfigure}{0.2\textwidth}
        \centering
        \adjustbox{raise ={1ex}{\height}}{\rotatebox{90}{\textbf{$t=\SI{1.3}{\second}$}}}
    \end{subfigure}
    \begin{subfigure}{0.30\textwidth}
    \includegraphics[width=1\textwidth]{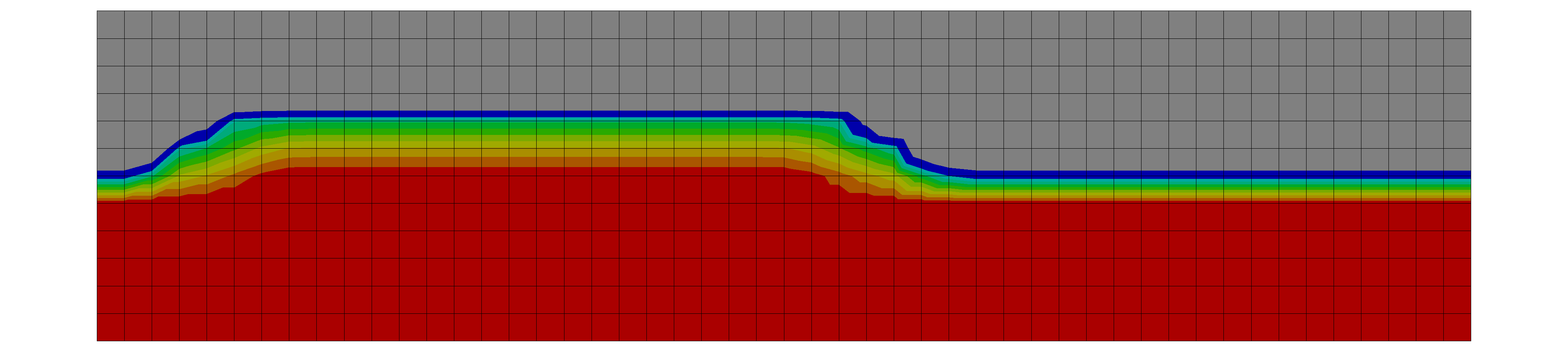} 
    \end{subfigure}
    \hspace{0.1\textwidth}
    \begin{subfigure}{0.30\textwidth}
    \centering
    \adjustbox{raise ={-2ex}}{
    \includegraphics[width=1\textwidth]{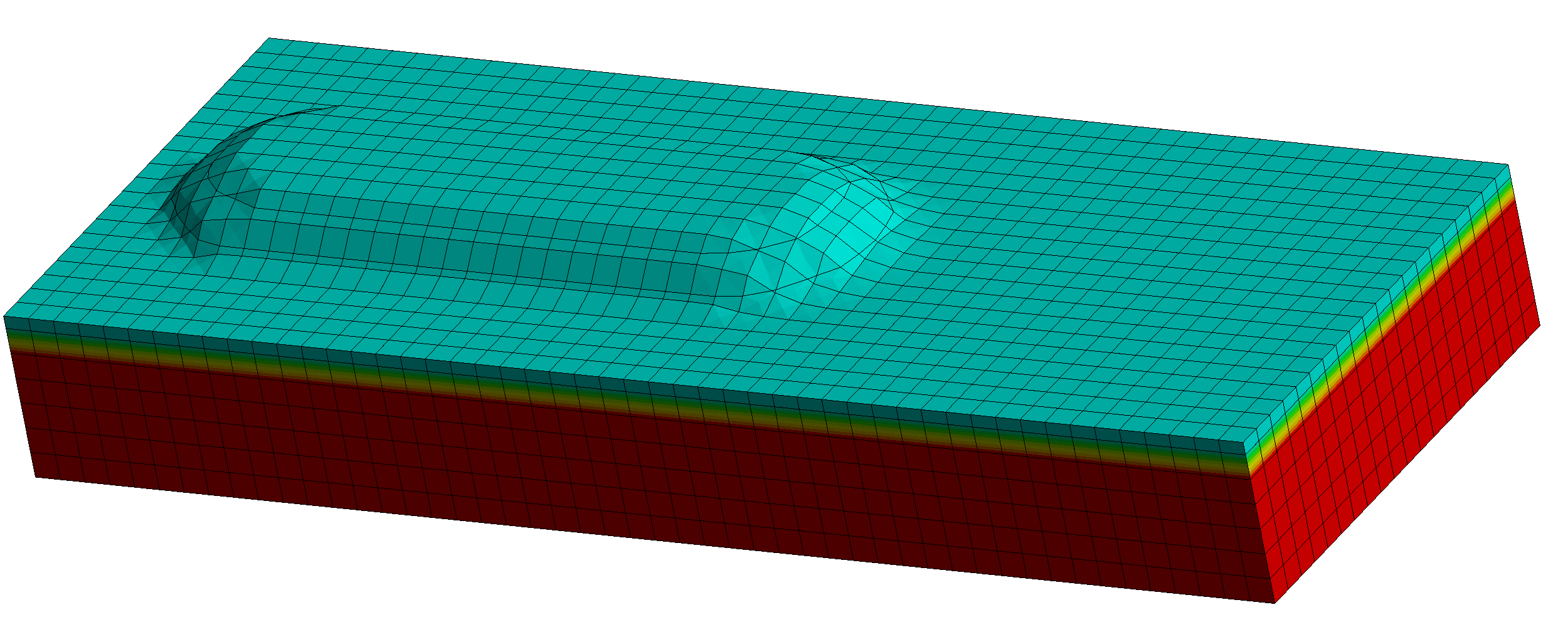}}
    \end{subfigure}
    \hfill
    \vspace{0.2em}
    
%%%% t=2.15s%%%%%%%%%%%%%%%%%%%%%%%%
    \begin{subfigure}{0.2\textwidth}
        \centering
        \adjustbox{raise ={1ex}{\height}}{\rotatebox{90}{\textbf{$t=\SI{2.15}{\second}$}}}
    \end{subfigure}
    \begin{subfigure}{0.30\textwidth}
    \includegraphics[width=1\textwidth]{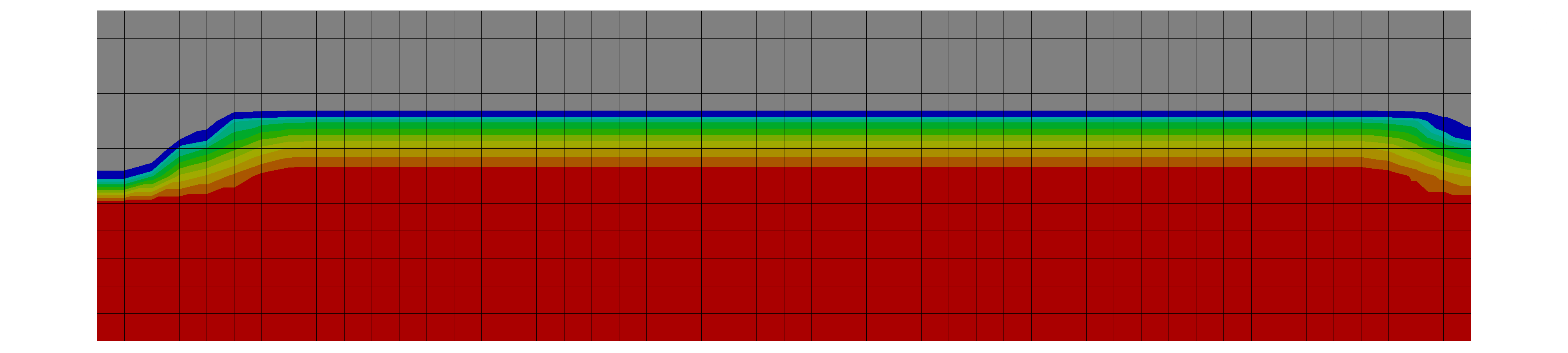} 
    \end{subfigure}
    \hspace{0.1\textwidth}
    \begin{subfigure}{0.30\textwidth}
    \centering
    \adjustbox{raise ={-2ex}}{
    \includegraphics[width=1\textwidth]{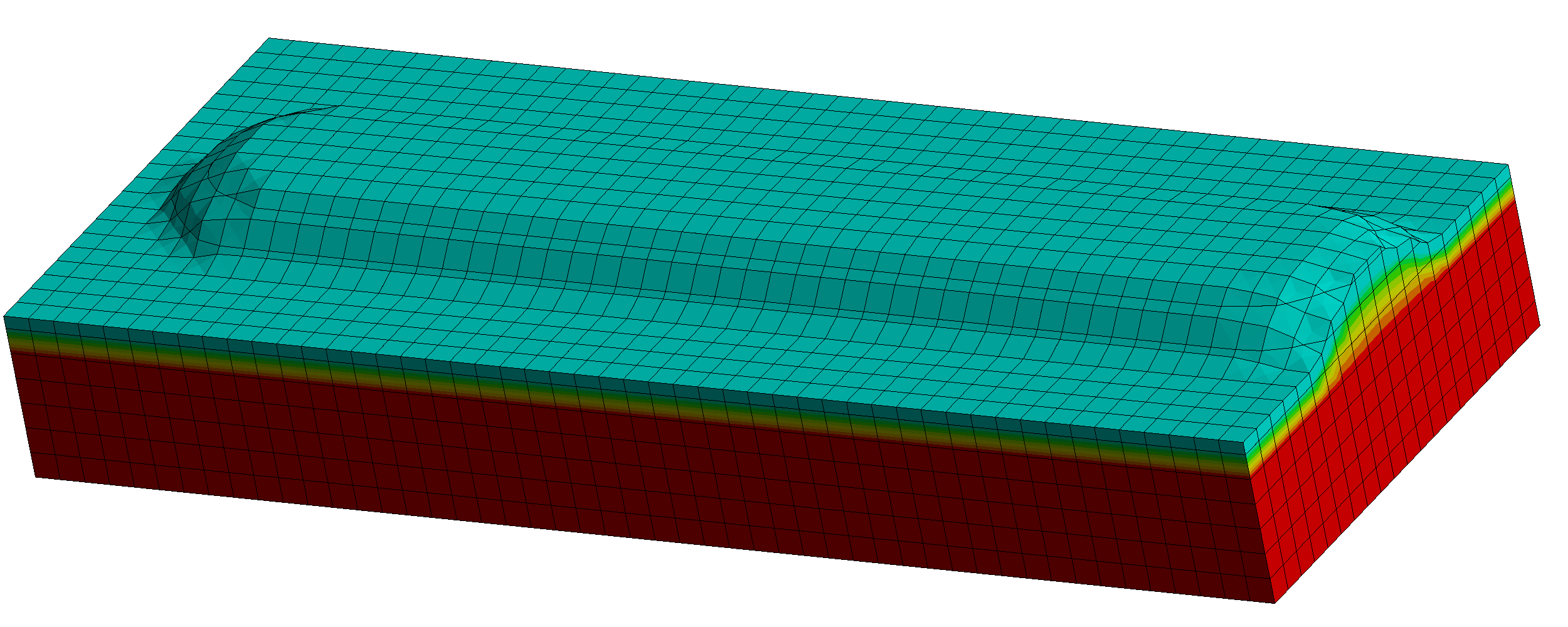}}
    \end{subfigure}
    \hfill
    \vspace{0.2em}
    
%%%% t=2.3s%%%%%%%%%%%%%%%%%%%%%%%%
    \begin{subfigure}{0.2\textwidth}
        \centering
        \adjustbox{raise ={1ex}{\height}}{\rotatebox{90}{\textbf{$t=\SI{2.3}{\second}$}}}
    \end{subfigure}
    \begin{subfigure}{0.30\textwidth}
    \includegraphics[width=1\textwidth]{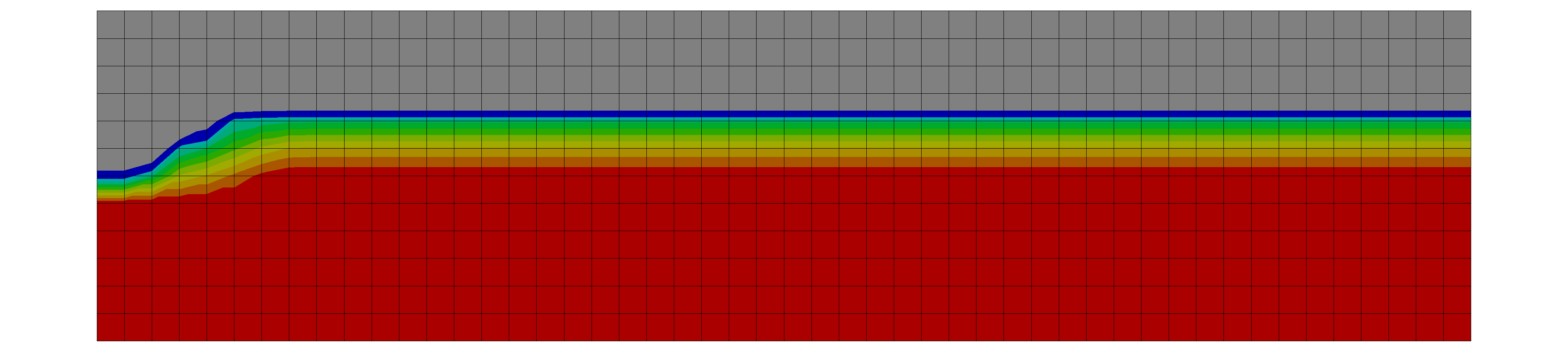} 
    \end{subfigure}
    \hspace{0.1\textwidth}
    \begin{subfigure}{0.30\textwidth}
    \centering
    \adjustbox{raise ={-2ex}}{
    \includegraphics[width=1\textwidth]{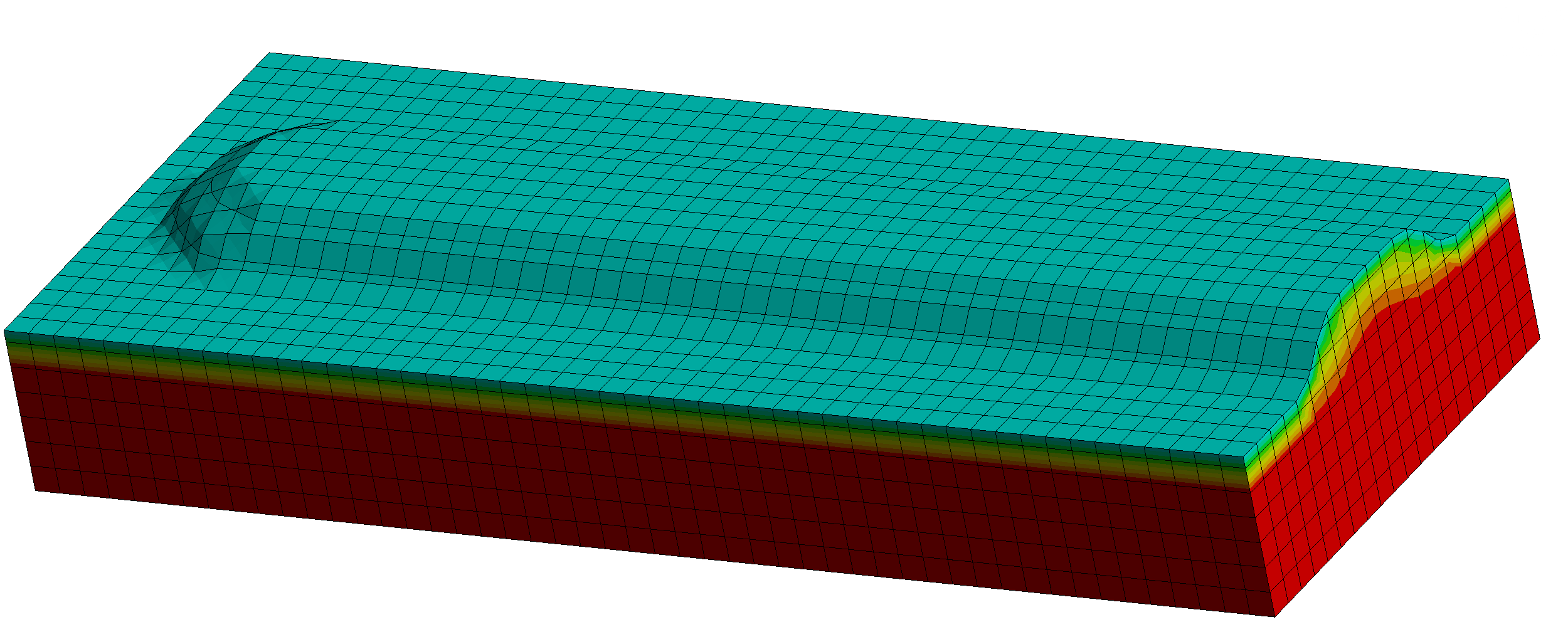}}
    \end{subfigure}
    \hfill
    \vspace{0.2em}

    %%%% legends
    \begin{subfigure}{1\textwidth}
    \centering
    \input{Figures/FEM/Legend_0-1}
    \end{subfigure}
    \caption{FEM results for the printed material $\phi$ in the blue cross section plane \protect\tikz[baseline]{\protect\filldraw(0,0)[color=blue,opacity=0.2] rectangle(0.3,0.3) ;} and as body view at different times, i.e. $t=\SI{0}{\second}$ to $t=\SI{2.3}{\second}$. The gray color \protect\tikz[baseline]{\protect\filldraw(0,0)[color=gray] rectangle(0.3,0.3) ;} indicates the air phase.}
    \label{Fig: FEM Result Phi}
\end{figure}

The simulation of the additive manufacturing process is fundamentally governed by the progressive accumulation of material along the prescribed deposition path. \autoref{Fig: FEM Result Phi} visualizes this evolution. Consistent with the mixture energy formulation defined in \autoref{EQ: mixed free energy}, the topology of the deposited track is not delineated by a sharp boundary, but rather by the continuous evolution of the solid volume fraction $\phi$. Consequently, the cross-sectional views reveal a diffuse interface zone between the bulk glass ($\phi=1$) and the surrounding air ($\phi=0$). Physically, this transition region captures the sintering and wetting zone where the material consolidates. The deposition follows the path P, resulting in a continuous track that is fully generated by $t=\SI{2.3}{\second}$.

\begin{figure}
\centering
%%%%%%%% t=0,39 and t=0,5 and t=1,3%%%%%%
    \begin{subfigure}{0.30\textwidth}
    \centering
    \includegraphics[width=1\textwidth]{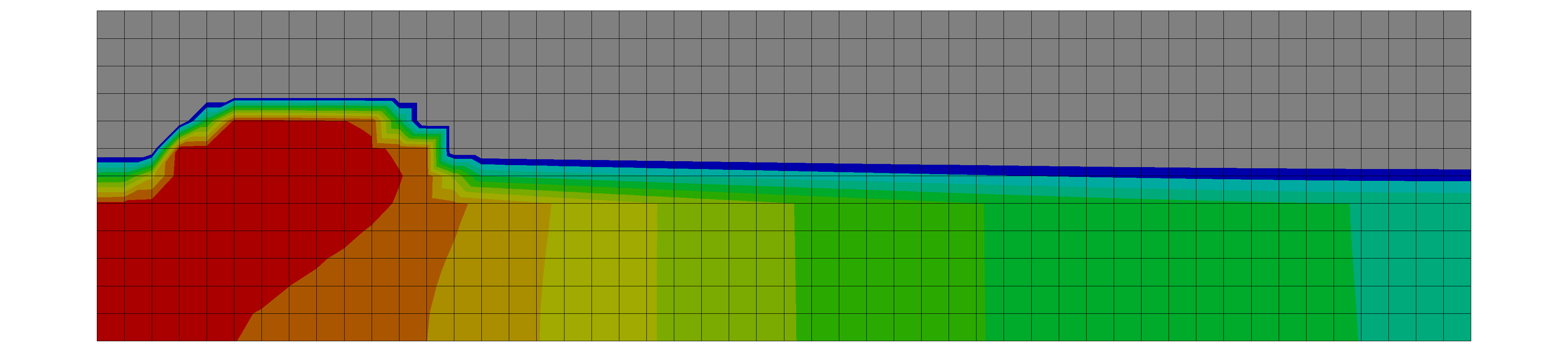}
    \end{subfigure}
    \begin{subfigure}{0.30\textwidth}
    \centering
    \includegraphics[width=1\textwidth]{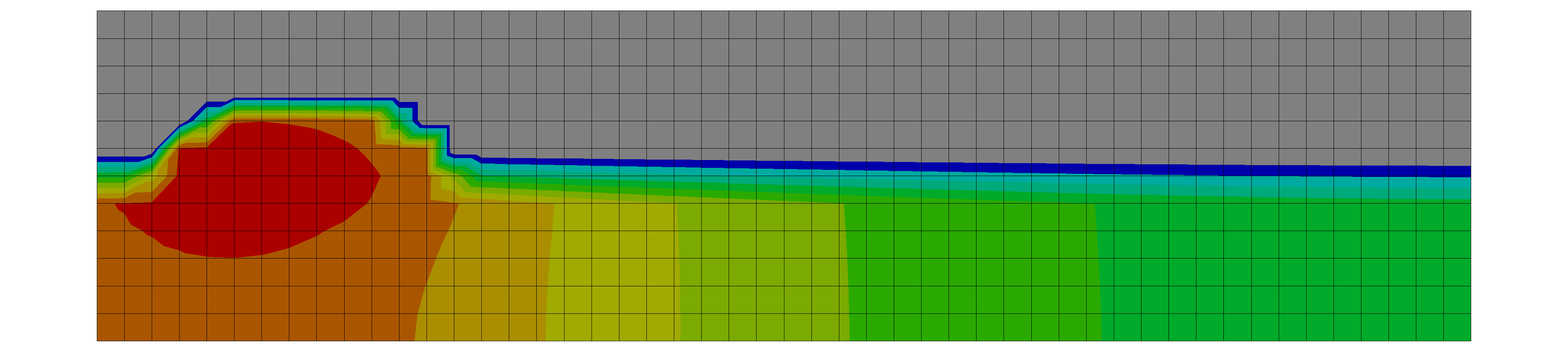}
    \end{subfigure}
    \begin{subfigure}{0.30\textwidth}
    \centering
    \includegraphics[width=1\textwidth]{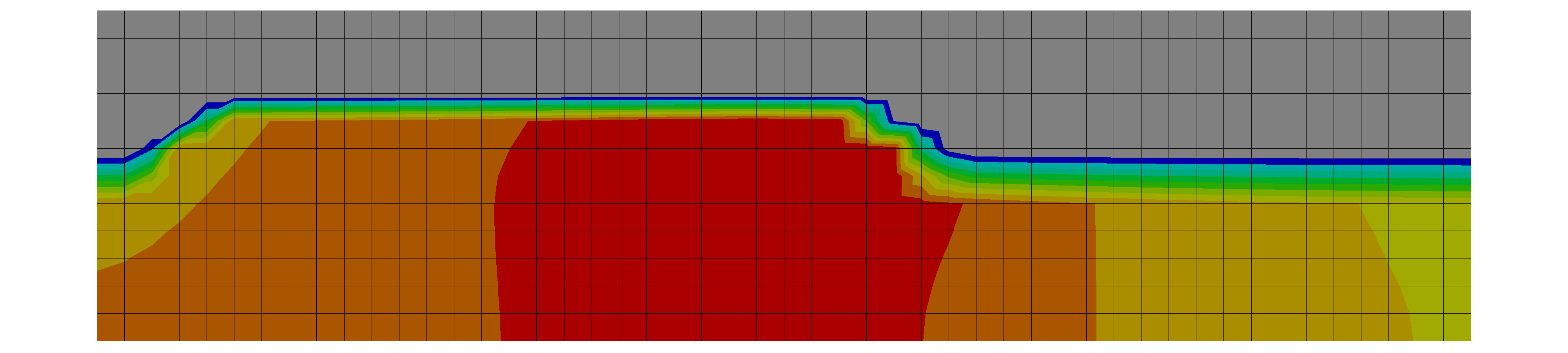}
    \end{subfigure}
    \hfill
    \begin{subfigure}{0.30\textwidth}
    \centering
    \input{Figures/FEM/Temperature/Legend/Legend_Temp_t0-39}
    \vspace{-15pt}
    \caption{Temperature $t=\SI{0.39}{\second}$} 
    \label{Fig: FEM temperature at t=0.39}
    \end{subfigure}
    \begin{subfigure}{0.30\textwidth}
    \centering
    \input{Figures/FEM/Temperature/Legend/Legend_Temp_t0-5}
    \vspace{-15pt}
    \caption{Temperature $t=\SI{0.5}{\second}$} 
    \label{Fig: FEM temperature at t=0.5}
    \end{subfigure}
    \begin{subfigure}{0.30\textwidth}
    \centering
    \input{Figures/FEM/Temperature/Legend/Legend_Temp_t1-3}
    \vspace{-15pt}
    \caption{Temperature $t=\SI{1.3}{\second}$} 
    \label{Fig: FEM temperature at t=1.3}
    \end{subfigure}
    \hfill
    
    \vspace{10pt}
    %\hfill
%%%%%%%% t=2,15, t=2,55 and t=4 %%%%%%%%%%%%%%%
    \begin{subfigure}{0.30\textwidth}
    \centering
    \includegraphics[width=1\textwidth]{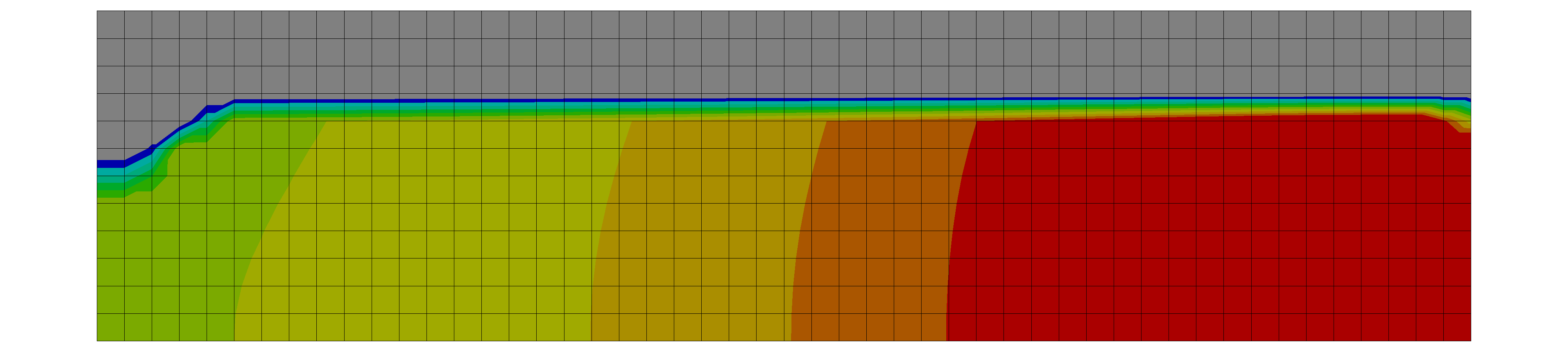}
    \end{subfigure}
    \begin{subfigure}{0.30\textwidth}
    \centering
    \includegraphics[width=1\textwidth]{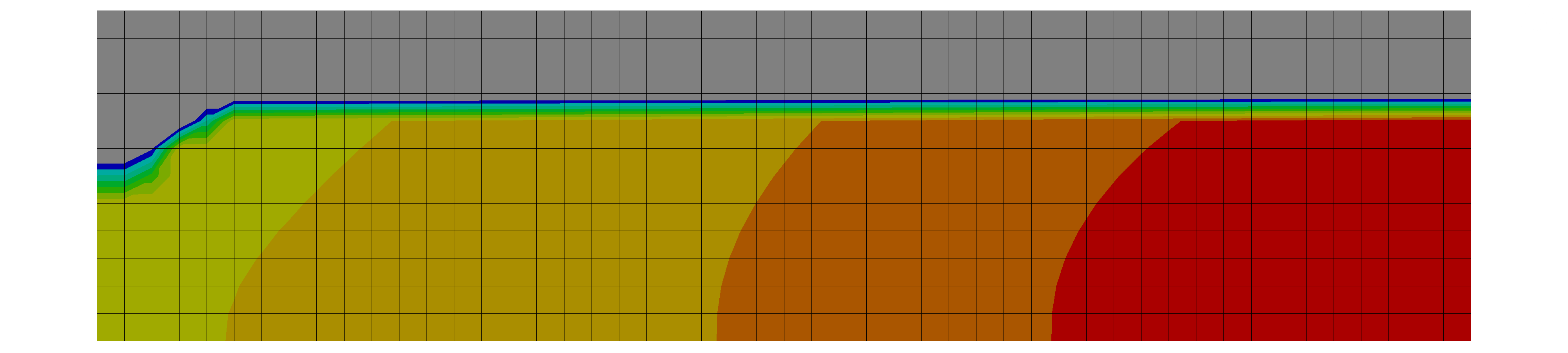}
    \end{subfigure}
    \begin{subfigure}{0.30\textwidth}
    \centering
    \includegraphics[width=1\textwidth]{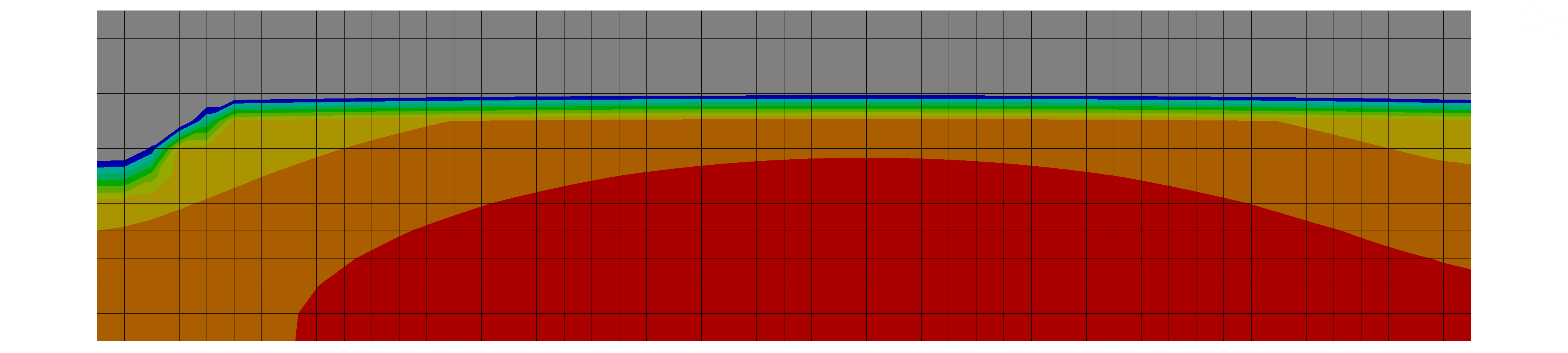}
    \end{subfigure}
    \hfill
    \begin{subfigure}{0.30\textwidth}
    \centering
    \input{Figures/FEM/Temperature/Legend/Legend_Temp_t2-15}
    \vspace{-15pt}
    \caption{Temperature $t=\SI{2.15}{\second}$}
    \label{Fig: FEM temperature at t=2.15}
    \end{subfigure}
    \begin{subfigure}{0.30\textwidth}
    \centering
    \input{Figures/FEM/Temperature/Legend/Legend_Temp_t2-55}
    \vspace{-15pt}
    \caption{Temperature $t=\SI{2.55}{\second}$}
    \label{Fig: FEM temperature at t=2.55}
    \end{subfigure}
    \begin{subfigure}{0.30\textwidth}
    \centering
    \input{Figures/FEM/Temperature/Legend/Legend_Temp_t4-0}
    \vspace{-15pt}
    \caption{Temperature $t=\SI{4}{\second}$}
    \label{Fig: FEM temperature at t=4}
    \end{subfigure}
    \hfill
    
%%%%%%%% Diagramm%%%%%%
    \begin{subfigure}{1\textwidth}
    \vspace{5pt}
    \centering
    \begin{tikzpicture}
\begin{axis}[
    width=0.8\textwidth,
    height=0.4\textwidth,
    xlabel={Time [$\SI{}{\second}$]},
    ylabel={Temperature $\theta$ [$\SI{}{\kelvin}$]},
    grid=major,
    %xmode=log,
    enlarge x limits=false,
    legend pos=north east,
    %legend cell align={left},
    %legend style={font=\footnotesize},
    xmin=0, %xmax=500000,
    ymin=300, ymax=3000,
    ytick={500,1000,2716},
    yticklabel style={
 /pgf/number format/fixed,
 /pgf/number format/precision=0,
 /pgf/number format/fixed zerofill,
 /pgf/number format/1000 sep={}
},
        extra y ticks={1683.15, 2003.15},
        extra y tick labels={$\theta^{\mathrm{g}}$,$\theta^{\mathrm{m}}$},
        extra y tick style={
            grid=major,
            %grid style={dashed}, % Gilt für alle extra ticks, man kann es auch pro tick setzen
        },
]
	
% FEM Temperatur Minimum
\addplot[blue, thick,solid, mark=none] table [x index=0,y index=1,col sep=semicolon] {Diagrams/FEM_Temperature.txt};
\label{Line: FEM Temperature Diagramm Min}
% FEM Temperatur Maxmimum
\addplot[red, thick, solid, mark=none] table [x index=0,y index=2,col sep=semicolon] {Diagrams/FEM_Temperature.txt}; 
\label{Line: FEM Temperature Diagramm Max}
% FEM Temperatur Average
\addplot[black, thick,solid, mark=none] table [x index=0,y index=3,col sep=semicolon] {Diagrams/FEM_Temperature.txt};
\label{Line: FEM Temperature Diagramm Aver}

% Linie für die Schmelztemperatur
%\draw[black, thick, dotted] (axis cs:0,2003.15) -- (axis cs:4.3,2003.15)    
%node at (axis cs:4,2143.15) {$\theta^{\mathrm{m}}$};
%;
% Linie für die Glastransition
%\draw[black, thick, dotted] (axis cs:0,1683.15) -- (axis cs:4.3,1683.15)    
%node at (axis cs:4,1543.15) {$\theta^{\mathrm{g}}$};
%;
%Subfigure a
\filldraw (axis cs:0.39,1959) circle (2.5pt);
\node at (axis cs:0.18,1959) {(\ref{Fig: FEM temperature at t=0.39})};

%Subfigure b
\filldraw (axis cs:0.5,2118) circle (2.5pt);
\node at (axis cs:0.5,2330) {(\ref{Fig: FEM temperature at t=0.5})};

%Subfigure c
\filldraw (axis cs:1.3,2309) circle (2.5pt);
\node at (axis cs:1.3,2480) {(\ref{Fig: FEM temperature at t=1.3})};

%Subfigure d
\filldraw (axis cs:2.15,2731) circle (2.5pt);
\node at (axis cs:2.42,2731) {(\ref{Fig: FEM temperature at t=2.15})};

%Subfigure e
\filldraw (axis cs:2.55,1474) circle (2.5pt);
\node at (axis cs:2.75,1474) {(\ref{Fig: FEM temperature at t=2.55})};

%Subfigure f
\filldraw (axis cs:4,302) circle (2.5pt);
\node at (axis cs:3.9,480) {(\ref{Fig: FEM temperature at t=4})};

\legend{Minimum, Maximum, Average};

\end{axis}

\end{tikzpicture}
    \caption{Temperature progression over the whole domain $\Omega$} 
    \label{Fig: FEM temperature diagramm}
    \end{subfigure}    
    \caption{FEM results for temperature evolution $\theta$ in the blue cross section plane \protect\tikz[baseline]{\protect\filldraw(0,0)[color=blue,opacity=0.2] rectangle(0.3,0.3) ;}  at different times, i.e., $t=\SI{0}{\second}$ to $t=\SI{4}{\second}$ and temperature over the whole domain $\Omega$ as diagram. The gray color \protect\tikz[baseline]{\protect\filldraw(0,0)[color=gray] rectangle(0.3,0.3) ;} indicates the air phase.}
    \label{Fig: FEM Result Temperature}
\end{figure}

The thermal field, which acts as the driving force for both phase transformation kinetics and mechanical deformation, evolves in direct correlation with this material deposition. \autoref{Fig: FEM Result Temperature} details the spatio-temporal temperature distribution. The process can be categorized into three distinct thermal regimes:
\begin{enumerate}
    \item Initial Dwell and Melt Pool Formation ($0 < t \leq \SI{0.5}{\second}$): The simulation initiates with a static heating phase. As shown in \autoref{Fig: FEM temperature at t=0.5}, the maximum temperature rises sharply to $\SI{2118}{\kelvin}$ before the heat source begins its trajectory. This dwell time is critical for overcoming the thermal inertia of the substrate, establishing an initial melt pool, and ensuring sufficient adhesion between the deposited material and the base layer.
    \item Quasi-Stationary Deposition ($t > \SI{0.5}{\second}$): Upon the onset of laser movement, a quasi-steady thermal state is established around the melt pool. The cross-sectional snapshots (\autoref{Fig: FEM temperature at t=0.5} to \autoref{Fig: FEM temperature at t=2.15} ) illustrate the intense energy localization induced by the volumetric Gaussian source. As the source traverses the domain, the local temperature within the deposition zone consistently exceeds the melting temperature, peaking at $\SI{2731}{\kelvin}$ at $t=\SI{2.15}{\second}$. This heated state ensures a fully liquid, low-viscosity regime, facilitating stress-free material redistribution in the liquid state. Crucially, these snapshots highlight the development of severe spatial thermal gradients. While the deposition zone experiences extreme heating, the substrate and deposed zone remain at significantly lower temperatures. This differential thermal field is the primary mechanism driving the thermal expansion that leads to residual stress accumulation (discussed in \autoref{Sec: FEM Residual stresses and deformation}).
    \item Rapid Quenching and Vitrification ($t > \SI{2.5}{\second}$): The cessation of the energy input marks the transition to the cooling phase. \autoref{Fig: FEM temperature diagramm} reveals a precipitous drop in temperature, with $\theta^{\mathrm{max}}$ falling from the heated liquid state to below the glass transition temperature in less than $\SI{0.4}{\second}$. This corresponds to an average cooling rate exceeding $\SI{2600}{\kelvin\per\second}$. Referring to the TTT investigations in \autoref{Sec: Material Point Results}, this rate is of critical importance since it is orders of magnitude faster than the critical cooling rate required for crystallization. Consequently, the model predicts that the liquid structure is kinetically trapped, ensuring that the solidification process proceeds via vitrification rather than crystallization.
\end{enumerate}

\subsubsection{Phase Transformation Kinetics}

\begin{figure}[htbp]
  \centering
  
  %% --- Spaltenüberschriften ---
  \begin{subfigure}{0.04\textwidth}
    \quad % Leerraum über der Zeit-Spalte
  \end{subfigure}\hfill
  \begin{subfigure}{0.30\textwidth}
    \centering\textbf{Crystalline}
  \end{subfigure}\hfill
  \begin{subfigure}{0.30\textwidth}
    \centering\textbf{Amorphous}
  \end{subfigure}\hfill
  \begin{subfigure}{0.30\textwidth}
    \centering\textbf{Liquid}
  \end{subfigure}
  
  \vspace{0.2em}

  %%%% t=0.39s %%%%%%%%%%%%%%%%%%%%%%%%%%%%%%%%%%%%%%
  \begin{subfigure}{0.04\textwidth}
    \centering
    \adjustbox{raise ={1ex}{\height}}{\rotatebox{90}{\textbf{$t=\SI{0.39}{\second}$}}}
  \end{subfigure}\hfill
  \begin{subfigure}{0.30\textwidth}
    \includegraphics[width=1\textwidth]{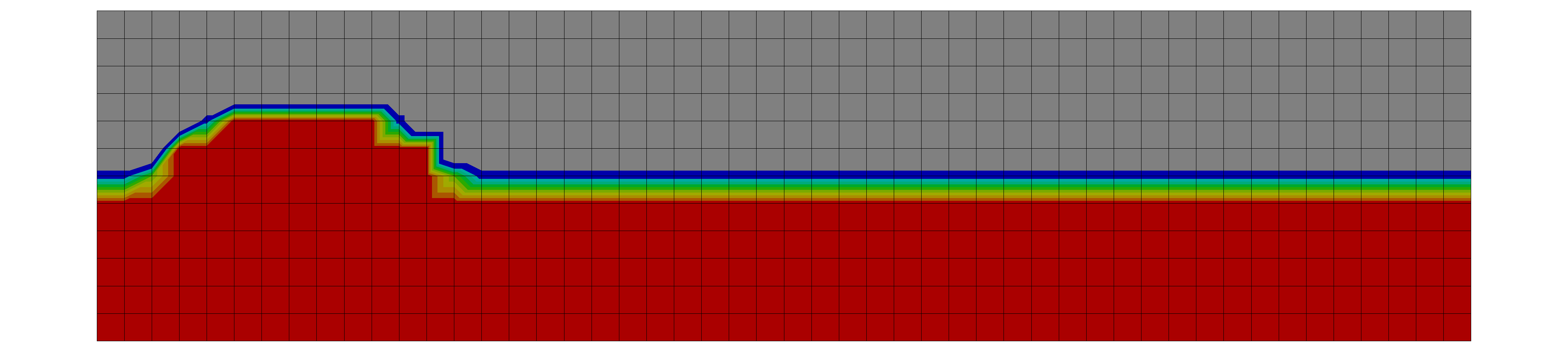}
    \caption{} \label{Fig: FEM Lambda Cry t=0,39}
  \end{subfigure}\hfill
  \begin{subfigure}{0.30\textwidth}
    \includegraphics[width=1\textwidth]{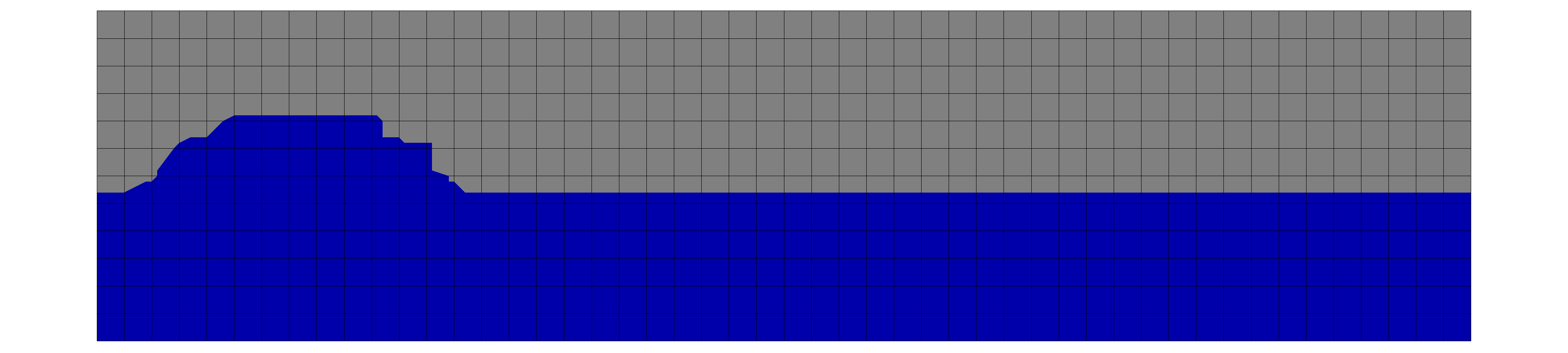}
    \caption{} \label{Fig: FEM Lambda Amo t=0,39}
  \end{subfigure}\hfill
  \begin{subfigure}{0.30\textwidth}
    \includegraphics[width=1\textwidth]{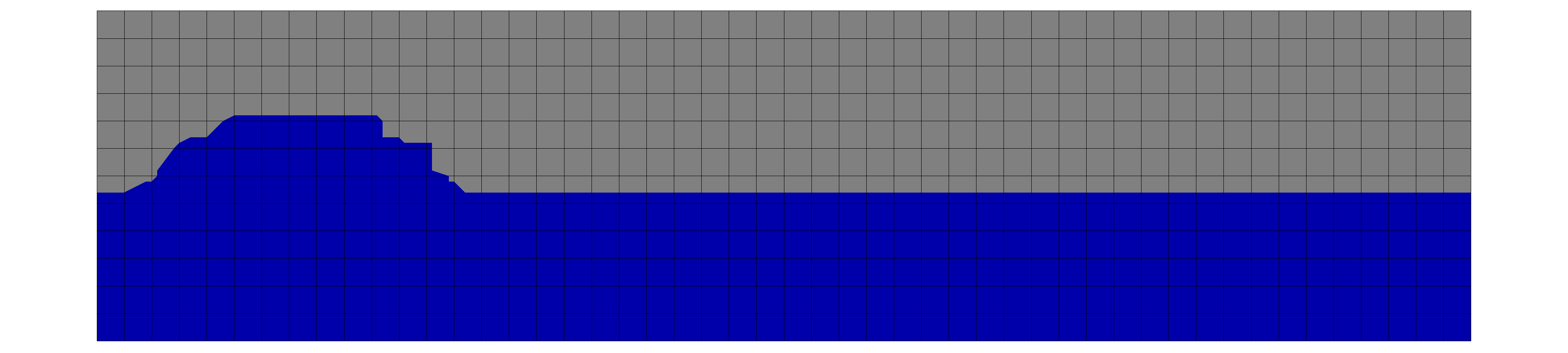}
    \caption{} \label{Fig: FEM Lambda Liq t=0,39}
  \end{subfigure}
  
  \vspace{0.2em}

  %%%% t=0.5s %%%%%%%%%%%%%%%%%%%%%%%%%%%%%%%%%%%%%%%
  \begin{subfigure}{0.04\textwidth}
    \centering
    \adjustbox{raise ={1ex}{\height}}{\rotatebox{90}{\textbf{$t=\SI{0.5}{\second}$}}}
  \end{subfigure}\hfill
  \begin{subfigure}{0.30\textwidth}
    \includegraphics[width=1\textwidth]{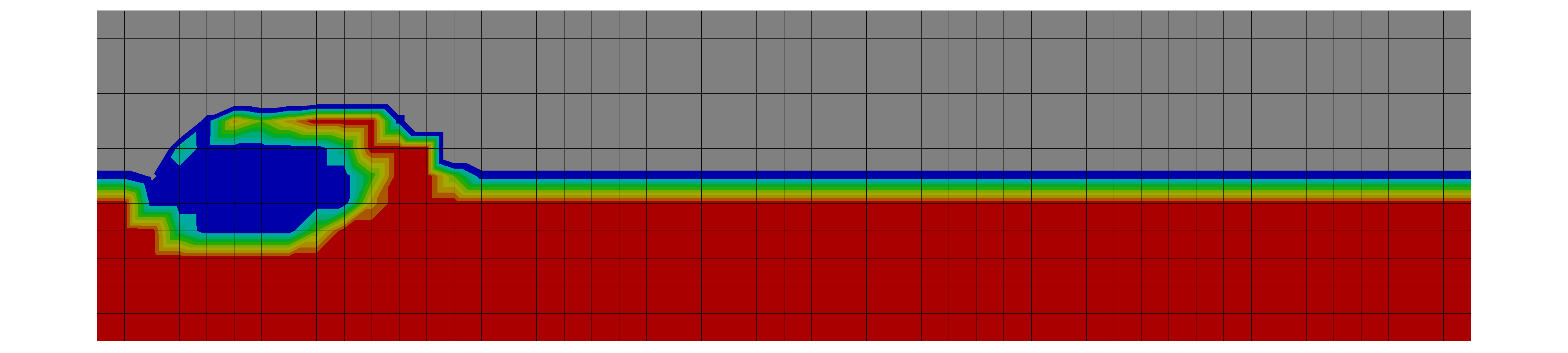}
    \caption{} \label{Fig: FEM Lambda Cry t=0,5}
  \end{subfigure}\hfill
  \begin{subfigure}{0.30\textwidth}
    \includegraphics[width=1\textwidth]{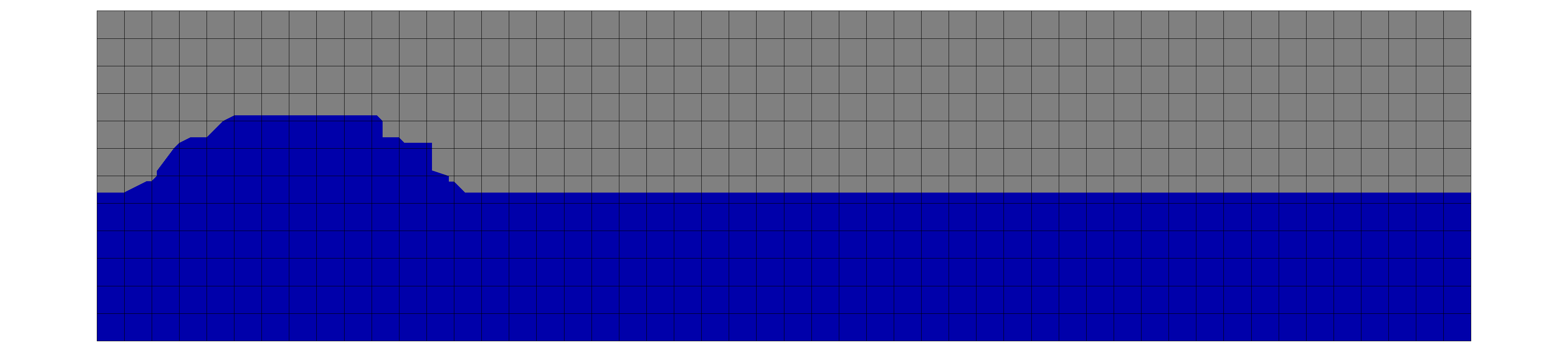}
    \caption{} \label{Fig: FEM Lambda Amo t=0,5}
  \end{subfigure}\hfill
  \begin{subfigure}{0.30\textwidth}
    \includegraphics[width=1\textwidth]{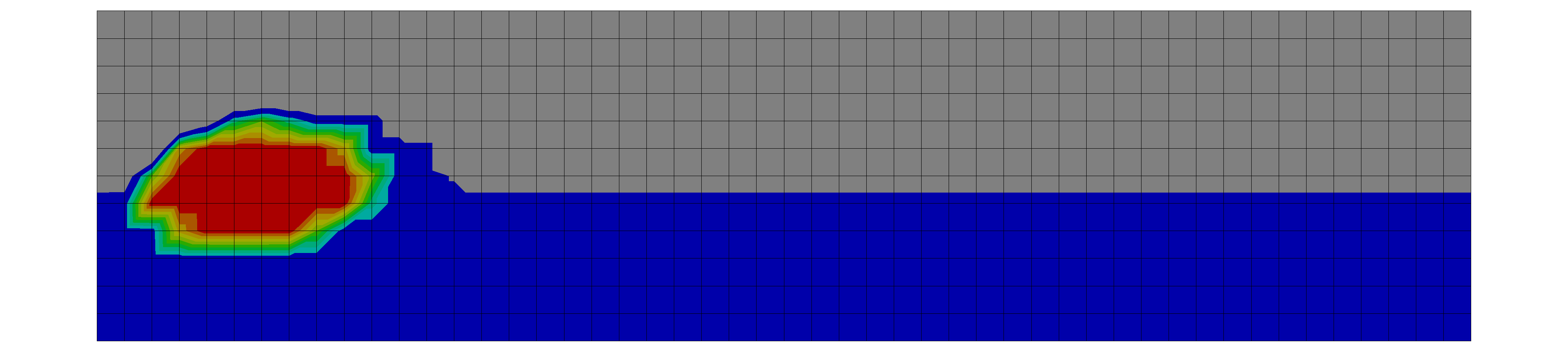}
    \caption{} \label{Fig: FEM Lambda Liq t=0,5}
  \end{subfigure}
  
  \vspace{0.2em}

  %%%% t=1.3s %%%%%%%%%%%%%%%%%%%%%%%%%%%%%%%%%%%%%%%
  \begin{subfigure}{0.04\textwidth}
    \centering
    \adjustbox{raise ={1ex}{\height}}{\rotatebox{90}{\textbf{$t=\SI{1.3}{\second}$}}}
  \end{subfigure}\hfill
  \begin{subfigure}{0.30\textwidth}
    \includegraphics[width=1\textwidth]{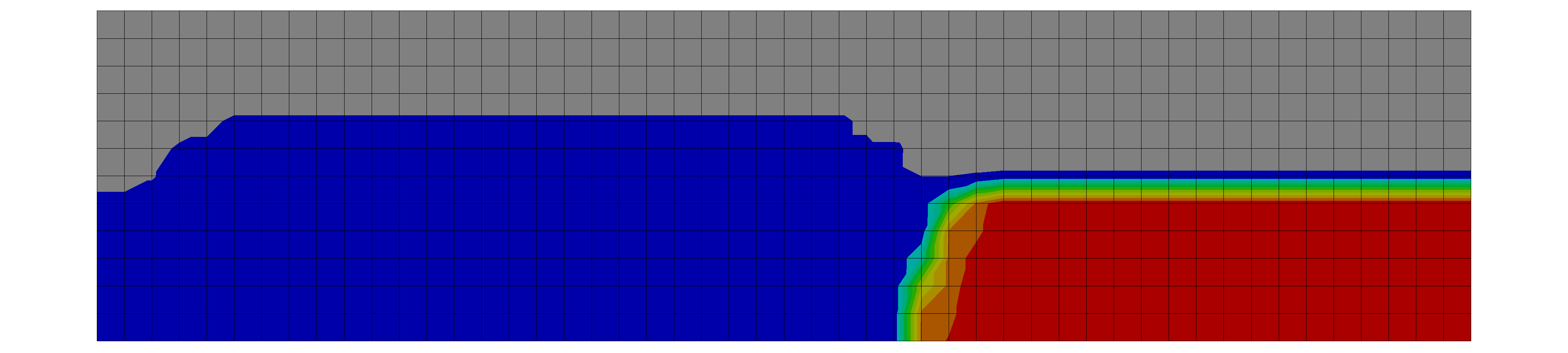}
    \caption{} \label{Fig: FEM Lambda Cry t=1,3}
  \end{subfigure}\hfill
  \begin{subfigure}{0.30\textwidth}
    \includegraphics[width=1\textwidth]{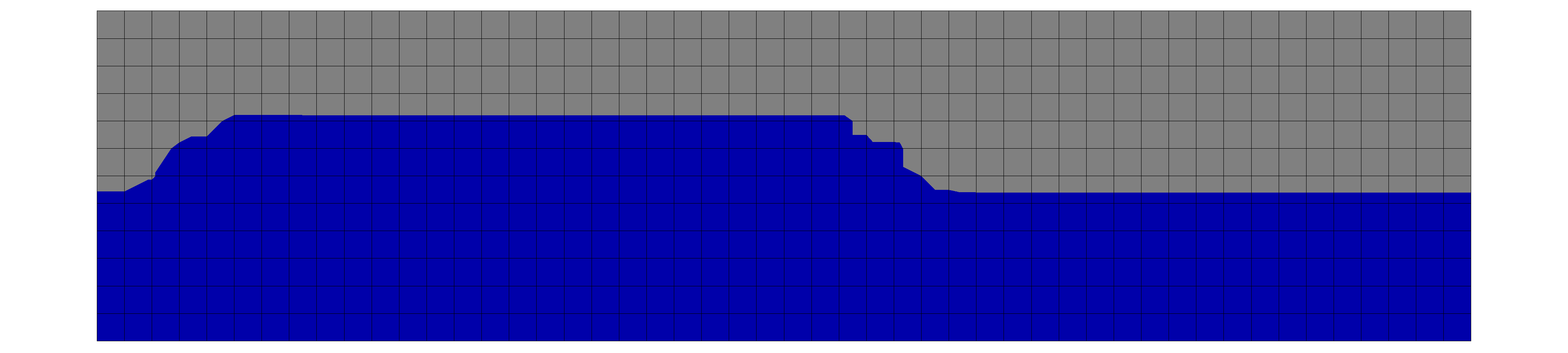}
    \caption{} \label{Fig: FEM Lambda Amo t=1,3}
  \end{subfigure}\hfill
  \begin{subfigure}{0.30\textwidth}
    \includegraphics[width=1\textwidth]{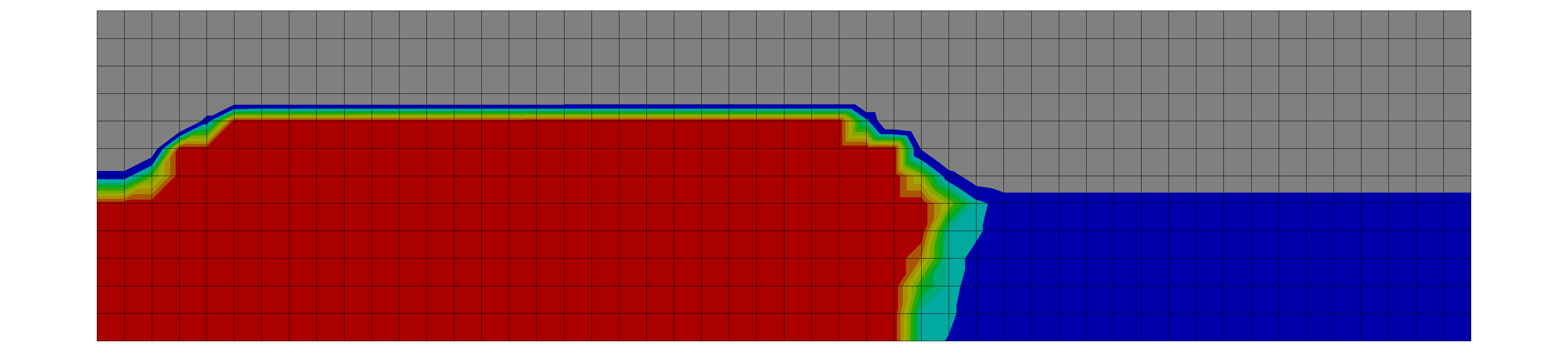}
    \caption{} \label{Fig: FEM Lambda Liq t=1,3}
  \end{subfigure}
  
  \vspace{0.2em}

  %%%% t=2.15s %%%%%%%%%%%%%%%%%%%%%%%%%%%%%%%%%%%%%%
  \begin{subfigure}{0.04\textwidth}
    \centering
    \adjustbox{raise ={1ex}{\height}}{\rotatebox{90}{\textbf{$t=\SI{2.15}{\second}$}}}
  \end{subfigure}\hfill
  \begin{subfigure}{0.30\textwidth}
    \includegraphics[width=1\textwidth]{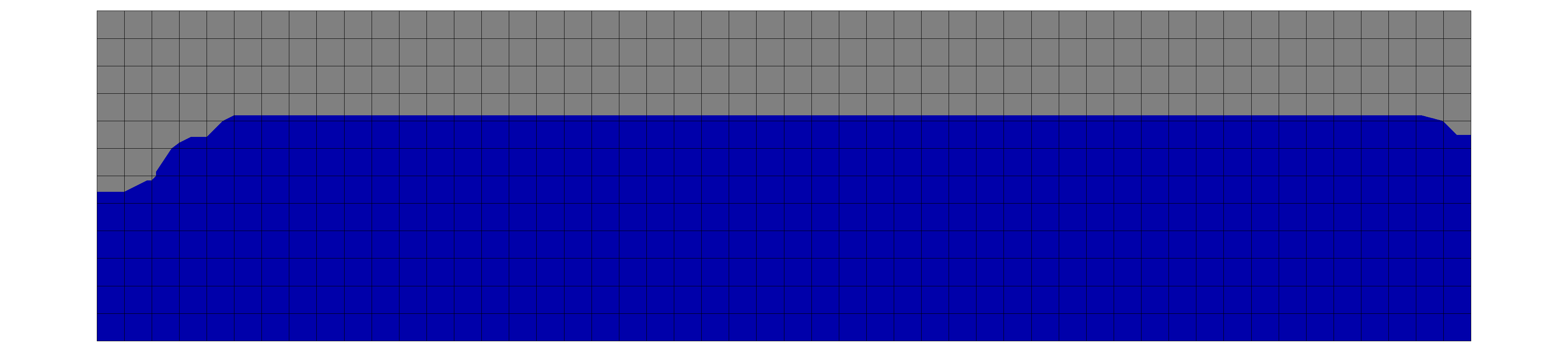}
    \caption{} \label{Fig: FEM Lambda Cry t=2,15}
  \end{subfigure}\hfill
  \begin{subfigure}{0.30\textwidth}
    \includegraphics[width=1\textwidth]{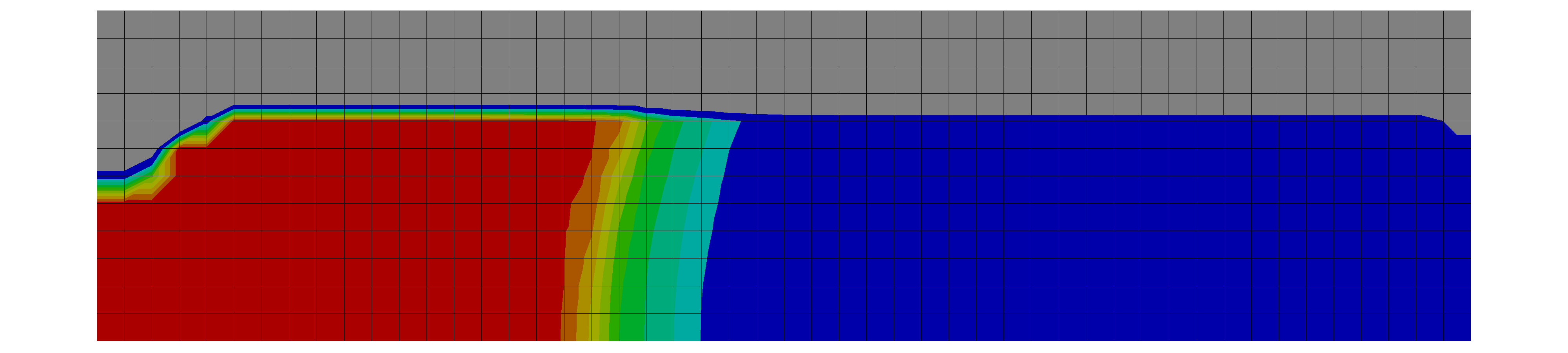}
    \caption{} \label{Fig: FEM Lambda Amo t=2,15}
  \end{subfigure}\hfill
  \begin{subfigure}{0.30\textwidth}
    \includegraphics[width=1\textwidth]{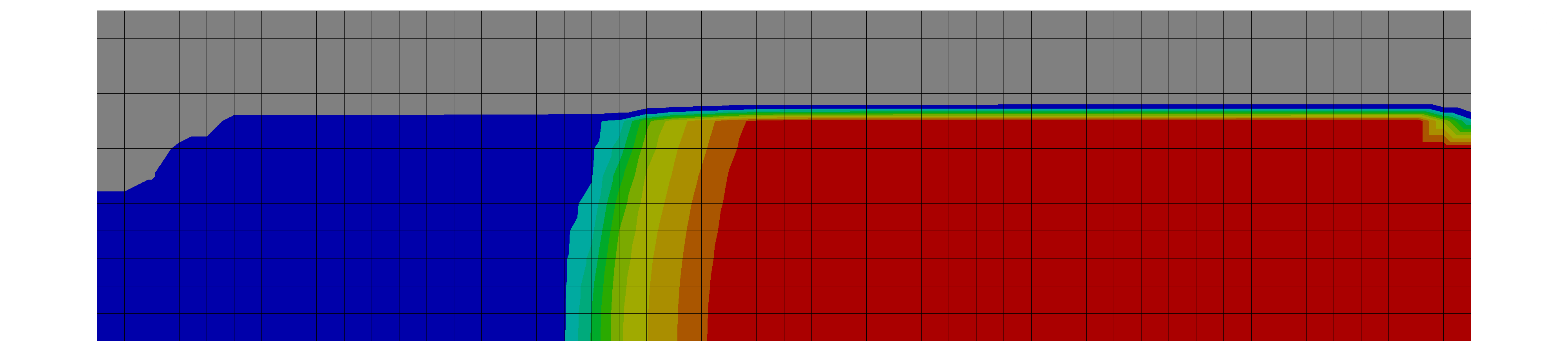}
    \caption{} \label{Fig: FEM Lambda Liq t=2,15}
  \end{subfigure}
  
  \vspace{0.2em}

  %%%% t=2.55s %%%%%%%%%%%%%%%%%%%%%%%%%%%%%%%%%%%%%%
  \begin{subfigure}{0.04\textwidth}
    \centering
    \adjustbox{raise ={1ex}{\height}}{\rotatebox{90}{\textbf{$t=\SI{2.55}{\second}$}}}
  \end{subfigure}\hfill
  \begin{subfigure}{0.30\textwidth}
    \includegraphics[width=1\textwidth]{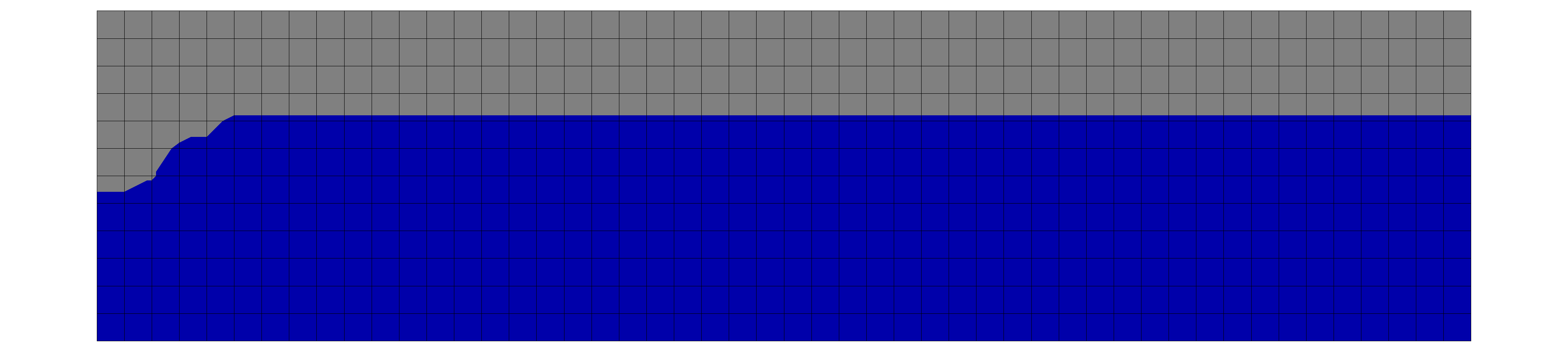}
    \caption{} \label{Fig: FEM Lambda Cry t=2,55}
  \end{subfigure}\hfill
  \begin{subfigure}{0.30\textwidth}
    \includegraphics[width=1\textwidth]{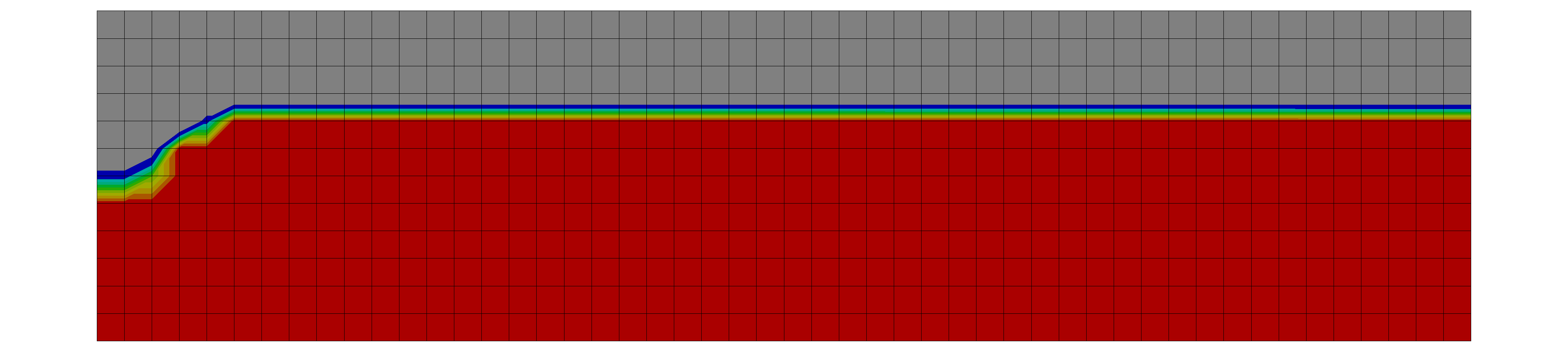}
    \caption{} \label{Fig: FEM Lambda Amo t=2,55}
  \end{subfigure}\hfill
  \begin{subfigure}{0.30\textwidth}
    \includegraphics[width=1\textwidth]{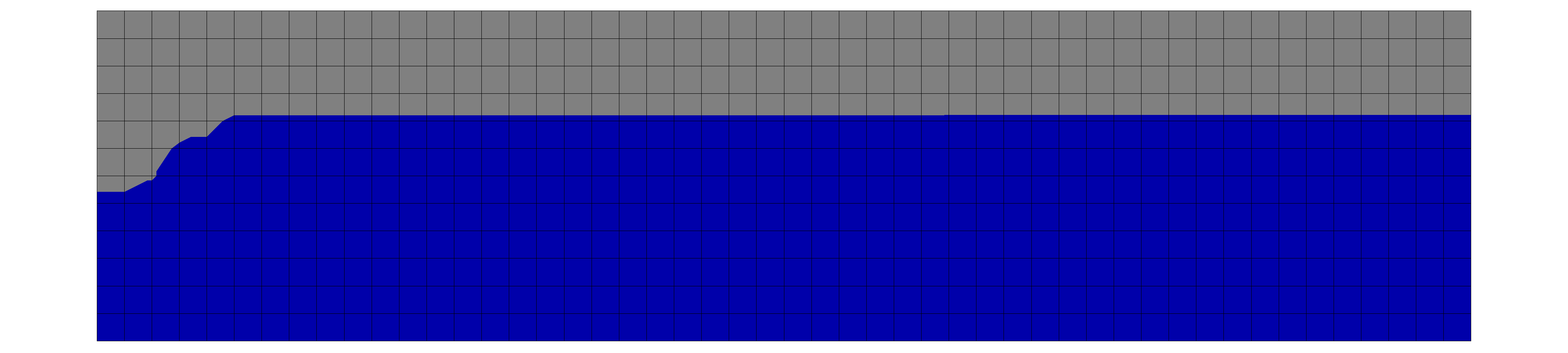}
    \caption{} \label{Fig: FEM Lambda Liq t=2,55}
  \end{subfigure}
  
  \vspace{0.2em}

  %%%% t=4s %%%%%%%%%%%%%%%%%%%%%%%%%%%%%%%%%%%%%%%%%
%  \begin{subfigure}{0.04\textwidth}
%    \centering
%    \adjustbox{raise ={5ex}{\height}}{\rotatebox{90}{\textbf{$t=\SI{4}{\second}$}}}
%  \end{subfigure}\hfill
%  \begin{subfigure}{0.30\textwidth}
%    \includegraphics[width=1\textwidth]{Figures/FEM/Lambda/Cry_t4-0.png}
%    \caption{} \label{Fig: FEM Lambda Cry t=4}
%  \end{subfigure}\hfill
%  \begin{subfigure}{0.30\textwidth}
%    \includegraphics[width=1\textwidth]{Figures/FEM/Lambda/Amo_t4-0.png}
%    \caption{} \label{Fig: FEM Lambda Amo t=4}
%  \end{subfigure}\hfill
%  \begin{subfigure}{0.30\textwidth}
%    \includegraphics[width=1\textwidth]{Figures/FEM/Lambda/Liq_t4-0.png}
%    \caption{} \label{Fig: FEM Lambda Liq t=4}
%  \end{subfigure}

%  \vspace{0.2em}

  %%%% legends %%%%%%%%%%%%%%%%%%%%%%%%%%%%%%%%%%%%%%
 \begin{subfigure}{0.04\textwidth}
    \quad % Leerraum über der Zeit-Spalte
  \end{subfigure}\hfill
  \begin{subfigure}{0.93\textwidth}
    \centering
    \input{Figures/FEM/Legend_0-1}
  \end{subfigure}
  
  \vspace{1em}
  \caption{FEM results for evolution of the phase transformation $\bflambda$ in the blue cross section plane \protect\tikz[baseline]{\protect\filldraw(0,0)[color=blue,opacity=0.2] rectangle(0.3,0.3) ;} at different times, i.e., $t=\SI{0}{\second}$ to $t=\SI{2.55}{\second}$. The gray color \protect\tikz[baseline]{\protect\filldraw(0,0)[color=gray] rectangle(0.3,0.3) ;} indicates the air phase.}
  \label{Fig: FEM Result Lambda}
\end{figure}

\begin{figure}
  \centering
    %%%% t=0.35s
    \begin{subfigure}{1\textwidth}
    \includegraphics[width=1\textwidth]{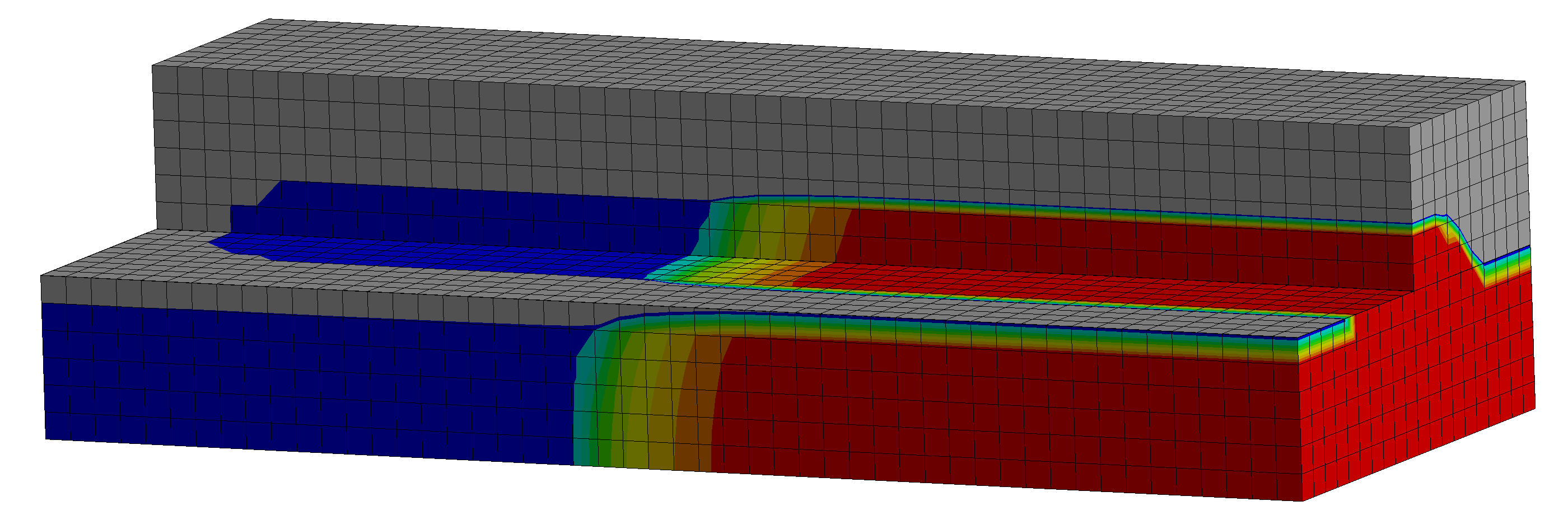}
    \end{subfigure}
    %%%% legends
    \begin{subfigure}{1\textwidth}
    \centering
    \input{Figures/FEM/Legend_0-1}
    \end{subfigure}
  \caption{Three-quarter sectional view of the liquid phase volume fraction at the end of the heat printing process $t=\SI{2.3}{\second}$, showing solidification parts of the printed track and the heterogeneous microstructure distribution. The color scale ranges from \protect\tikz[baseline]{\protect\filldraw(0,0)[color=blue] rectangle(0.3,0.3) ;} (fully solid) to \protect\tikz[baseline]{\protect\filldraw(0,0)[color=red] rectangle(0.3,0.3) ;} (fully liquid). \protect\tikz[baseline]{\protect\filldraw(0,0)[color=gray] rectangle(0.3,0.3) ;} color indicates the air phase.}
  \label{Fig: FEM Result Lambda2}
\end{figure}

The evolution of the microstructural state is a direct consequence of the thermal history described in the previous section. \autoref{Fig: FEM Result Lambda}  illustrates the spatio-temporal distribution of the crystalline, amorphous, and liquid volume fractions within the cross-sectional plane. The visual sequence concludes at $t=\SI{2.55}{\second}$ because the phase transformation is fully completed at this point; consequently, the microstructural state at the final simulation time of $t=\SI{4}{\second}$ is intentionally omitted, as it is visually identical to the state at $t=\SI{2.55}{\second}$. The simulation results reveal three critical kinetic regimes: substrate remelting, steady-state deposition, and vitrification:

\begin{enumerate}
    \item Substrate Remelting ($0 < t \leq \SI{0.5}{\second}$): During the initial static heating, the crystalline feedstock is destabilized. Crucially, \autoref{Fig: FEM Lambda Cry t=0,5} and \autoref{Fig: FEM Lambda Liq t=0,5} reveal that the phase transformation is not confined to the deposited material. The thermal energy penetrates the substrate surface, inducing a localized melting of the crystalline base. This remelting zone is vital for engineering applications, as it ensures a monolithic material bond between the printed track and the substrate, preventing delamination.
    
    \item Steady-State Deposition ($t > \SI{0.5}{\second}$): As the heat source traverses the domain, the material fully transforms into the liquid phase \autoref{Fig: FEM Lambda Cry t=2,15}. \autoref{Fig: FEM Result Lambda2} provides a volumetric view of this process at $t=\SI{2.3}{\second}$, revealing the characteristic elongated morphology of the melt pool. The \textit{comet-tail} shape, prominently visible as the red liquid phase region trailing the heat source in \autoref{Fig: FEM Result Lambda2}, is a manifestation of the material's thermal inertia and the finite speed of the deposition. Notably, the transition from liquid (\protect\tikz[baseline]{\protect\filldraw(0,0)[color=red] rectangle(0.3,0.3) ;}) to solid (\protect\tikz[baseline]{\protect\filldraw(0,0)[color=blue] rectangle(0.3,0.3) ;}) in \autoref{Fig: FEM Result Lambda} is not discrete. The intermediate gradient regions represent a thermodynamic mushy zone. In this non-equilibrium volume, the kinetic rates of transformation compete with the cooling rate, confirming the model's ability to capture the diffuse interface intrinsic to continuum phase field approximations.
    
    \item Vitrification and Latent Heat Absence: Upon cooling, the model predicts a complete divergence from the initial state. While the substrate remains crystalline \autoref{Fig: FEM Lambda Cry t=2,55}, the deposited track transforms entirely into an amorphous glass structure \autoref{Fig: FEM Lambda Amo t=2,55}. This confirms that the local cooling rates exceeded the critical threshold for crystallization. A significant thermodynamic observation here is the steepness of the vitrification front, which is prominently visible during the active cooling phase in \autoref{Fig: FEM Lambda Amo t=2,15} and \autoref{Fig: FEM Lambda Liq t=2,15}. Unlike crystallization solidification, where the release of latent heat buffers the temperature drop, the second-order glass transition involves no latent heat release. Consequently, the melt pool collapses rapidly into the vitreous state without thermal arrest, resulting in the homogenous amorphous structure observed in the final time step
\end{enumerate}

\subsubsection{Residual Stresses and Deformation}
\label{Sec: FEM Residual stresses and deformation}

\begin{figure}
  \centering
\begin{tikzpicture}
\begin{axis}[
    width=0.8\textwidth,
    height=0.4\textwidth,
    xlabel={Time [$\SI{}{\second}$]},
    ylabel={Stress $\sigma$ [$\SI{}{\mega\pascal}$]},
    grid=major,
    %xmode=log,
    enlarge x limits=false,
    legend pos=north west,
    %legend cell align={left},
    %legend style={font=\footnotesize},
    xmin=0, %xmax=500000,
    yticklabel style={
 /pgf/number format/fixed,
 /pgf/number format/precision=0,
 /pgf/number format/fixed zerofill,
 /pgf/number format/1000 sep={}
},
%        extra y ticks={1683.15, 2003.15},
%        extra y tick labels={$\theta^{\mathrm{g}}$,$\theta^{\mathrm{m}}$},
%        extra y tick style={
%            grid=major,
            %grid style={dashed}, % Gilt für alle extra ticks, man kann es auch pro tick setzen
%        },
]
	
% FEM Temperatur Minimum
\addplot[red, thick,solid, mark=none, dashed] table [x index=0,y index=1,col sep=semicolon] {Diagrams/FEM_Stress.txt};
\label{Line: FEM Stress Mises Max}
\addplot[red, thick,solid, mark=none] table [x index=0,y index=3,col sep=semicolon] {Diagrams/FEM_Stress.txt};
\label{Line: FEM Stress Hyd Max}
\addplot[blue, thick,solid, mark=none] table [x index=0,y index=2,col sep=semicolon] {Diagrams/FEM_Stress.txt};
\label{Line: FEM Stress Hyd Min}
\end{axis}

\begin{axis}[
            width=0.8\textwidth,
            height=0.4\textwidth,
            xlabel={},
            ylabel={Temperature $\theta$ [$\SI{}{\kelvin}$]},
            ylabel near ticks,
            ymin=300,
            ytick={\pgfkeysvalueof{/pgfplots/ymin}, \pgfkeysvalueof{/pgfplots/ymax}},
            axis y line*=right,
            axis x line=none,
            xmin=0,
            legend style={at={(0.05,-0.25)},anchor=north west,legend columns=-1
            },
               enlarge x limits=false,
            yticklabel style={/pgf/number format/1000 sep=,/pgf/number format/precision=0},
        extra y ticks={1683.15, 2003.15},
        extra y tick labels={$\theta^{\mathrm{g}}$,$\theta^{\mathrm{m}}$},
        extra y tick style={
            grid=major,
            grid style={dashed}, % Gilt für alle extra ticks, man kann es auch pro tick setzen
        },
            ]
            \addlegendimage{/pgfplots/refstyle=Line: FEM Stress Hyd Max}\addlegendentry{Line: FEM Stress Hyd Max};
            \addlegendimage{/pgfplots/refstyle=Line: FEM Stress Hyd Max}\addlegendentry{Line: FEM Stress Hyd Max};
            \addlegendimage{/pgfplots/refstyle=Line: FEM Stress Hyd Min}\addlegendentry{Line: FEM Stress Hyd Min};
        \addplot[black, solid, mark=none] table [x index=0,y index=2,col sep=semicolon] {Diagrams/FEM_Temperature.txt};
                %\draw [gray, dashed] ({axis cs:0, 2108.41}) -- ({axis cs:4e8, 2108.41});
\legend{$\sigma^{\mathrm{vM}}$ Max, $\sigma^{\mathrm{hyd}}$ Max, $\sigma^{\mathrm{hyd}}$ Min, $\theta$ Max};
        \end{axis}

\end{tikzpicture}
    \caption{Temporal evolution of the maximum von Mises stress $\sigma^{\mathrm{vM}}$ and the hydrostatic stresses $\sigma^{\mathrm{hyd}}$ within the entire computational domain $\Omega$. }
    \label{Fig: FEM Result Diagramm Stresses}
\end{figure}
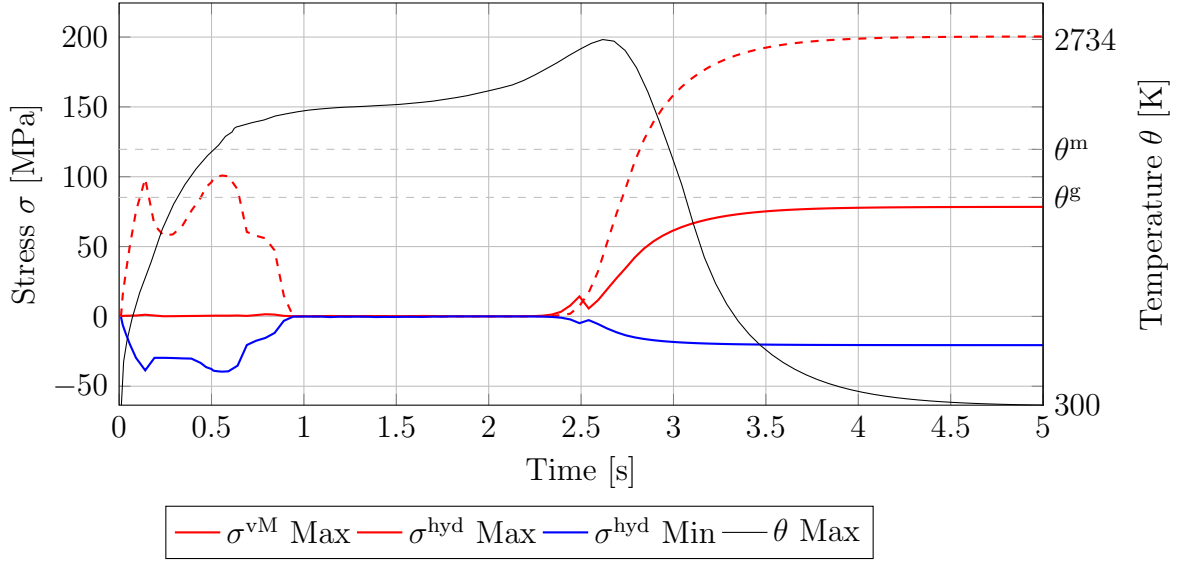

\begin{figure}
    \centering
%%%% t=0.39s%%%%%%%%%%%%%%%%%%%%%%%%
    \begin{subfigure}{0.30\textwidth}
    \includegraphics[width=1\textwidth]{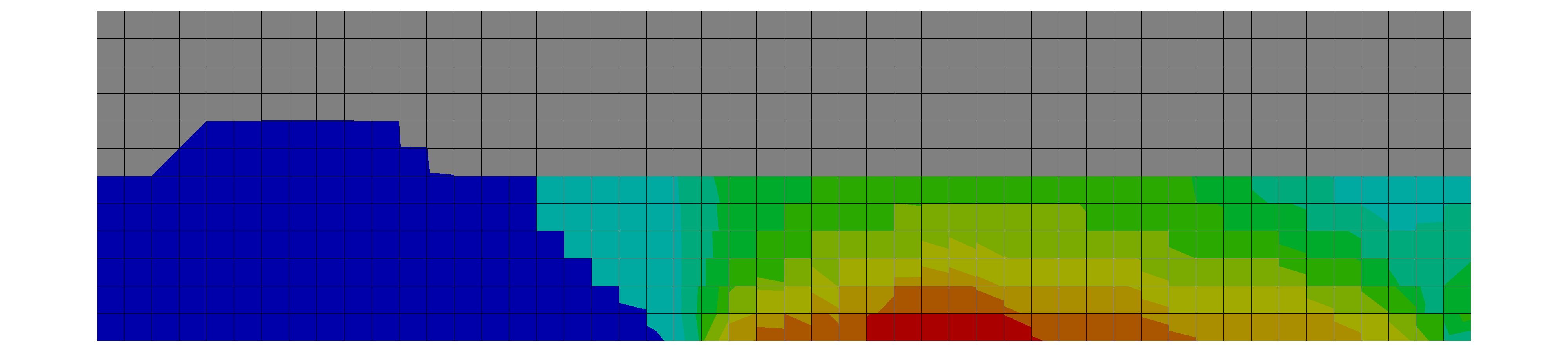} 
    \end{subfigure}
    \begin{subfigure}{0.30\textwidth}
    \centering
    \includegraphics[width=0.9\textwidth]{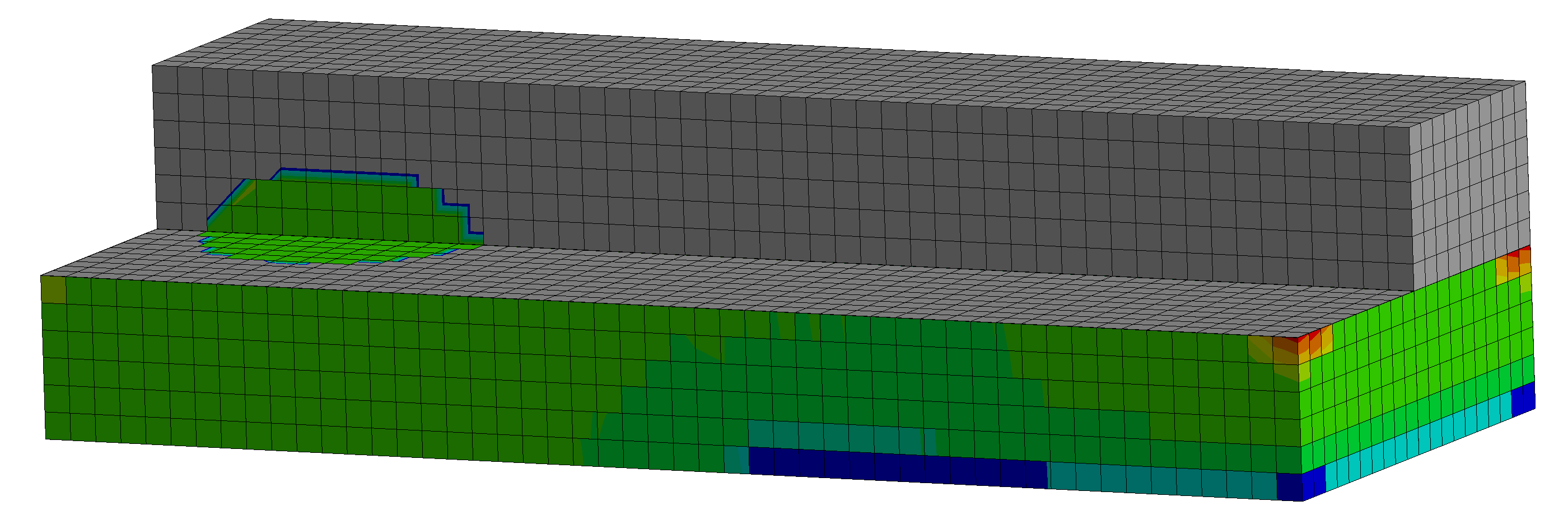}
    \end{subfigure}
    \begin{subfigure}{0.30\textwidth}
    \includegraphics[width=1\textwidth]{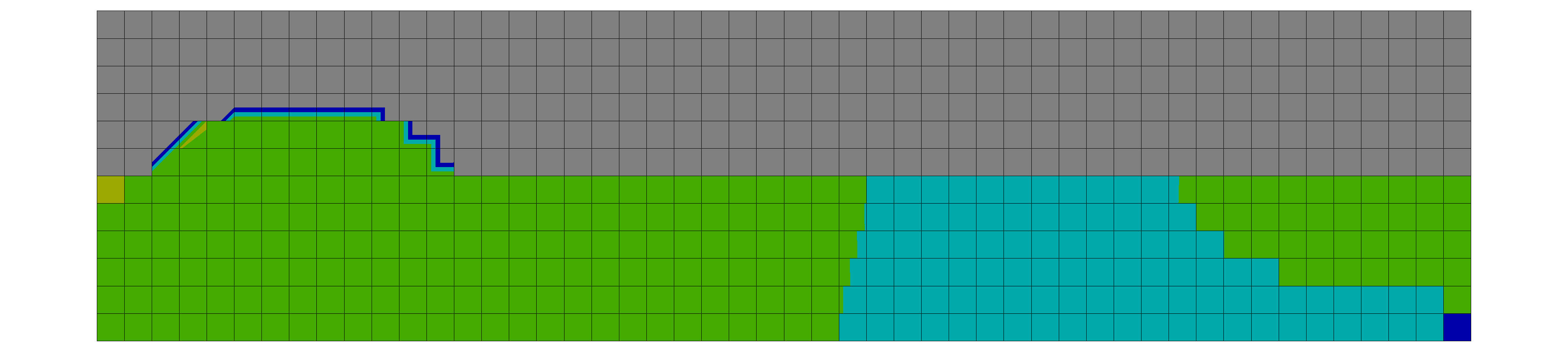}
    \end{subfigure}
    \hfill
    \begin{subfigure}{0.30\textwidth}
    \centering
    \input{Figures/FEM/Stress/Legend/Legend_Stress_t0-39}
    \caption{Von Mises stress $t=\SI{0.39}{\second}$} 
    \end{subfigure}
    \begin{subfigure}{0.62\textwidth}
    \centering
    \input{Figures/FEM/Stress/Legend/Legend_Hyd_t0-39}
    \caption{Hydrostatic stress $t=\SI{0.39}{\second}$} 
    \label{Fig: FEM hydrostatic stress t=0.39}
    \end{subfigure}
    \hfill  
%%%% t=0.5s%%%%%%%%%%%%%%%%%%%%%%%%
    \begin{subfigure}{0.30\textwidth}
    \includegraphics[width=1\textwidth]{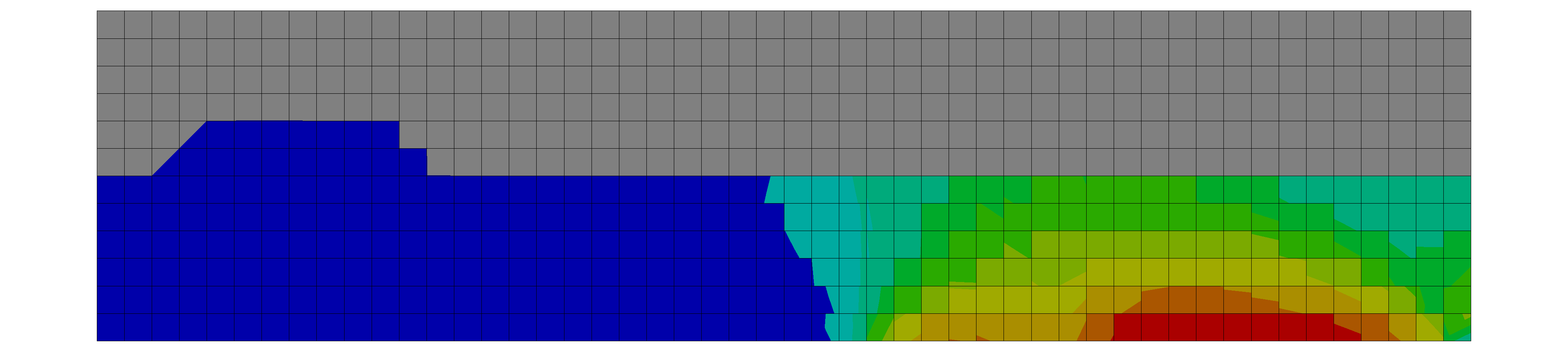} 
    \end{subfigure}
    \begin{subfigure}{0.30\textwidth}
    \centering
    \includegraphics[width=0.9\textwidth]{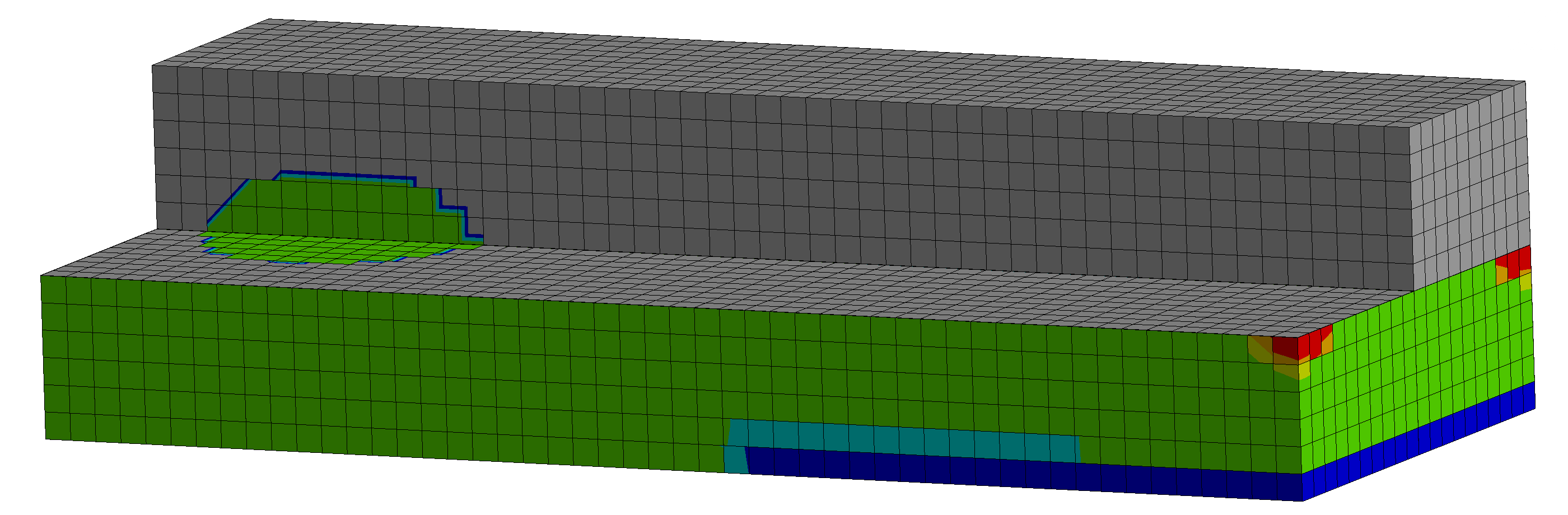}
    \end{subfigure}
    \begin{subfigure}{0.30\textwidth}
    \includegraphics[width=1\textwidth]{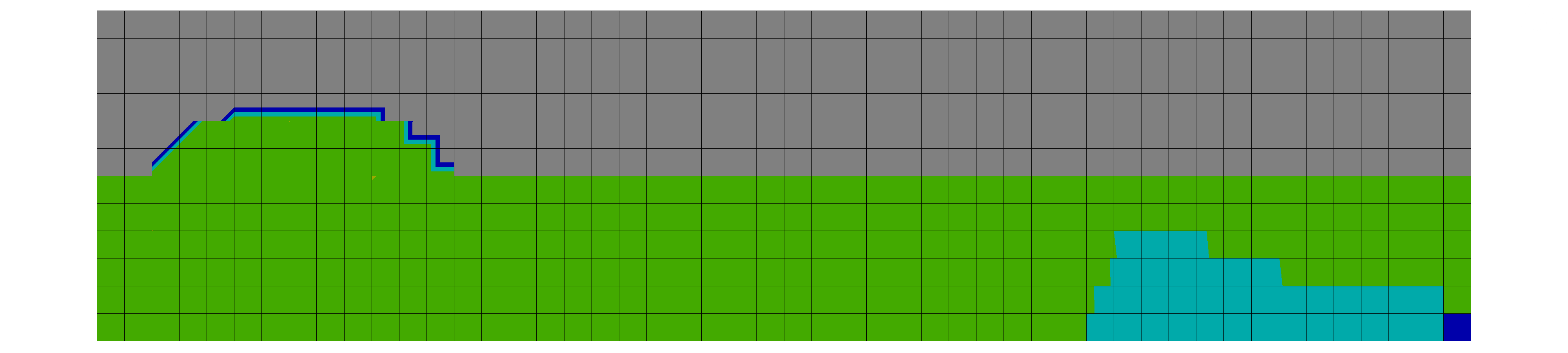}
    \end{subfigure}
    \hfill
    \begin{subfigure}{0.30\textwidth}
    \centering
    \input{Figures/FEM/Stress/Legend/Legend_Stress_t0-5}
    \caption{Von Mises stress $t=\SI{0.5}{\second}$} 
    \end{subfigure}
    \begin{subfigure}{0.62\textwidth}
    \centering
    \input{Figures/FEM/Stress/Legend/Legend_Hyd_t0-5}
    \caption{Hydrostatic stress $t=\SI{0.5}{\second}$} 
    \label{Fig: FEM hydrostatic stress t=0,5}
    \end{subfigure}
    \hfill
%%%% t=1.3s%%%%%%%%%%%%%%%%%%%%%%%%
    \begin{subfigure}{0.30\textwidth}
    \includegraphics[width=1\textwidth]{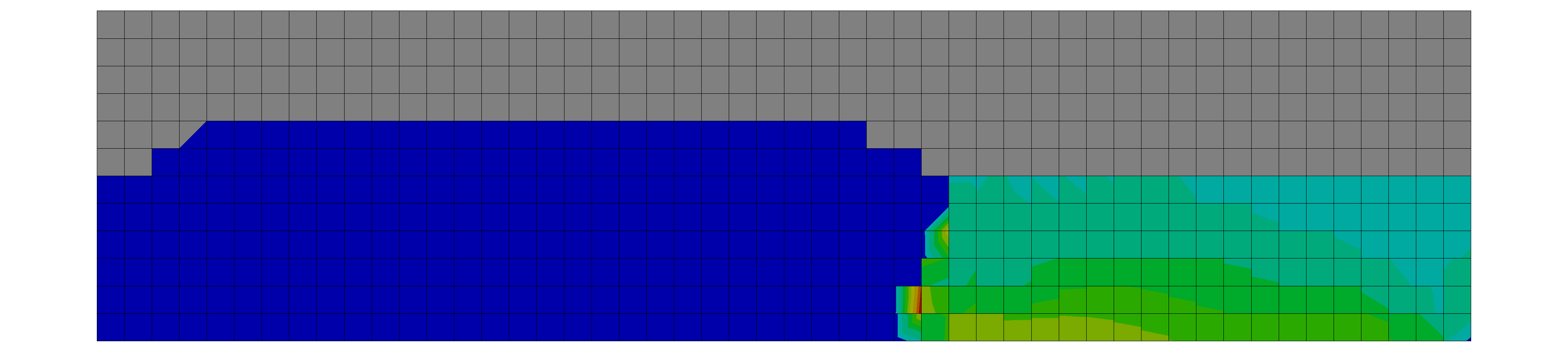} 
    \end{subfigure}
    \begin{subfigure}{0.30\textwidth}
    \centering
    \includegraphics[width=0.9\textwidth]{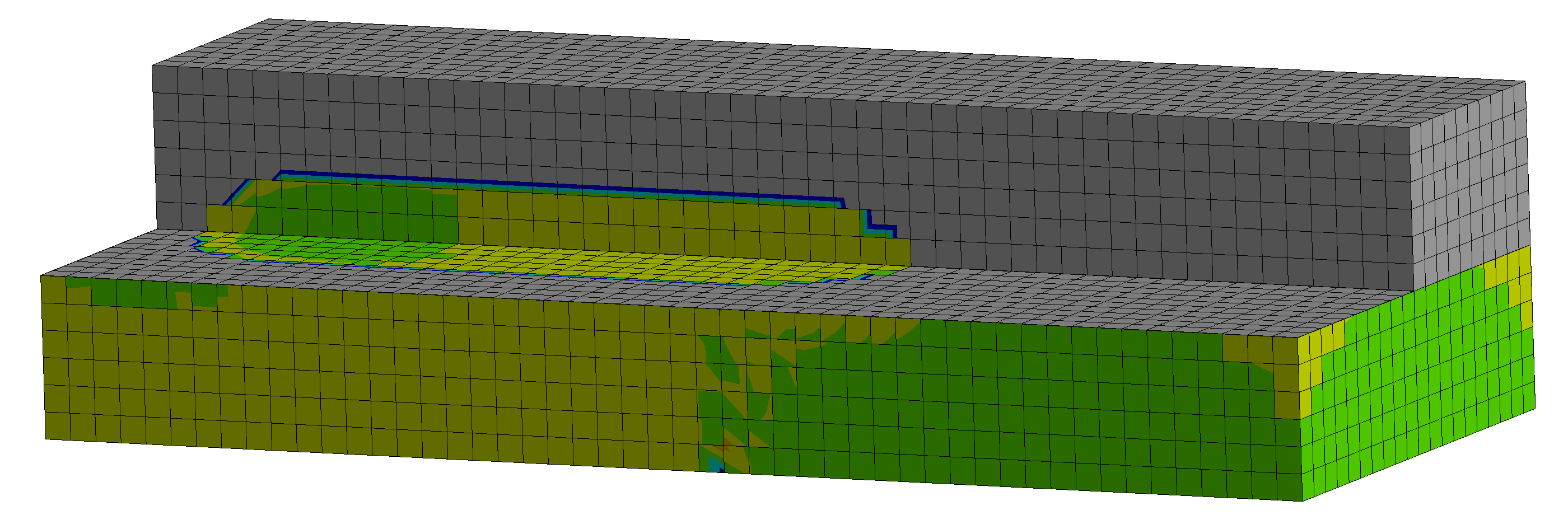}
    \end{subfigure}
    \begin{subfigure}{0.30\textwidth}
    \includegraphics[width=1\textwidth]{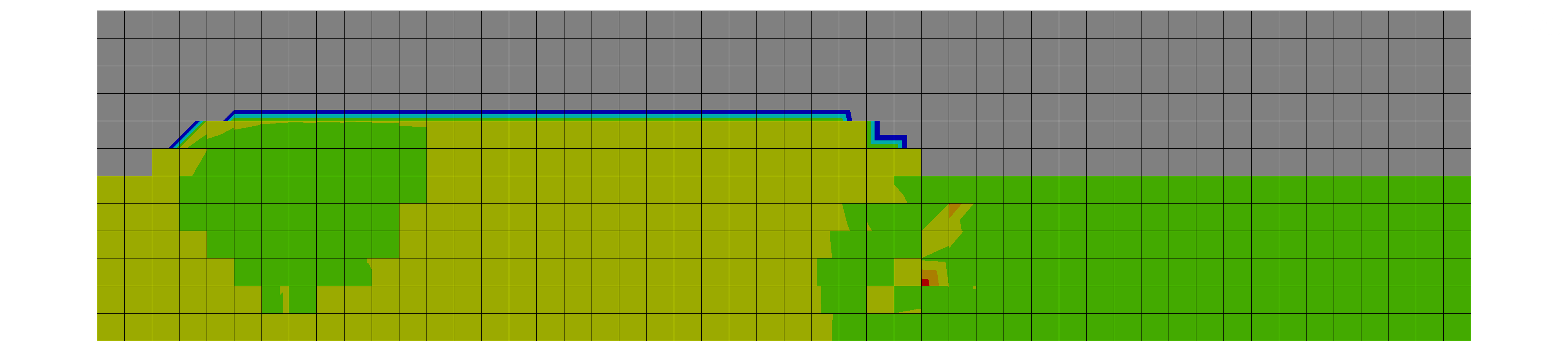}
    \end{subfigure}
    \hfill
    \begin{subfigure}{0.30\textwidth}
    \centering
    \input{Figures/FEM/Stress/Legend/Legend_Stress_t1-3}
    \caption{Von Mises stress $t=\SI{1.3}{\second}$} 
    \end{subfigure}
    \begin{subfigure}{0.62\textwidth}
    \centering
    \input{Figures/FEM/Stress/Legend/Legend_Hyd_t1-3}
    \caption{Hydrostatic stress $t=\SI{1.3}{\second}$} 
    \label{Fig: FEM hydrostatic stress t=1.3}
    \end{subfigure}
    \hfill
%%%% t=2.15s%%%%%%%%%%%%%%%%%%%%%%%%
    \begin{subfigure}{0.30\textwidth}
    \includegraphics[width=1\textwidth]{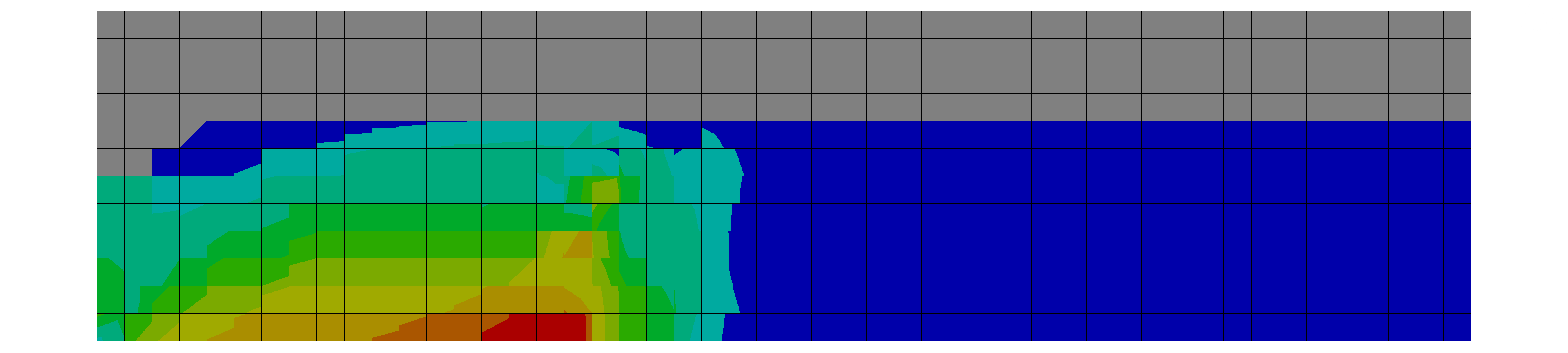} 
    \end{subfigure}
    \begin{subfigure}{0.30\textwidth}
    \centering
    \includegraphics[width=0.9\textwidth]{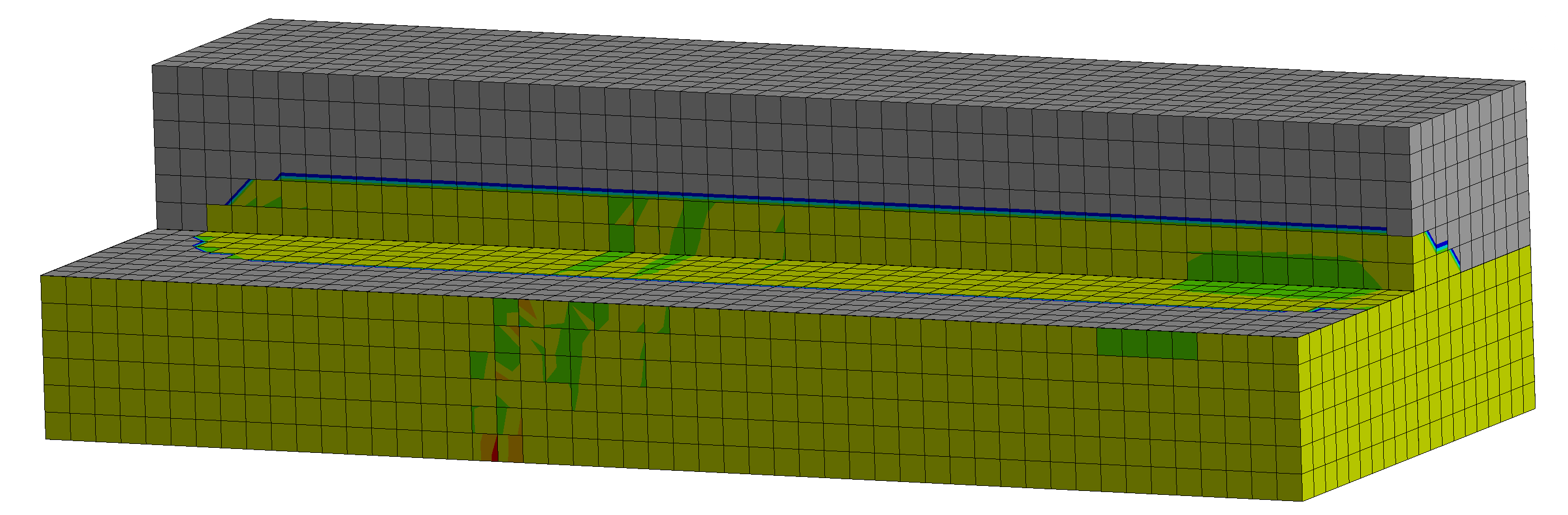}
    \end{subfigure}
    \begin{subfigure}{0.30\textwidth}
    \includegraphics[width=1\textwidth]{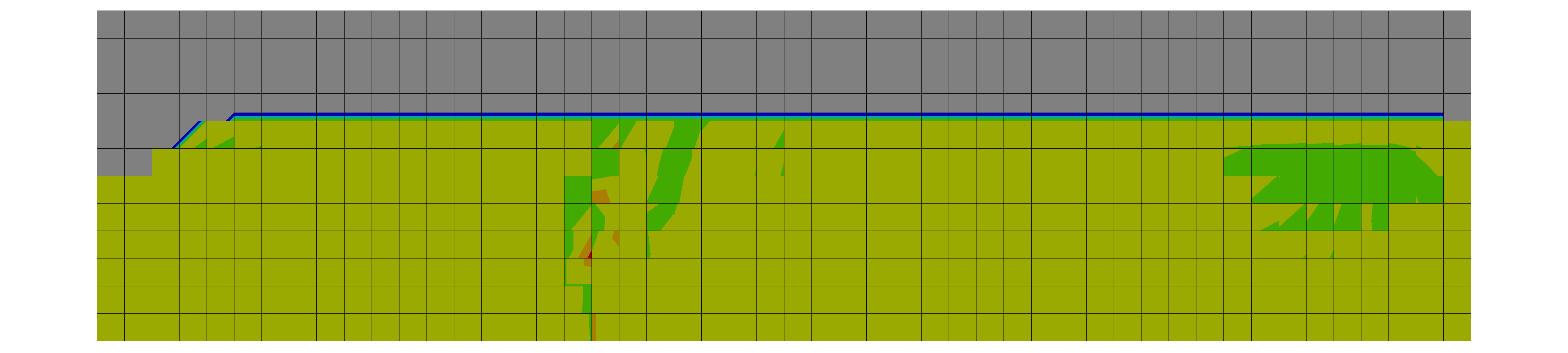}
    \end{subfigure}
    \hfill
    \begin{subfigure}{0.30\textwidth}
    \centering
    \input{Figures/FEM/Stress/Legend/Legend_Stress_t2-15}
    \caption{Von Mises stress $t=\SI{2.15}{\second}$} 
    \end{subfigure}
    \begin{subfigure}{0.62\textwidth}
    \centering
    \input{Figures/FEM/Stress/Legend/Legend_Hyd_t2-15}
    \caption{Hydrostatic stress $t=\SI{2.15}{\second}$} 
    \label{Fig: FEM hydrostatic stress t=2.15}
    \end{subfigure}
    \hfill
%%%% t=2.55s%%%%%%%%%%%%%%%%%%%%%%%%
    \begin{subfigure}{0.30\textwidth}
    \includegraphics[width=1\textwidth]{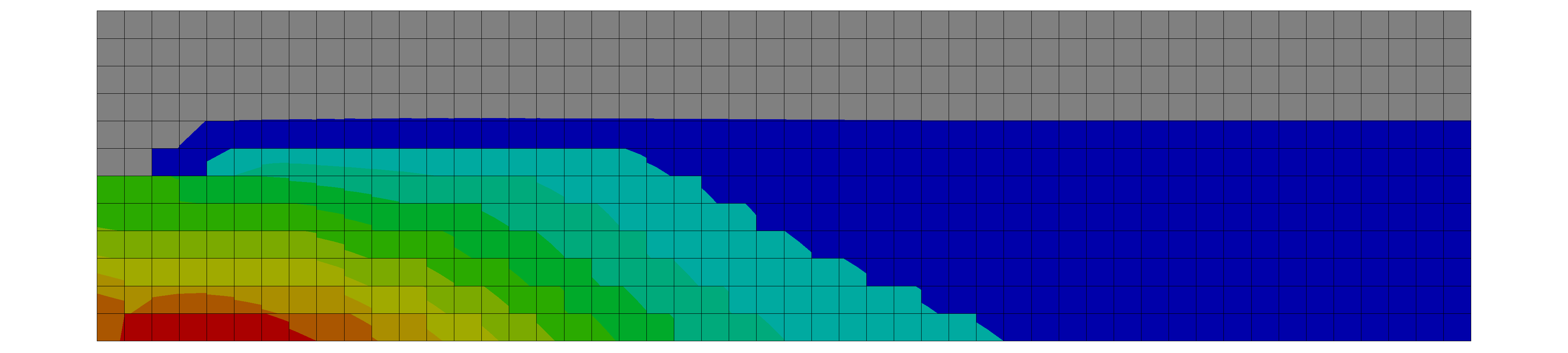} 
    \end{subfigure}
    \begin{subfigure}{0.30\textwidth}
    \centering
    \includegraphics[width=0.9\textwidth]{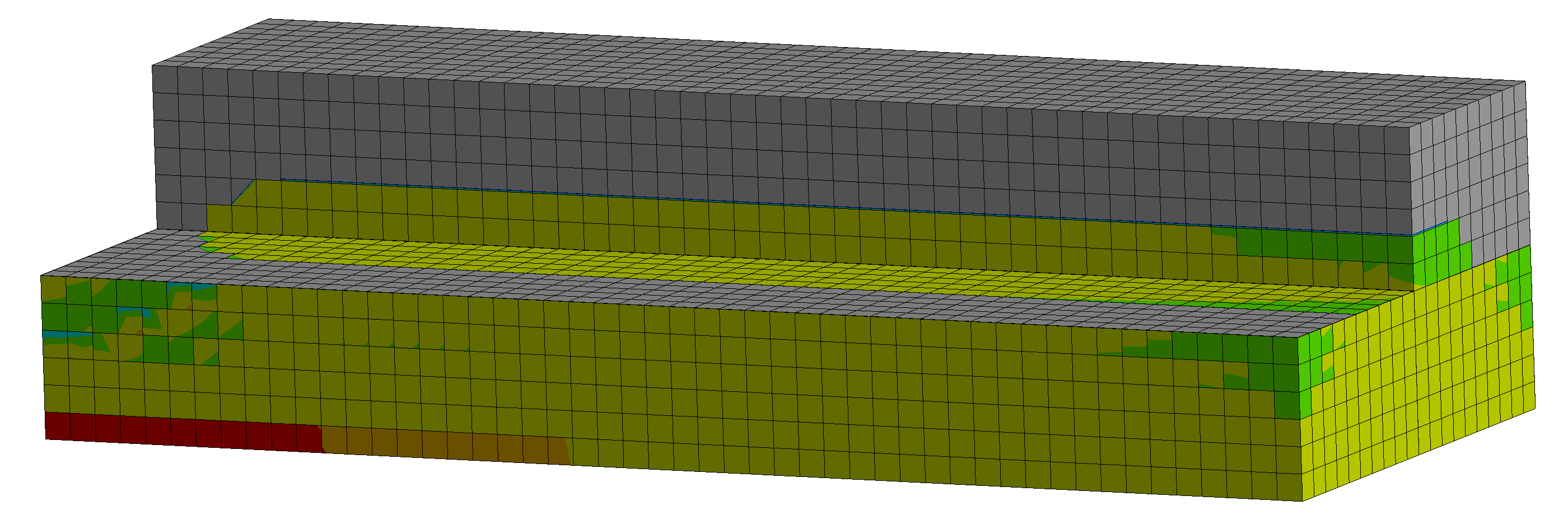}
    \end{subfigure}
    \begin{subfigure}{0.30\textwidth}
    \includegraphics[width=1\textwidth]{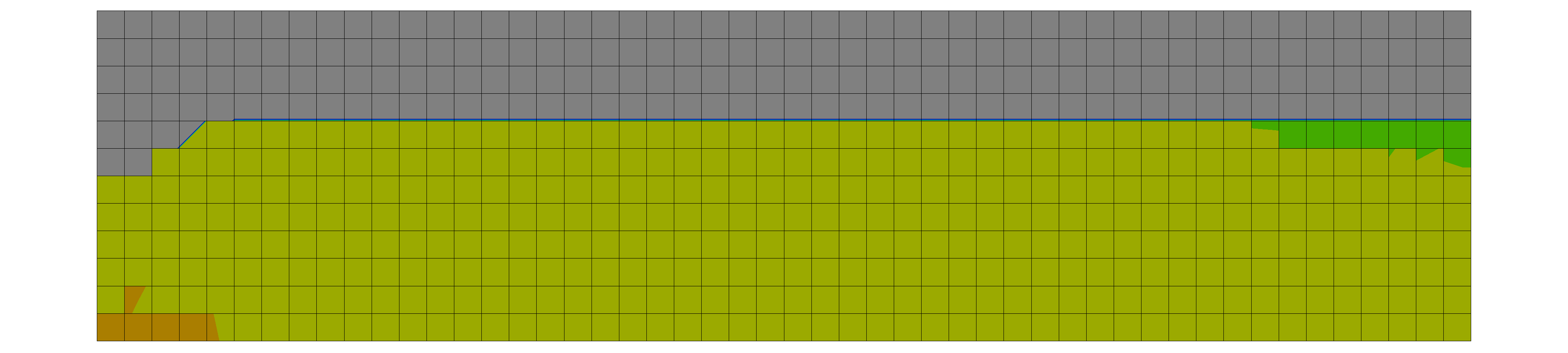}
    \end{subfigure}
    \hfill
    \begin{subfigure}{0.30\textwidth}
    \centering
    \input{Figures/FEM/Stress/Legend/Legend_Stress_t2-55}
    \caption{Von Mises stress $t=\SI{2.55}{\second}$} 
    \end{subfigure}
    \begin{subfigure}{0.62\textwidth}
    \centering
    \input{Figures/FEM/Stress/Legend/Legend_Hyd_t2-55}
    \caption{Hydrostatic stress $t=\SI{2.55}{\second}$} 
    \label{Fig: FEM hydrostatic stress t=2.55}
    \end{subfigure}
    \hfill
%    \end{figure}
%%%% t=4s%%%%%%%%%%%%%%%%%%%%%%%%
%\clearpage   
%\begin{figure}[H]\ContinuedFloat
%    \centering    
    \begin{subfigure}{0.30\textwidth}
    \includegraphics[width=1\textwidth]{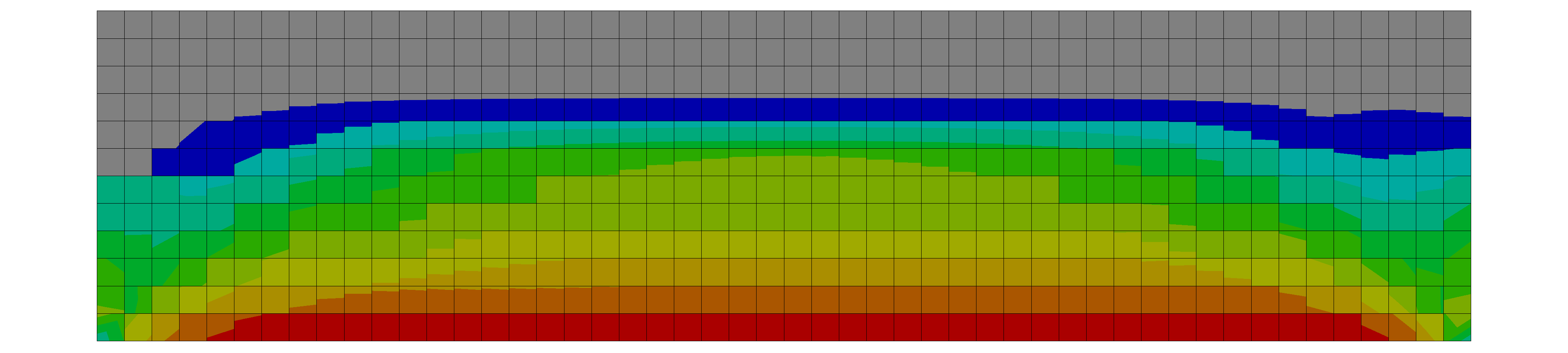} 
    \end{subfigure}
    \begin{subfigure}{0.30\textwidth}
    \centering
    \includegraphics[width=0.9\textwidth]{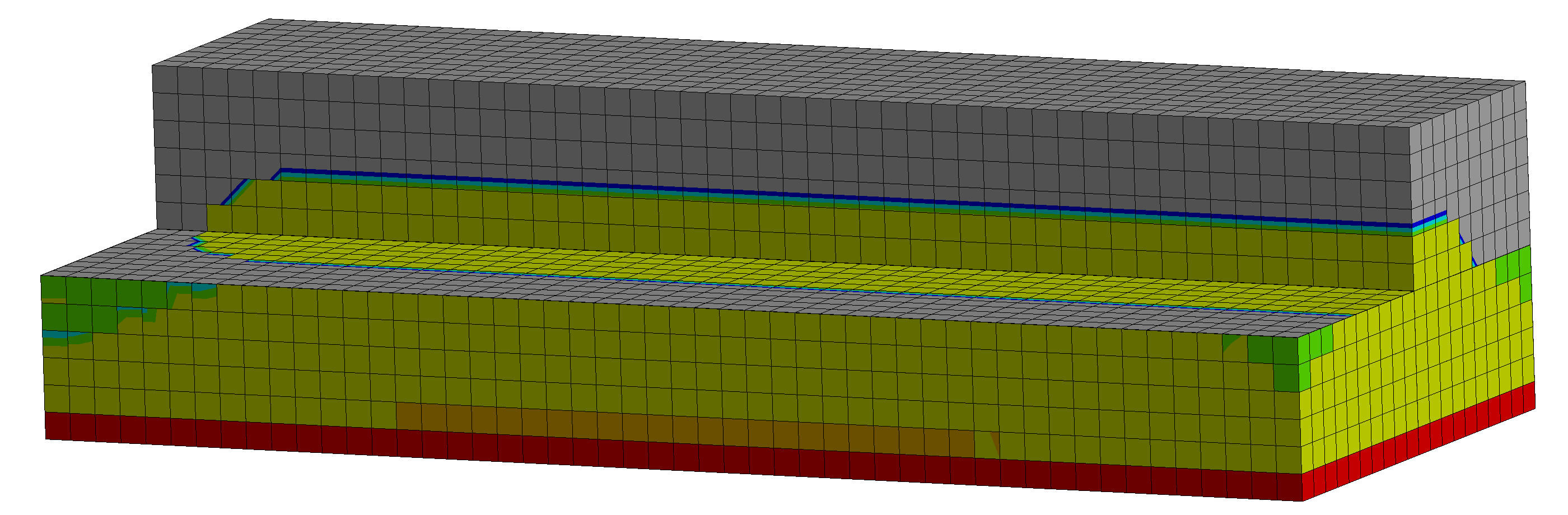}
    \end{subfigure}
    \begin{subfigure}{0.30\textwidth}
    \includegraphics[width=1\textwidth]{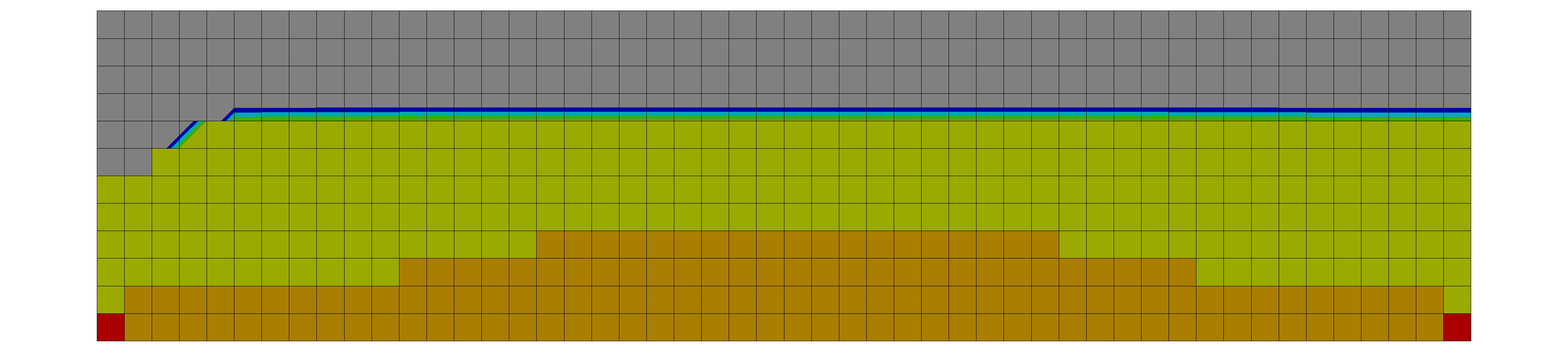}
    \end{subfigure}
    \hfill
    \begin{subfigure}{0.30\textwidth}
    \centering
    \input{Figures/FEM/Stress/Legend/Legend_Stress_t4-0}
    \caption{Von Mises stress $t=\SI{4}{\second}$} 
    \end{subfigure}
    \begin{subfigure}{0.62\textwidth}
    \centering
    \input{Figures/FEM/Stress/Legend/Legend_Hyd_t4-0}
    \caption{Hydrostatic stress $t=\SI{4}{\second}$} 
    \label{Fig: FEM hydrostatic stress t=4}
    \end{subfigure}
    \hfill
    \caption{FEM results for von Mises stress hydrostatic stress $\sigma^{\mathrm{hyd}}$ in the blue cross section plane \protect\tikz[baseline]{\protect\filldraw(0,0)[color=blue,opacity=0.2] rectangle(0.3,0.3) ;} and the von Mises stress in bottom view at different times, i.e. $t=\SI{0}{\second}$ to $t=\SI{300}{\second}$. The gray color \protect\tikz[baseline]{\protect\filldraw(0,0)[color=gray] rectangle(0.3,0.3) ;} indicates the air phase.}
    \label{Fig: FEM Result Stress}
\end{figure}

The mechanical response is governed by the intricate interplay between thermal contraction, phase-change induced density variations, and the relaxation capability of the viscous deformation. It is important to note that all stress values presented in this section represent unaveraged Gauss point results, ensuring that local extrema within the finite elements are captured without smoothing artifacts. \autoref{Fig: FEM Result Stress} and \autoref{Fig: FEM Result Diagramm Stresses} presents the evolution of the von Mises equivalent stress $\sigma^{\mathrm{vM}}$
\begin{equation}
    \sigma^{\mathrm{vM}}=\sqrt{\frac{\left(\sigma_{11}-\sigma_{22}\right)^2+\left(\sigma_{22}-\sigma_{33}\right)^2+\left(\sigma_{33}-\sigma_{11}\right)^2}{2}} \quad ,
\end{equation}
alongside the hydrostatic stress $\sigma^{\mathrm{hyd}}$ as
\begin{equation}
   \sigma^\mathrm{hyd}=\frac{1}{3}\left(\sigma_{11}+\sigma_{22}+\sigma_{33}\right) \,. 
\end{equation}

The history of the stress state, correlated with the temperature data, reveals three mechanically distinct regimes (\autoref{Fig: FEM Result Diagramm Stresses}):
\begin{enumerate}
    \item Heating and Initial Constraint ($0 < t \leq \SI{0.5}{\second}$): In the initial phase, the laser dwells at the starting point. The local temperature rises rapidly. The rapid thermal expansion of the heated zone is kinematically constrained by the surrounding cold substrate. Since the material is still below the melting point in the far-field, the viscosity remains high, preventing relaxation. This leads to an initial spike in the von Mises stress, driven essentially by the elastic mismatch between the expanding hot spot and the rigid clamp.
    \item Viscous Relaxation ($0.5 < t \leq \SI{2.3}{\second}$): During the deposition process, the local temperature in the process zone consistently exceeds the glass transition temperature. Consequently, the viscosity drops to its minimum limit. Despite the extreme thermal gradients, the stresses in the melt pool relax almost instantaneously. The  relaxation time in this regime is smaller than the process time, allowing the viscous deformation FCv to accommodate the expansive strains fully.
    \item Quenching and Densification ($t > \SI{2.3}{\second}$): The critical mechanical process occurs during cooling. As the temperature drops, the viscosity increases exponentially, freezing the microstructure. From this point, two distinct physical mechanisms drive the accumulation of residual stress:
    \begin{itemize}
        \item Thermal Contraction: The shrinkage due to the temperature drop.
        \item Phase-Specific Densification: The material undergoes a volumetric change as the liquid structure rearranges into the amorphous solid. Since viscous relaxation is no longer possible due to the high viscosity, these combined volumetric contractions are locked in as elastic residual stresses.
    \end{itemize}
\end{enumerate}

The spatial distribution of the von Mises stress (\autoref{Fig: FEM Result Stress}) highlights the influence of boundary conditions. The maximum equivalent stress is observed at the very end of the simulation ($t=\SI{4}{\second}$) and is predominantly located within the solid phases. The stress concentration reaches its peak at the bottom fixation. This arises because the substrate is perfectly constrained; as the entire component cools and attempts to shrink, the reaction forces at the clamp generate peak shear stresses.

To understand the volumetric nature of these forces, we analyze the hydrostatic stress distribution shown in the images below. The contour plots follow this convention:
\begin{itemize}
    \item The warm colors (\protect\tikz[baseline]{\protect\filldraw(0,0)[color=ANSYS_Red] rectangle(0.3,0.3) ;}, \protect\tikz[baseline]{\protect\filldraw(0,0)[color=ANSYS_Orange] rectangle(0.3,0.3) ;}, \protect\tikz[baseline]{\protect\filldraw(0,0)[color=ANSYS_Yellow] rectangle(0.3,0.3) ;}) indicate positive values, corresponding to hydrostatic tension.
    \item The cool colors (\protect\tikz[baseline]{\protect\filldraw(0,0)[color=ANSYS_Blue] rectangle(0.3,0.3) ;}, \protect\tikz[baseline]{\protect\filldraw(0,0)[color=ANSYS_LightBlue] rectangle(0.3,0.3) ;}, \protect\tikz[baseline]{\protect\filldraw(0,0)[color=ANSYS_Green] rectangle(0.3,0.3) ;}) indicate negative values, corresponding to hydrostatic pressure (compression).
\end{itemize}
Based on this color-coding, the temporal evolution of the hydrostatic stress state reveals the following characteristic regimes:

\begin{enumerate}
    \item Static heating ($0 < t \leq \SI{0.5}{\second}$): As the laser dwells at the starting point, a state dominated by hydrostatic compression emerges. The rapidly expanding hot spot is kinematically constrained by the surrounding cold substrate. Consequently, the material within the process zone is unable to expand freely and experiences significant hydrostatic pressure, while tensile reaction forces in the far field remain negligible.
    
    \item Steady-state deposition ($0.5 < t \leq \SI{2.3}{\second}$): The relaxation capability of the material becomes evident. The "mushy zone" and the liquid melt pool are clearly visible as stress-free regions, confirming that the liquid phase cannot support deviatoric or hydrostatic stresses due to the low viscosity.

    \item Quenching and Densification ($t > \SI{2.3}{\second}$): Following the solidification, the stress state inverts into a mixed mode with predominant tensile contributions. The deposited track, undergoing both thermal contraction and phase-change densification, pulls against the rigid substrate. This generates high hydrostatic tension within the vitreous layer. 
\end{enumerate}

\begin{figure}
    \centering
%%%%%%%% t=0,39and t=0,5 and t=1,3%%%%%%
    \begin{subfigure}{0.31\textwidth}
    \includegraphics[width=1\textwidth]{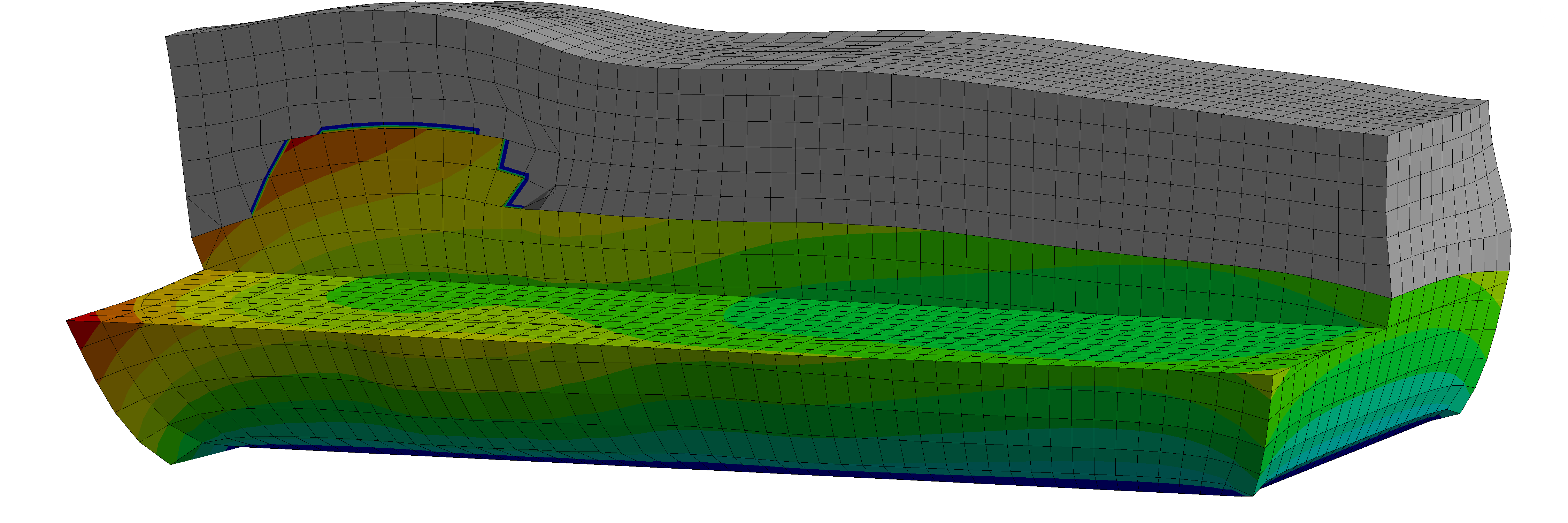}
    \end{subfigure}
    \begin{subfigure}{0.31\textwidth}
    \includegraphics[width=1\textwidth]{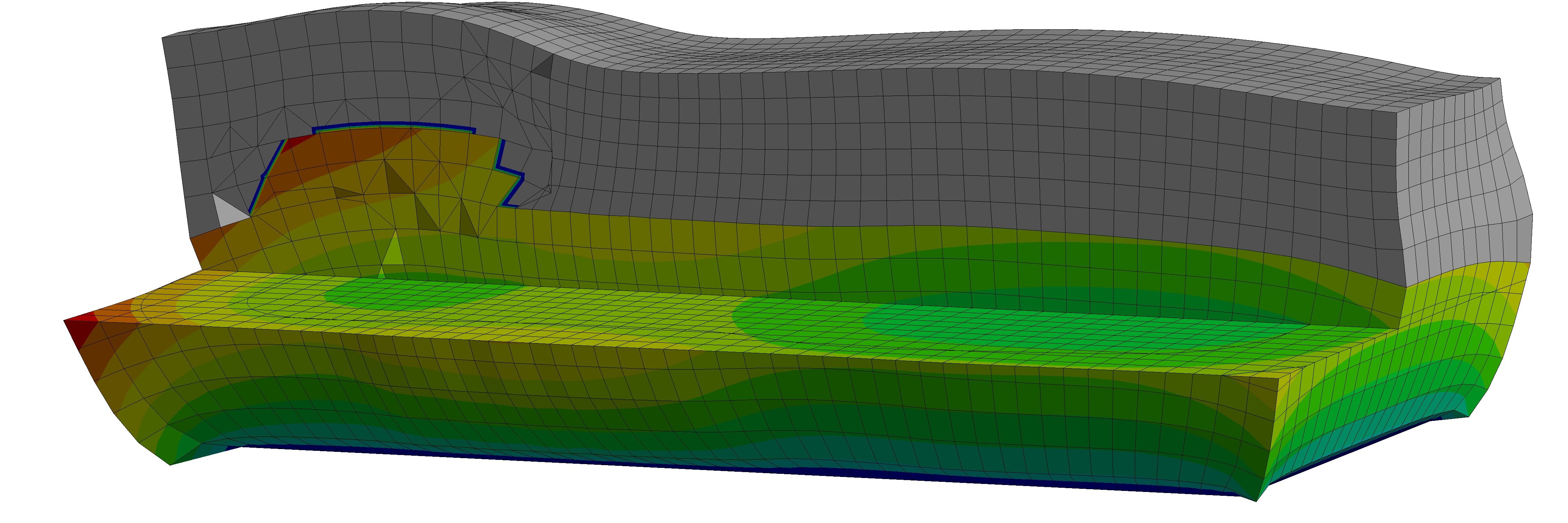}
    \end{subfigure}
    \begin{subfigure}{0.31\textwidth}
    \includegraphics[width=1\textwidth]{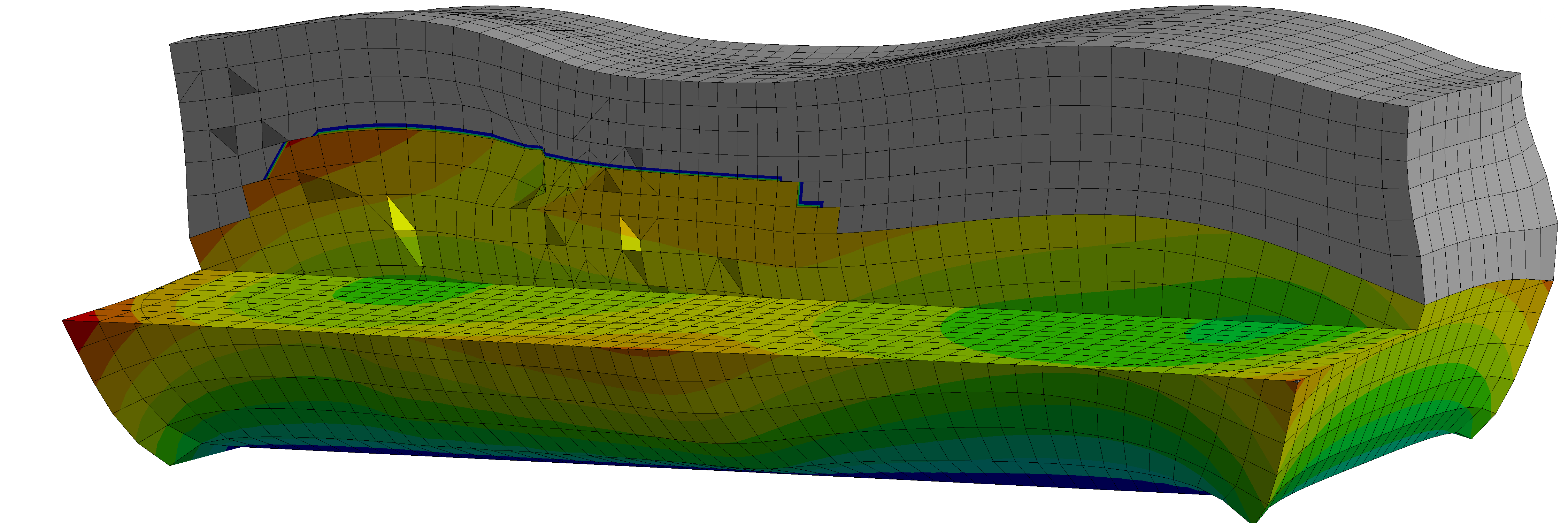}
    \end{subfigure}
    \hfill
    \begin{subfigure}{0.31\textwidth}
    \centering
    \input{Figures/FEM/Deformation/Legend/Legend_Defo_t0-39}
    \caption{Deformation $t=\SI{0.39}{\second}$} 
    \end{subfigure}
    \begin{subfigure}{0.31\textwidth}
    \centering
    \input{Figures/FEM/Deformation/Legend/Legend_Defo_t0-5}
    \caption{Deformation $t=\SI{0.5}{\second}$} 
    \end{subfigure}
    \begin{subfigure}{0.31\textwidth}
    \centering
    \input{Figures/FEM/Deformation/Legend/Legend_Defo_t1-3}
    \caption{Deformation $t=\SI{1.3}{\second}$} 
    \end{subfigure}
    \hfill
%%%%%%%% t=2,15, t=2,55 and t=4 
    \begin{subfigure}{0.31\textwidth}
    \includegraphics[width=1\textwidth]{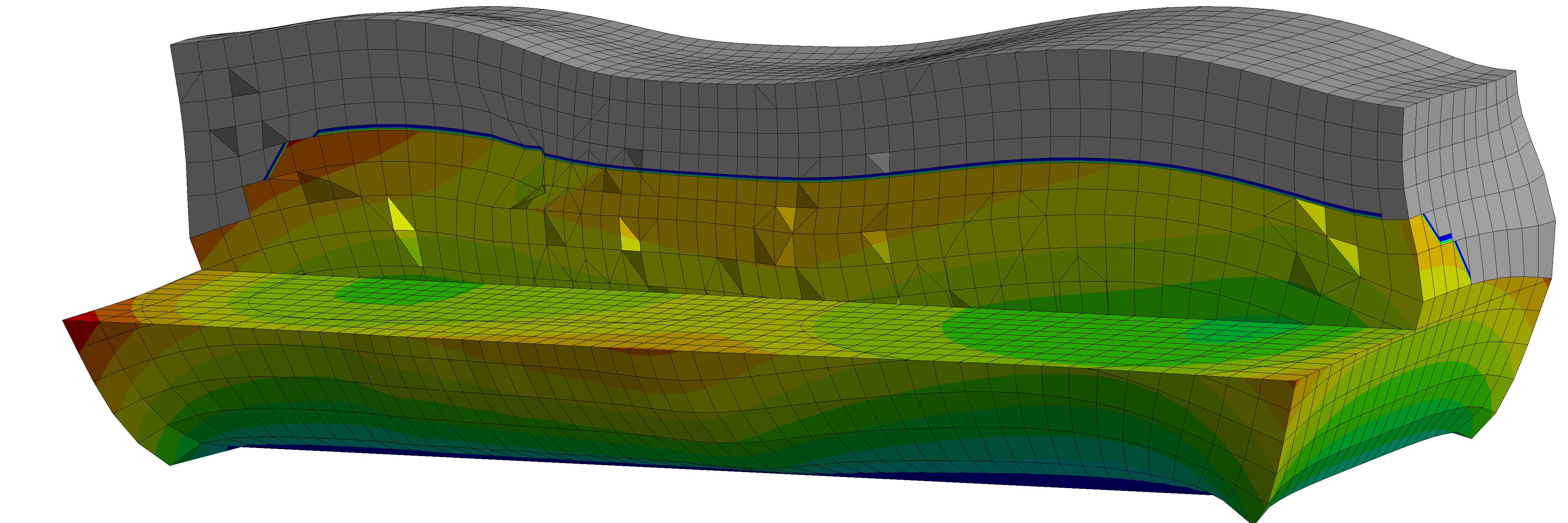}
    \end{subfigure}
    \begin{subfigure}{0.31\textwidth}
    \includegraphics[width=1\textwidth]{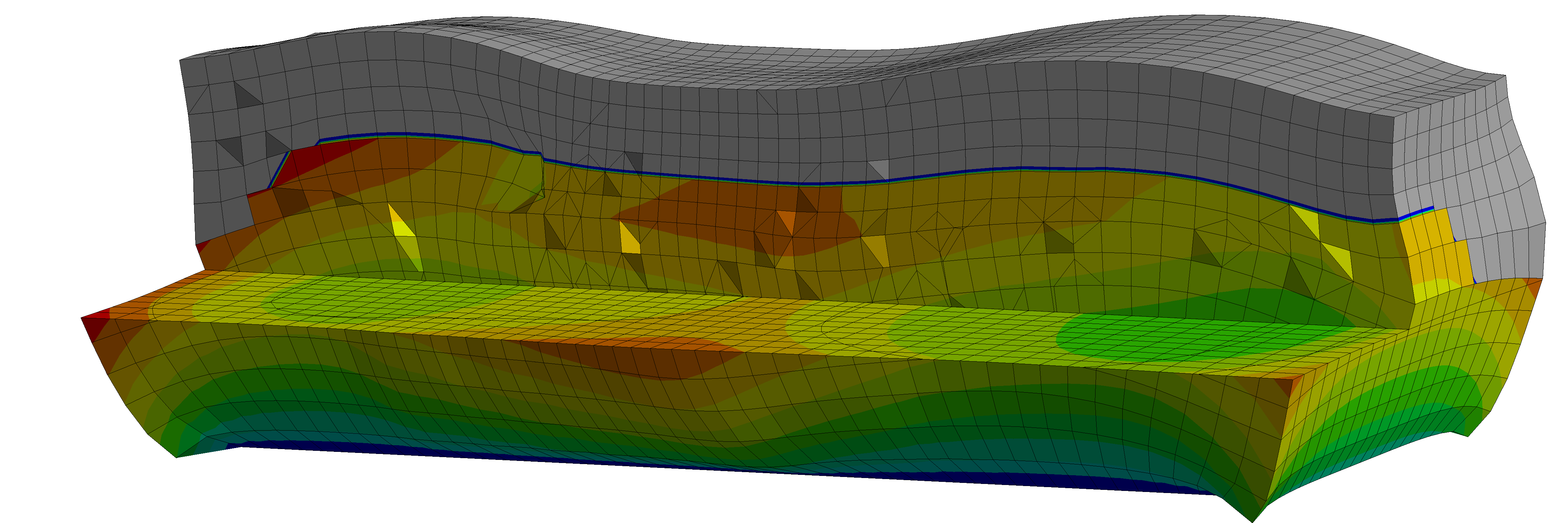}
    \end{subfigure}
    \hfill
    \begin{subfigure}{0.31\textwidth}
    \includegraphics[width=1\textwidth]{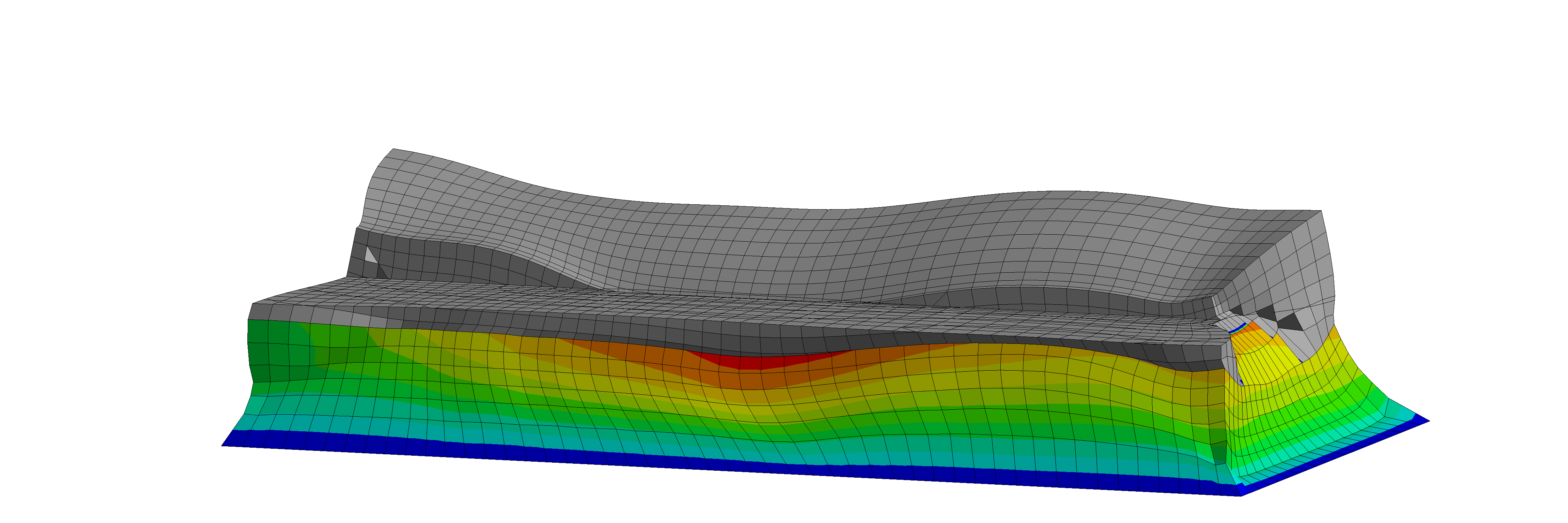}
    \end{subfigure}
    \hfill
    \begin{subfigure}{0.31\textwidth}
    \centering
    \input{Figures/FEM/Deformation/Legend/Legend_Defo_t2-15}
    \caption{Deformation $t=\SI{2.15}{\second}$} 
    \end{subfigure}
    \begin{subfigure}{0.31\textwidth}
    \centering
    \input{Figures/FEM/Deformation/Legend/Legend_Defo_t2-55}
    \caption{Deformation $t=\SI{2.55}{\second}$} 
    \end{subfigure}
    \begin{subfigure}{0.31\textwidth}
    \centering
    \input{Figures/FEM/Deformation/Legend/Legend_Defo_t4-0}
    \caption{Deformation $t=\SI{4}{\second}$} 
    \end{subfigure}
    \hfill

    \caption{FEM results for the Deformation with a scale factor of 1000 at different times, i.e. $t=\SI{0}{\second}$ to $t=\SI{4}{\second}$. The gray color \protect\tikz[baseline]{\protect\filldraw(0,0)[color=gray] rectangle(0.3,0.3) ;} indicates the air phase.}
    \label{Fig: FEM Result Defo}
\end{figure}

The temporal evolution of the displacement field \autoref{Fig: FEM Result Defo} reveals two opposing kinematic mechanisms.
During the heating and deposition phase ($t \leq \SI{2.3}{\second}$), a distinct thermal expansion is observable. The material within the heat-affected zone expands volumetrically, pushing against the cooler surroundings. This results in a temporary, localized swelling of the structure in the process zone.
However, the final deformation state ($t = \SI{4}{\second}$) is dominated by the cooling phase. As the component returns to ambient temperature, the solidified top layer undergoes significantly greater contraction than the substrate. This differential shrinkage generates an eccentric contractile force relative to the neutral axis. Consequently, a global bending moment is induced, causing the structure to collapse inward and the edges of the prism to curl upward (warpage). The final deflection amplitude of $\SI{0.132}{\mu\metre}$ serves as a quantitative measure of the accumulated eigenstrains that could not be dissipated through viscous relaxation.

\section{Conclusion and Outlook}\label{chapt:Conclusion}
This work presented a comprehensive multi-physics framework for the simulation of glass-forming materials, specifically tailored to capture the non-equilibrium conditions inherent to additive manufacturing processes. By utilizing the extended Hamilton principle, a unified variational formulation was derived that rigorously couples thermal conduction, finite strain mechanics, viscoelastic deformation, and the phase transformations between crystalline, amorphous, and liquid states. This variational approach ensured thermodynamic consistency throughout the modeling of the distinct first-order melting and second-order glass transitions.

The numerical implementation, facilitated by the Neighbored Element Method (NEM) for the thermal field and a robust finite element integration for the mechanical response, proved highly efficient in handling the strong non-linearities of the coupled system. 

The material point simulations validated the model’s kinetic behavior, accurately reproducing the rate-dependent competition between crystallization and vitrification characteristic of Time-Temperature-Transformation (TTT) diagrams. Furthermore, the transition from local kinetics to macroscopic boundary value problems revealed the critical impact of the thermal history on structural integrity.

The three-dimensional rectangular prism simulation elucidated how the rapid increase in viscosity at the glass transition temperature leads to the freezing of inelastic deformation. The results highlighted that residual stresses are not merely a result of thermal contraction, but are significantly driven by the phase-specific densification during vitrification. The analysis of the hydrostatic stress state revealed a critical inversion from compressive stresses during the heating phase to significant hydrostatic tension in the vitrified track upon cooling. This tensile state, identified in the final solidified structure, poses a direct risk for cold cracking. Additionally, the model successfully predicted the macroscopic warpage resulting from the eccentric contractile forces, demonstrating its capability to capture the complex spatial distribution of phases-ranging from a fully vitrified structure to residual crystallinity in the substrate, and the resulting geometric distortions.

Future research will focus on the experimental validation of the predicted residual stress distributions and the resultant optical birefringence in printed components. Additionally, the framework offers a promising basis for integrating fracture mechanics to predict cold cracking probabilities, or for optimizing process parameters to tailor the refractive index profiles in functionally graded glass elements.

\section*{Acknowledgments}
The authors gratefully acknowledge the funding by the German Research Foundation (Deutsche Forschungsgemeinschaft, DFG) under Germany's Excellence Strategy within the Cluster of Excellence PhoenixD 2 (Project ID 390833453).

\appendix

\addcontentsline{toc}{chapter}{Bibliography}
\bibliographystyle{plain}
\bibliography{bib}

\end{document}